\documentclass[apj]{emulateapj}
\usepackage{graphicx}
\usepackage{color}
\usepackage{hyperref}
\hypersetup{colorlinks=true,citecolor=blue}

\usepackage{natbib}

\usepackage{amsmath}
\def \bea {\begin{eqnarray}}     
\def \ena {\end{eqnarray}}          
\def \bee {\begin{equation}}
\def \ene {\end{equation}}

\def \gas {{\rm gas}}
\def	\th		{\rm {\rm th}}
\def	\H		{\rm {H}}
\def	\K		{~\rm {K}}
\def	\cm		{\,\rm {cm}}
\def	\yr		{\,\rm {yr}}
\def	\B		{\rm {B}}
\def	\erg	{\, \rm {erg}}
\def	\s		{\, \rm {s}}
\def	\g		{\, \rm {g}}
\def	\rad	{\rm {rad}}
\def	\gas	{\rm {gas}}

\def    \ba     {{\bf a}}

\def    \ma {\hat{\bf a}}               
\def    \me {\hat{\bf e}}

\def	\mum	{\,{\mu \rm{m}}}

\def    \apjl  		{\rm {ApJL}}
\def    \apj  		{\rm {ApJ}}
\def    \mnras  	{\rm {MNRAS}}
\def    \araa  		{\rm {ARA\&A}}
\def    \aa  		{\rm {A\&A}}
\def    \apjl  		{\rm {ApJL}}

\begin{document}

\title{Alignment and rotational disruption of dust}
\author{A. Lazarian}
\affil{Department of Astronomy, University of Wisconsin-Madison, USA; \href{mailto:alazarian@facstaff.wisc.edu}{alazarian@facstaff.wisc.edu}}
\affil{Korea Astronomy and Space Science Institute, Daejeon 34055, South Korea}
\affil{Center for Computation Astrophysics, Flatiron Institute, 162 5th Ave, New York, NY 10010}
\author{Thiem Hoang}
\affil{Korea Astronomy and Space Science Institute, Daejeon 34055, South Korea; \href{mailto:thiemhoang@kasi.re.kr}{thiemhoang@kasi.re.kr}}
\affil{Korea University of Science and Technology, 217 Gajeong-ro, Yuseong-gu, Daejeon, 34113, South Korea}

\date{Draft version \today} 

\shorttitle{Grain alignment and rotational disruption}
\shortauthors{Lazarian \& Hoang}

\begin{abstract}
We reveal a deep connection between alignment of dust grains by RAdiative torques (RATs) and MEchanical Torques (METs) and rotational disruption of grains introduced by \cite{Hoangetal:2019}. The disruption of grains happens if they have attractor points corresponding to high angular momentum (high-J). We introduce  {\it fast disruption} for grains that are directly driven to the high-J attractor on a timescale of spin-up, and  {\it slow disruption} for grains that are first moved to the low-J attractor and gradually transported to the high-J attractor by gas collisions. The enhancement of grain magnetic susceptibility via iron inclusions expands the parameter space for high-J attractors and increases percentage of grains experiencing the disruption. The increase in the magnitude of RATs or METs can increase the efficiency of fast disruption, but counter-intuitively, decreases the effect of slow disruption by forcing grains towards low-J attractors, whereas the increase in gas density accelerates disruption by faster transporting grains to the high-J attractor. We also show that disruption induced by RATs and METs depends on the angle between the magnetic field and the anisotropic flow. We find that pinwheel torques can increase the efficiency of {\it fast disruption} but may decrease the efficiency of {\it slow disruption} by delaying the transport of grains from the low-J to high-J attractors via gas collisions. The selective nature of the rotational disruption opens a possibility of observational testing of grain composition as well as physical processes of grain alignment.

\end{abstract}

\section{Introduction}
The mechanism of interstellar grain alignment may be claimed to be a problem of the longest standing in astrophysics. The alignment of non-spherical dust grains was first discovered in 1949 (\citealt{Hiltner:1949}; \citealt{Hall:1949}). A sequence of theoretical works consistently failed in explaining this puzzle (see \citealt{Lazarian:2003} for the history of the problem). 

The key to understanding why it was so difficult to crack the problem is related to it happened to be multi-facet one with a number of physical processes involved and with different regimes for the alignment for grains of different sizes and composition (see \citealt{Lazarianetal:2015} and \citealt{Anderssonetal:2015} for reviews). The major mechanism of alignment is based on the radiative torques that were suggested in 1976 (\citealt{Dolginov:1976}), but was ignored for two decades.

The efficiency of radiative torques in spinning up grains was confirmed in \cite{{DraineWeingartner:1996},{DraineWeingartner:1997}} (hereafter DW96 and DW97). The analytical theory of grain alignment by radiative torques was formulated in \cite{LazarianHoang:2007a}, henceforth LH07) where the abbreviation RATs was coined for RAdiative Torques. 

The RAT effect can be understood in terms of differential scattering of photons with left and right-hand circular polarization by an {\it irregular} grain. The requirement of a grain being irregular is essential for it to have helicity and to experience RATs (LH07). 

The idea that grains should be helical to experience RATs was proposed by \cite{Dolginov:1976}. However, subsequent numerical simulations showed that the model of a helical grain that they suggested does not produce any alignment or spin-up. A successful model of a helical grain was suggested in LH07, and this established grain helicity as the physical source of RATs. The properties of this simple Analytical MOdel (AMO) were shown to reproduce well the response of the actual irregular grains to the anisotropic radiation, thus resolving the mysterious dynamics of dust grains subject to the radiation. A recent numerical study by \cite{Herranenetal:2019} for an extended ensemble of grain shapes and compositions successfully reproduced the main  features of AMO. Therefore, due to its simplicity and the available analytical description, for this study we adopt AMO as our model of RATs. 

The understanding that grain irregularity produces grain helicity allowed \cite{LazarianHoang:2007b} to introduce a new type of grain alignment arising from the motion of grains with respect to the ambient gas. As the offshoot of the AMO, a model of the alignment of irregular grains by MEchanical Torques (METs) was suggested in \cite{LazarianHoang:2007b}. Unlike the earlier discussed stochastic processes of mechanical alignment of thermally rotating grains proposed by Gold (\citealt{Gold:1952}; \citealt{RobergeHanany:1995}; \citealt{{Lazarian:1994},{Lazarian:1995a}}) or cross-sectional or cross-over alignments proposed later for fast rotating grains (\citealt{Lazarian:1995a}; \citealt{LazarianEfroimsky:1999}), METs are much more efficient. This efficiency arises from METs similar to RATs as being regular torques.

The relative importance of the RAT and MET alignment is still not completely clear with the current research, suggesting that, while METs do have the domain for dominance, in most astrophysical settings, RATs are expected to dominate over METs. The latter statement is supported by calculations and numerical simulations which show that the action of METs on grains is less regular as compared to RATs (see \citealt{DasWeingartner:2016}; \citealt{Hoangetal:2018}). Therefore, in what follows, we shall focus our discussion on RATs and then generalize our results on the grains aligned by METs. 

In a separate development, rotational disruption of dust grains induced by RATs, which was termed RAdiative Torque Disruption (RATD), was identified as an essential process controlling the distribution of grain sizes (\citealt{{Hoangetal:2019},{Hoang:2019}}; \citealt{HirashitaHoang:2020}). In addition, grain surface chemistry in star-forming regions was also shown to be dependent on the rotational state of the grains (\citealt{{HoangTung:2019},{Hoang:2019b}}; see \citealt{Hoang:2020} for a review). 
 
 The processes of grain rotation were studied mostly in relation with grain alignment. Stochastic torques of large amplitude can induce fast rotation. For instance, \cite{Gold:1952} considered purely stochastic torques arising from the interaction of grains with gas and showed that the grains can get fast rotation moving with respect to the gas with supersonic velocities. This mechanism was found to be efficient in rotational disruption of nanoparticles (\citealt{HoangTram:2019}; \citealt{HoangLee:2020}). 
 
 Regular torques can induce faster rotation compared to the stochastic torques of comparable amplitude. For instance, it was demonstrated by \cite{Purcell:1979} that grains in the typical interstellar conditions are rotating with rotational velocities significantly larger than the thermal ones. The variations of the grain absorption surface coefficient, photoemission of electrons and formation of H$_2$ atoms on the selected catalytic sites on grain surface were identified by Purcell as the key sources of grain suprathermal rotation.
 
 Finally, RATs and METs can result in fast grain rotation if grains subject to anisotropic radiation fluxes or mechanical flows get attractor points of high angular momentum (hereafter high-J attractor). Unlike \cite{Purcell:1979}'s torques, RATs are much better defined as there is less dependence of RATs on the unknown details of grain surface properties. The existence of the grain spin-up and alignment by RATs is inevitable if the grains are not significantly smaller than the wavelength, $\lambda$, of the impinging radiation. 
 
\cite{DraineWeingartner:1996} discovered that the amplitudes of uncompensated RATs for the typical interstellar radiation field (ISRF) can be comparable or larger than the amplitudes of the uncompensated \cite{Purcell:1979} torques. However, as it was shown by the subsequent research (e.g. \citealt{DraineWeingartner:1997}), this did not mean that placed in the radiation field grains will necessarily rotate with high velocities. This is a significant difference between RATs and the Purcell's torques. The latter are fixed in the grain frame, while the former act in the laboratory frame. Therefore, the change of the grain orientation with respect to the radiation direction changes both the torque directions and their amplitude. This results in the complex grain dynamics that was clarified only after the introduction of AMO in LH07.\footnote{The uncompensated RATs arising from isotropic radiation are similar to the Purcell's torques, but their amplitude is orders of magnitude less than those arising from anisotropic radiation (\citealt{DraineWeingartner:1997}). Therefore, in this paper, we disregard this type of RATs.} 

It was shown in LH07 and confirmed in subsequent studies (\citealt{HoangLazarian:2008}; \citealt{HoangLazarian:2009a}) that the grains aligned by RATs can have both high and low angular momentum attractors, which are defined as high-J and low-J attractor, respectively. For typical ISRF, the rotation of $0.1\mum$ grains is much faster than thermal rotation for the alignment in the high-$J$ attractor points, and it is subthermal for grains in the low-$J$ attractor points, assuming that grain randomization is ignored at low-J attractors. When taking into account gas randomization, grains at low-J attractors are gradually transported to high-J attractors, which are more stable (\citealt{{HoangLazarian:2008},{HoangLazarian:2016a}}). Evidently, for fast grain disruption that occurs in less than a damping time, only the alignment  high-$J$ points is important. For grain alignment without high-J attractors, grains driven to low-J are gradually randomized by gas collisions. If grains are aligned with only low-J attractors, the efficiency of both grain alignment efficiency is rather low (\citealt{HoangLazarian:2008}), which implies inefficient rotational disruption.
 
The grain irregularity that induces uncompensated RATs also induces METs, as was suggested in \cite{LazarianHoang:2007b} and proven in numerical calculations in \cite{Hoangetal:2018} (see also \citealt{DasWeingartner:2016}). The spin-up by METs is much more efficient than the spin-up by stochastic \cite{Gold:1952}'s torques. Therefore, it is natural that both RATs and METs present the prime interest for the processes for studies of the grain rotational disruption and the rotational modification of grain chemistry.
 
 In what follows, in Section \ref{precession}, we discuss the precession processes of grains about the magnetic field, electric field, and radiation direction, which defines the axis of grain alignment. We then discuss AMO and properties of grain alignment by RATs and magnetic relaxation in Section \ref{sec:AMO}. The strength of RATs and the effect of spin-up and spin-down as well as gas randomization is discussed in Section \ref{sec:spinup_down}. Rotational disruption of grains by RATs is discussed in Section \ref{sec:rot_disr}, and the time-dependence alignment and disruption is discussed in Section \ref{sec:timedepend}. Discussion of alignment and disruption by METs is presented in Section \ref{sec:METs}. Effect of grain composition on alignment and disruption is discussed in Section \ref{sec:graincomp}. The role of pinwheel torques is discussed in Section \ref{sec:pinwheel}. Extended discussion and a summary of our findings is presented in Sections \ref{sec:discuss} and \ref{sec:summary}, respectively.
 
 \section{Grain Precession and Alignment Axis}
\label{precession}

Grain alignment is traditionally considered with respect to magnetic fields. However, as we discuss below, it can proceed with respect to other axes, including the electric field, radiation, and gas flow.  

\subsection{Precession induced by magnetic and electric torques}
A dust grain with magnetic moment $\mu$ experiences Larmor precession about the ambient magnetic field. The Larmor precession rate is given by
\begin{equation}
\Omega_{\rm B}=\frac{\mu B}{I_{\|}\omega}, 
\label{eq:tl}
\end{equation}
where $I_{\|}$ is the grain maximal moment of inertia, $\omega$ is the grain angular velocity, and $B$ is the magnetic field strength. For simple shapes, the moments of inertia are readily available. For instance, for an oblate spheroid,
\begin{equation}
I_{\|}=\frac{8\pi}{15}\rho a_1a_2^4,\ I_{\perp}=\frac{4\pi}{15}\rho a_1a_2^2(a_1^2+a_2^2),
\label{inertia}
\end{equation}
where $\rho$ is the grain mass density, $a_1,\ a_2$ denote the semi-minor and semi-major grain axes. Realistic grains are irregular, so usually one defines the effective grain size $a$ as the radius of the equivalent spherical grain of the same volume. For irregular grains, the maximum moment of inertia is denoted by $I_{1}=8\pi \rho \alpha_{1} a^{5}/15$ with $\alpha_{1}$ is the factor of unity order (DW96). For the sake of simplicity, the use of expressions similar to one given by Equation (\ref{inertia}) is justified.

For paramagnetic grains, it was shown by \cite{Dolginov:1976} that the magnetic moment is induced by the Barnett effect (\citealt{Barnett:1915}; \citealt{Landau:1960}) and reads,
\begin{equation}
\mu_{\rm Bar}=\frac{\chi(0)V\hbar}{g\mu_{\rm B}}\omega, \label{eq:mubar}
\end{equation}
where $V$ is the volume of the dust grain, $\hbar$ is the Planck constant divided by $2\pi$, $\mu_{\rm B}=e\hbar/2m_ec\approx 9.274\times 10^{-21}$ erg G$^{-1}$ is the Bohr magneton, $m_e$ is the mass of the electron, $c$ is the speed of light, and $g$ is the $g$-factor, which is $g\approx 2$ for electrons. The zero frequency magnetic susceptibility, $\chi(0)$, for paramagnetic materials, is given by
\begin{equation}
    \chi(0)=\frac{n_p \mu_B^2}{3kT_{d}},
\end{equation}
where $n_p=f_p n$ is the density of paramagnetic species, and $T_{d}$ is the grain temperature. 

The zero-frequency susceptibility increases substantially if grains have strongly magnetic inclusions, as first discussed in \cite{JonesSpitzer:1967}. For such a grain with superparamagnetic inclusions \citep{Morrish:2001},
\begin{equation}
 \chi_{\rm sp} (0) = 1.2 \times 10^{-2} N_{\rm incl} f_{\rm sp} \left(\frac{15~{\rm K}}{T_{d}}\right),
 \label{chi_sp}
 \end{equation}
$f_{\rm sp}$ is the fraction of atoms that are within the superparamagnetic clusters, $N_{\rm incl}$ is the number of atoms per cluster that can vary from 10 to $10^6$ (\citealt{Kneller:1963p6410}). 

The period of the Larmor precession of the grain angular momentum ${\bf J}$ in the magnetic field can be obtained by combining Eqs. (\ref{eq:tl}) and (\ref{eq:mubar}):
\begin{eqnarray}
t_{\rm L}&=&\frac{4\pi}{5} \frac{g_e\mu_{\rm B}}{\hbar}\rho_s s^{-2/3}a^{2}B^{-1}\chi(0)^{-1}\\ \nonumber
&\approx& 1.3\ \hat{\rho}\hat{s}^{-2/3}a_{-5}^2\hat{B}^{-1}\hat{\chi}^{-1}\ {\rm yr},
\label{larmor_p}
\end{eqnarray}
where {\bf $s=a_{1}/a_{2}$ be the grain axial ratio, $a=a_{2}s^{1/3}$}, $a_{-5}= a/(10^{-5}\cm)$, the normalized value of the magnetic field is $\hat{B}=B/(5~\mu$G) and $\hat{\chi}=\chi(0)/(10^{-4})$. Naturally, the Larmor period can be significantly reduced if grains have superparamagnetic inclusions (see Equation (\ref{chi_sp}). 

In addition, a charged grain with potential $U$ performs gyro-rotation in the magnetic field over a time:
\begin{equation}
\omega_{\rm gyro}^{-1} \sim 2.4 \times 10^2 \hat{\rho} a_{-5}^2\left( \frac{U}{0.3\,
{\rm V}} \right)^{-1} \left( \frac{B}{5 \, \mu {\rm G}} \right)^{-1} \,
\yr,
\label{gyro}
\end{equation}
where $\hat{\rho}=\rho/(3 \g\cm^{-3})$. 

In the process of moving perpendicular magnetic field or precessing in it, the grain experiences the electric field with amplitude
\begin{equation}
E_{\rm ind}=\frac{V_{\rm grain,\bot}}{c} B,
\end{equation}
where $V_{\rm grain,\bot}$ is the grain velocity component perpendicular to the magnetic field. This field induces grain additional precession if the grain has electric dipole moment component parallel ${\bf p}_{el,J}$:
\begin{equation}
{\bf \Gamma}_{el}={\bf p}_{el,J} \times {\bf E}_{\rm ind}.
\end{equation}

The resulting precession rate $\Omega_{el}$ can be faster or slower than the rate of Larmor precession given by Equation (\ref{eq:tl}), and one can write
\begin{equation}
\Omega_{el}=\aleph \Omega_B,
\label{Omegael}
\end{equation}
where
\begin{equation}
\aleph=\frac{p_{el, J}V_{\rm grain, \bot}}{\mu c}, 
\label{aleph0}
\end{equation}
where the electric moment $p_{el}$:
\begin{equation}
\label{pel}
p_{el}= qa \kappa_{el} \approx 1.0 \times 10^{-15} \, U_{0.3} a_{-5}^2 \hat{\kappa}_{el} , {\rm statcoulomb}~{\rm cm}, 
\end{equation}
where $\hat{\kappa}_{el}=\kappa_{el}/10^{-2}$ is a parameter describing the charge distribution and $U_{0.3}=U/(0.3\rm V)$ is the electric potential of the grain. The importance of this type of precession for grain alignment was identified in \cite{Lazarian:2020}. 

For the grain velocities $V_{\rm grain,\bot}$ adopted from \cite{Yanetal:2004}, the estimate for $\aleph$ can be obtained from  \cite{Weingartner:2009}). For {\it carbonaceous grains}, one gets
\begin{equation}
\label{aleph_carb}
\aleph_{\rm carb} \approx 220 \, a_{-5} \hat{\kappa}_{el} U_{0.3} \left( \frac{\omega}{10^5 \s^{-1}} 
\right)^{-1} \left( \frac{T_d}{15 \K} \right) 
\left( \frac{V_{\rm grain,\perp}}{1~km\s^{-1}} \right)~~~.
\end{equation}
The difference of $\aleph$ for silicate and carbonaceous grains arises from the difference in the values of the
magnetic moments of these grains $\mu_{\rm carb}\approx 10^{-4} \mu_{\rm sil}$ (\citealt{LazarianDraine:2000}). Therefore for {\it silicate grains} one gets
\begin{equation}
   \aleph_{\rm sil}\approx 10^{-4} \aleph_{\rm carb}.
\end{equation}

Both $\kappa_{el}$ and $V_{\rm grain,\bot}$ are rather uncertain. In particular, in \cite{LazarianHoang:2019}, we argued that 
\cite{Yanetal:2004} may overestimate the actual velocities of dust grains due to the resonance interaction of charged 
grains with interstellar magnetohydrodynamic (MHD) turbulence. The lower limit for a $0.1\mum$ grain due to the hydrodynamic interactions in magnetized
gas is $0.1$ km/s (\citealt{LazarianYan:2002}). A detailed study of possible $\kappa_{el}$ can be found in \cite{JordanWeingartner:2009}.

\begin{figure}
\includegraphics[width=0.5\textwidth]{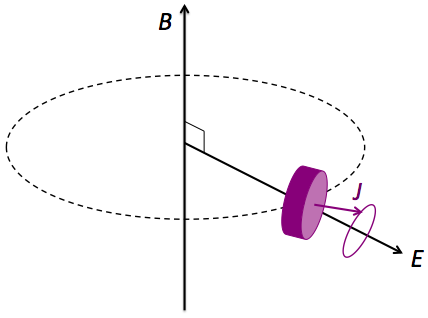}
\caption{Illustration of the electric field acting on a grain gyrating with respect to a magnetic field. The grain electric dipole moment along the angular momentum precesses around the electric field, whereas the grain is gyrating around the magnetic field (dashed line).}
\label{electric}
\end{figure}

An obvious feature that was missed by the earlier researchers is that for typical carbonaceous grain $\Omega_{el}\gg \Omega_B$,
i.e., that the grain actually precesses around the electric field and not around magnetic field. It was noted in \cite{Lazarian:2020} that this provides a significant difference in the alignment of the two types of grains that has not been considered so far, namely, the alignment direction of silicate and carbonaceous
grains is {\it perpendicular to each other}.

Figure \ref{electric} illustrates the precession of a disk-like grain with respect to the electric field arising from the grain gyro-rotation about the magnetic field.

\subsection{Precession induced by RATs and METs}
Subject to an anisotropic radiation field of energy density $u_{\rm rad}$ and anisotropy degree $\gamma$, dust grains experience {\it radiative} precession around the radiation direction due to RATs at a rate (see LH07): 
\begin{eqnarray}
\Omega_{\rm rad}&\approx& 0.057\ \hat{\rho}^{-1/2}\hat{s}^{1/3}a_{-5}^{-1/2}\hat{T_d}^{-1/2}f_{1}(\Theta)\nonumber \\
&\times&\left(\frac{u_{\rm rad}}{u_{\rm ISRF}}\right) \left(\frac{\lambda}{1.2\ \mu{\rm m}}\right) \left(\frac{\gamma \overline{|\mathbf{Q_{\Gamma}}|}}{0.01}\right)
\left(\frac{\omega_d}{\omega}\right) \ {\rm yr}^{-1},~~~
\label{rad_p}
\end{eqnarray}
where $\omega$ is normalized to the thermal angular velocity $\omega_{d}=\sqrt{2k T_{\rm d}/I_{\|}}$, and $\mathbf{Q_{\Gamma}}$ is given by Equation (\ref{eq:QRAT_avg}). The factor $f_1(\Theta)$ is a function of the angle $\Theta$ between the grain axis of major inertia and the direction of radiation defined in LH07 for a toy model of an oblate grain subject to the bombardment by particles.

Because the grain radiative precession vanishes in the limit of perfect alignment, there will always be some finite angle between the grain angular momentum and the radiation field anisotropy, changing the efficiency of RATs in terms of grain spin-up. Moreover, the radiative precession rate increases with increasing the radiation energy density $u_{\rm rad}$ but decreases with increasing the rotation rate $\omega$. 
 
The mechanical bombardment of grains by a regular flow of atoms/molecules induces METs \citep{LazarianHoang:2007b}. This process results in the precession of a toy oblate grain around the flow direction at a period:
\bea
    t_{\rm mech}&\approx& \frac{2\pi I_{\|}\omega}{\pi a^{3}s^{-2}n_{\rm H}m_{\rm H}V_{\rm grain}^{2}e(e^{2}-1)K(\Theta,e)\sin 2\Theta}\nonumber\\
    &\simeq& 36 \left(\frac{\omega}{\omega_{\rm th}}\right)\left(\frac{v_{\rm th}}{V_{\rm grain}}\right)^{2}\left(\frac{s}{0.5}\right)^{2}\frac{1}{\sin 2\Theta} ~\rm yr,
    \label{t_mex}
\ena
where $v_{\rm th}=(kT_{\gas}/m_{\H})^{1/2}$ with gas temperature $T_{\gas}$ is the gas thermal speed, $V_{\rm grain}$ is the grain velocity with respect to the medium. In general, one should also use in Equation (\ref{t_mex}) a function $f_2(\Theta)$ which should be defined empirically
for irregular grains e.g. using the numerical simulations similar to those in \cite{Hoangetal:2018}.

Naturally, in the presence of both RATs and METs, the precession induced by METs dominates provided that
\begin{equation}
    \Omega_{\rm mech}=\frac{2\pi}{t_{\rm mech}},
\end{equation}
is faster than $\Omega_{\rm rad}$ given by Equation (\ref{rad_p}). As we discuss in Section \ref{axis}, the axis of the fastest rotation determines the alignment axis. 

Depending on the magnetic field strength and the radiation flux interacting with irregular grains, for sufficiently large dust velocities, grains may be aligned with the direction of mechanical flow.  Grain motion with respect to the magnetic field arising from MHD turbulence is one of the causes of inducing grain precession induced by mechanical torques.

\subsection{Axis of grain alignment}
\label{axis}

Different torques acting on a grain induce its precession. The process that induces the fastest precession determines the grain {\it axis of alignment}.\footnote{Here grain alignment is implied the alignment of grain angular momentum.} The grain alignment can happen in the direction of the radiation/mechanical flow or it can happen in terms of magnetic field. What is actually happens depends on whether the rate of precession that RATs/METs induce is larger or smaller than the grain precession in the external magnetic field $B$.

In the sections above several precession rates are introduced. If we denote 
\begin{equation}
    \Omega_k=max[\Omega_{\rm rad}, \Omega_{\rm mech} ],
\end{equation}
then the direction of $\Omega_k$ determines the precession about either the direction of the radiation flux of a mechanical flow. Provided that $\Omega_k$ is larger than both $\Omega_B$ and $\Omega_{el}$, its direction defines the {\it axis of alignment}. This type of alignment was discussed in LH07 but was mostly ignored till the description of grain alignment in the accretion disks made it evident that radiation direction can define the axis of alignment (see \citealt{Tazakietal:2017}). 

The traditional grain alignment in the interstellar medium (ISM) is usually considered in terms of magnetic field. Therefore the dust polarization is accepted as a major tracer of magnetic fields in diffuse ISM and molecular clouds. The condition for this is that 
$\Omega_B$ is larger than both $\Omega_k$ and $\Omega_{el}$. While this is applicable to silicate grains, the considerations we provided above testify, that this condition is unlikely to be true for carbonaceous $0.1\mum$ grains. Those for typical interstellar conditions rotate with respect to the electric field with their axis of alignment being perpendicular to the magnetic field. 

In the setting that $\Omega_{el}$ is the fastest of $\Omega_B$ and $\Omega_k$ two limiting cases are possible. If $\Omega_k$ is much larger than the rate of gyro precession $\omega_{\rm gyro}$ given by Equation (\ref{gyro}), then the axis of alignment of the grain is the axis passing through the instantaneous position of grain in the plane perpendicular to magnetic field ${\bf B}$. In the opposite situation the averaging over gyro-rotation is required. The degrees of alignment are different in these two cases. 

For both RAT and MET alignment processes, the angle $\psi$ between the flow and the alignment axis is an important parameter that determines the degree of alignment and the rate of grain rotation (LH07; \citealt{HoangLazarian:2016a}). If $\Omega_k$ is the largest precession frequency, this angle $\psi=0$. If $\Omega_B$ is the largest, then $\psi_B$ is the angle between the magnetic field ${\bf B}$ and the direction of the flow. In case of $\Omega_E$ being dominant, the precession axis is performing the motion:
\begin{equation}
    \phi=\phi_0+\omega_{\rm gyro} t,
\end{equation}
and the angle $\psi_E$ between the electric field and the flow can be obtained from the trigonometric identity. 

In what follows, we focus on the alignment and disruption of grains with respect to directions of $\Omega_k$ and $\Omega_B$. 

 \section{AMO and grain alignment at high and low-$J$ attractors}\label{sec:AMO}
For simplicity, we first discuss grain alignment by RATs using AMO and disregard the effect of gas randomization and thermal fluctuations within the grain. The latter effects will be discussed in Section \ref{sec:spinup_down}. The physics of RAT alignment is briefly presented in \ref{sec:physical}.
 
 \subsection{Alignment and rotation induced by RATs}
 
\begin{figure}
\includegraphics[width=0.5\textwidth]{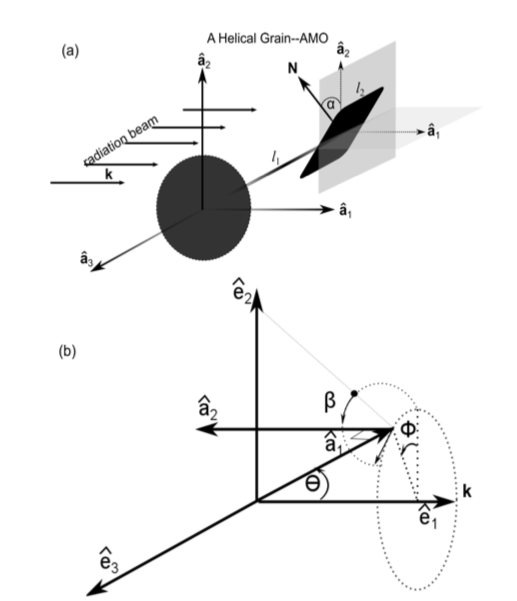}
\caption{Panel (a): The model used to derive analytical formulae of RATs  (AMO) and study the RAT alignment. The model has an oblate reflecting body with a mirror attached at an angle $\alpha$. The vector $\ma_1$ corresponds to the axis of the maximal moment of inertia, and two other principal axes are $\ma_{2}\ma_{3}$, and these vectors describe the grain model. The radiation beam is coming from the left, along the ${\bf k}$ direction. Panel (b): Orientation of the grain in the lab reference system $\me_{1}\me_{2}\me_{3}$ employed for calculating torque components. The axis ${\me}_{1}$ is chosen along the radiation direction ${\bf k}$. The axis ${\me}_{2}$ is perpendicular to $\me_{1}$, and $\me_{3}$ perpendicular to the plane $\me_{1}\me_{2}$. The grain orientation in the lab frame is described by three angles, $\Theta, \beta$, and $\Phi$. See more details in \cite{LazarianHoang:2007a}.}
\label{AMO}
\end{figure}

The AMO is the model of helical grain that is very simple, but, nevertheless, correctly describes the major properties of RAT alignment. It also provide the qualitative features of the MET alignment. The model consists of an ellipsoidal body with the attached mirror. The helicity of the grain is achieved by the mirror attached at an angle to the grain, as shown in the upper panel of Figure \ref{AMO}. In LH07, this angle was chosen to be 45 degrees. The AMO was confirmed by the extensive study of RATs applied to different shapes by \cite{Herranenetal:2019}.
RATs induced by a radiation beam on a helical grain can be calculated in the lab reference frame (see lower panel of Figure \ref{AMO}). The upper panel of Figure \ref{phase} illustrates the two torque components that act on a helical grain, $Q_{e1}(\Theta)$ and $Q_{e2}(\Theta)$, as function of the angle $\Theta$ made by the grain axis of maximum moment of inertia, $\ma_{1}$, and the radiation direction, ${\bf k}$. The component $Q_{e3}$ is present even for the axisymmetric grain, and it is responsible for the grain precession in the beam of radiation. It was shown in LH07 that the torques calculated with Discrete Dipole Approximation code (DDSCAT) (\citealt{DraineFlatau:1994}) for irregular shapes show a similar functional dependence as the torques calculated using our toy model. However, the relative amplitude of $Q_{e1}(\Theta)$ and $Q_{e2}(\Theta)$ changes from one shape to another and also varies with the wavelength. To account for this change, the ratio
\begin{equation}
q^{\rm max}=\frac{Q_{e1}^{\rm max}}{Q_{e2}^{\rm max}}
\label{qmax}
\end{equation}
was introduced, and the properties of alignment were studied as this ratio varied. With this model LH07 showed that the torques can both spin up the grain and slow down its rotation to the sub-thermal values. What situation is actually realized with a given grain depends on both the value of $q^{\rm max}$, which can be calculated numerically for a given grain and the radiation direction with respect to the axis of fastest grain precession, i.e., the axis of alignment. 
The latter, as we discussed in Section \ref{precession}, can be either the direction of the magnetic field or the direction of radiation for the case of RAT alignment.

\subsection{Spin-up and spin-down in the process of RAT alignment}

\begin{figure}
\includegraphics[width=0.5\textwidth]{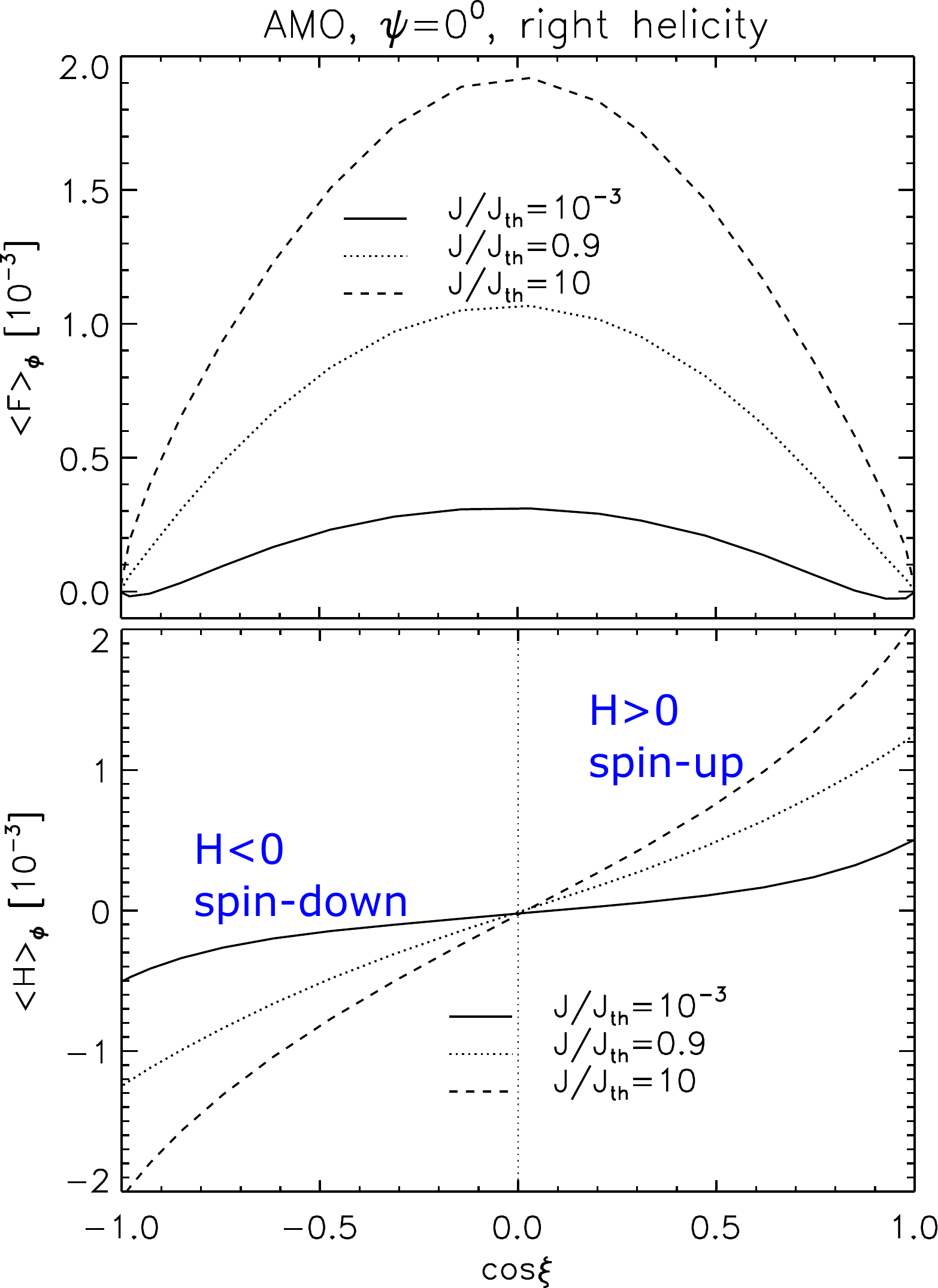}
\caption{Aligning torque component ($\langle F\rangle$, upper) and spin-up component ($\langle H\rangle$, lower) as functions of cosine of the angle $\xi$ between ${\bf J}$ and the magnetic field from AMO assuming the radiation direction is parallel to the magnetic field, which are obtained by averaging over thermal fluctuations with different ratios of $J$ to its thermal value and over the fast Larmor precession. The vertical dotted line in the lower panel divides two regions of spin-up and spin-down by RATs. Modified from \cite{HoangLazarian:2008}.}
\label{torques}
\end{figure}

Let us provide a closer look on the processes taking place at the high-$J$ and low-$J$ attractor points. Consider the torques acting on a AMO at $\psi =0$ that are depicted in Figure \ref{torques}, where $\langle H\rangle$ is the component of RATs projected onto the grain angular momentum that acts to spin up/down the grain, and $\langle F\rangle$ is the RAT component that is perpendicular to the angular momentum (DW97). Here the angle brackets denote the averaging over thermal fluctuation and the Larmor precession (\citealt{HoangLazarian:2008}), which are omitted in the following for simplicity. The grain is spun-up (spun down) for $H>0$ ($H<0$), and the angle $\xi$ between ${\bf J}$ and the magnetic field increases/decreases for $F>0$ ($F<0$). In the vicinity of low-$J$ attractors corresponding to $\cos\xi=-1$, the alignment torques $F$ are weak and $J$ is small. Therefore, the random forcing arising from gaseous bombardment can induce significant variations of $\cos \xi$ and move grains to the range where the acceleration torques are positive. This was first demonstrated in \cite{HoangLazarian:2008} for ordinary paramagnetic grains and in \cite{HoangLazarian:2016a} for grains with iron inclusions.

It is evident from there that in the vicinity of the stationary point corresponding to $\cos\xi=1$, the alignment torques $F$ vanish. At the same time, the spin-up torques are maximal and can drive grains to higher velocities. It is also obvious that for grains outside the stationary point $\cos\xi=1$, the alignment torques $F$ move grains towards the low-$J$ attractor point if $F>0$. For the case with high-J attractors (right panel), one has $F<0$ for $\cos\xi\sim 0.7-1$. Thus, a faction of grains falling in this range of angles are rapidly driven to high-J attractors, whereas grains outside this range are driven to the low-J attractor.

\begin{figure}
\includegraphics[width=0.5\textwidth]{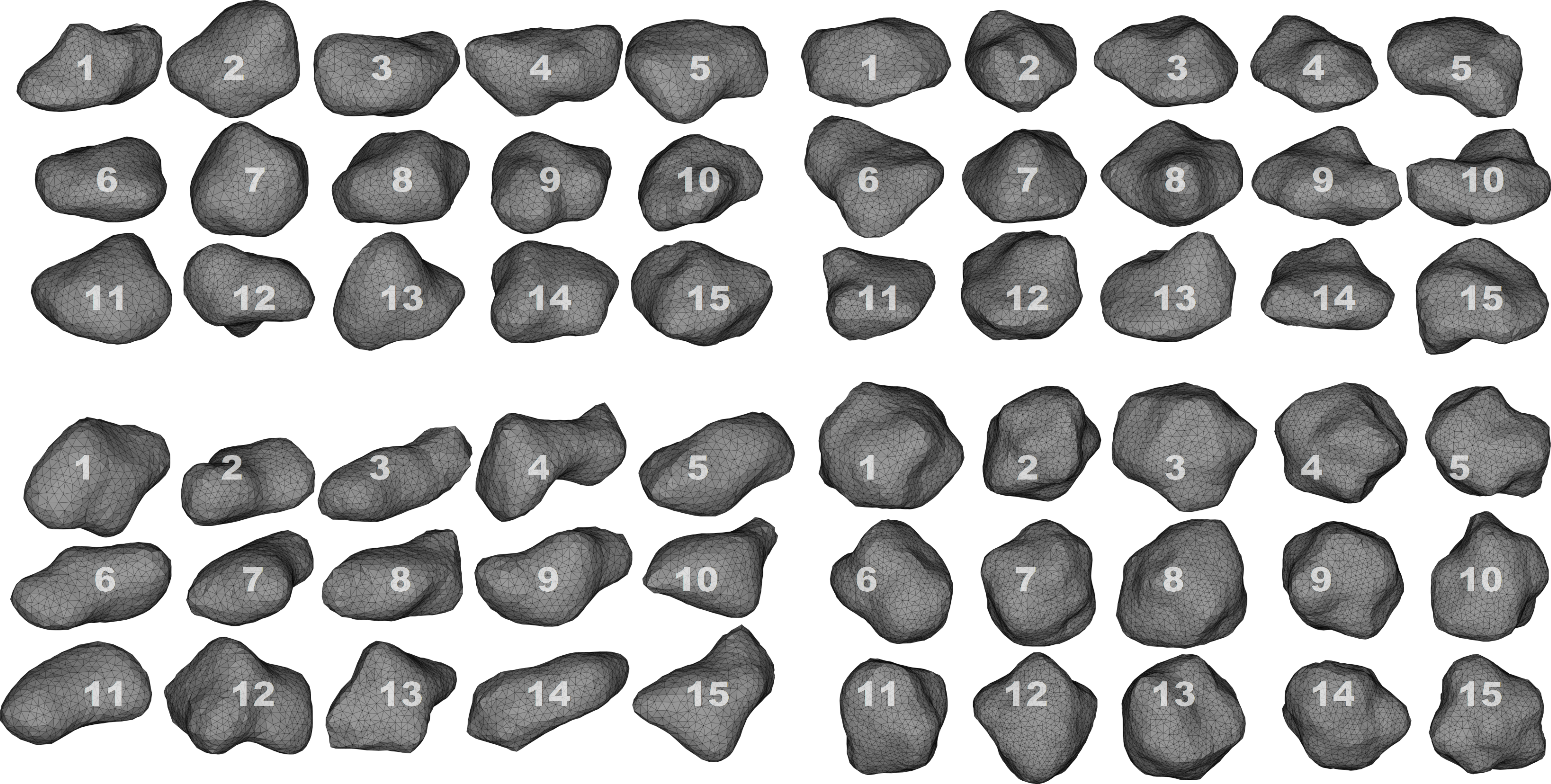}
\includegraphics[width=0.5\textwidth]{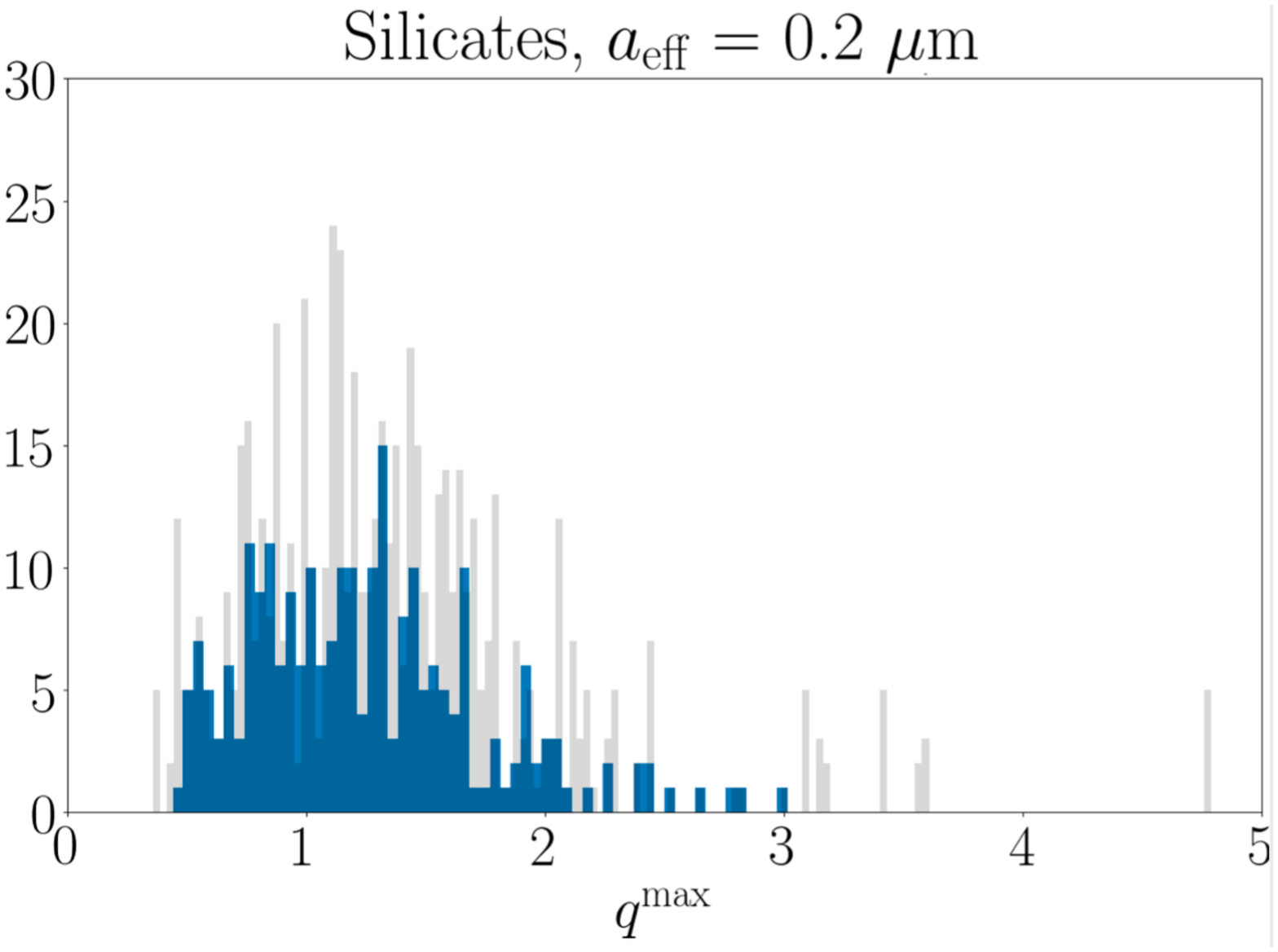}
\includegraphics[width=0.5\textwidth]{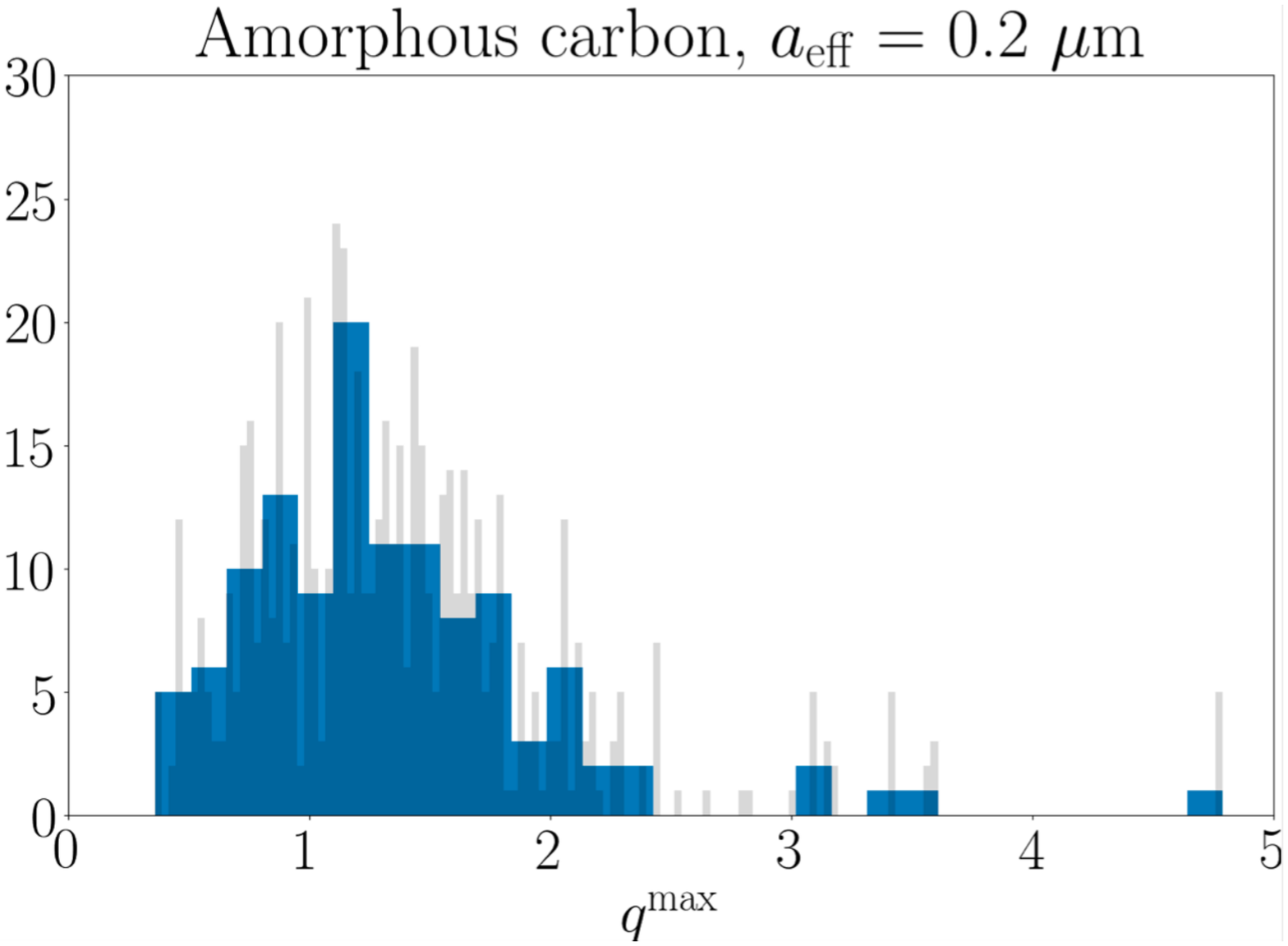}
\caption{Upper panel: Examples of grain shapes studied numerically. The second upper right group of 15 grains corresponds to "oblate grains". Middle panel: Histogram distribution of $q^{\max}$ obtained for an ensemble of silicate "oblate grains" subject to the ISRF is given by blue and for the entire distribution of grain shapes by gray. Lower panel: Same as the middle panel, but for carbonaceous grains. From \cite{Herranenetal:2019}}
\label{fig:herranen}
\end{figure}

For the alignment with respect to the magnetic field, the existence of high-J and low-J attractor points depends both on the grain shape and the angle $\psi$ between the radiation direction and the direction of the magnetic field. The corresponding parameter space for the existence of these high-J attractor points was identified in LH07 within the AMO with the predictions tested recently in \cite{Herranenetal:2019} using a distribution of random grain shapes. The results for $q^{\max}$ from that study are shown in Figure \ref{fig:herranen}.

It is only rotation in positions corresponding to high-J attractor points that is both stable and provides appreciable rotational rates that are important for grain disruption and can affect grain surface chemistry. Note that these rotational rates depend not only on the anisotropic radiation flux but also on $\psi$. At the same time, the grains at low-J rotation are bound to rotate slowly, with the typical rotational temperature corresponding to the grain temperature $T_d$, which in many situations is lower than the thermal velocity of grain rotation in gas with temperature $T_{\gas}>T_d$, i.e., grains at low-J attractor points can rotate {\it subthermally}.

Consider first grains with low-$J$ attractor points only. The phase trajectories for such grains are shown in lower panel of Figure \ref{phase}. It is evident that most of the grains are being moved to low-$J$ attractors without increasing much their rotational speed. Only a small fraction of grains corresponding to $\cos \xi\approx 1$ are moving towards the stationary points A and B, which are the repeller  points. The grains get in the vicinity of repeller points but cannot stay there. As a result, they are also turned to the low-$J$ attractor. 

The reason for this type of dynamics can be found if one decomposes the torques acting on grains into the alignment torques $F$ and spin-up torques $H$. This decomposition for the AMO is presented in Figure \ref{torques}. 

The trajectories to the low-J attractor points correspond to grain spin-down and the alignment there is unstable to randomization, e.g., from gaseous bombardment. As a result of the randomization $\cos \xi$ can get different values, including the values of corresponding to the trajectories that go to the high-J attractors. Eventually, after many cycles of being moved to low-J attractor, randomized with a small probability to be moved to the high-J attractor, most of the grains end up at the high-J attractor points. This corresponds both to perfect alignment, which will be discussed in detail in Section \ref{sec:spinup_down}. 


Figure \ref{phase} corresponds to the initial grain velocities significantly larger than the thermal velocity. At the same time, for the studies of spin-up, it is advantageous to start with thermal distribution of grain angular momenta. This setting is illustrated in Figure \ref{fig:MAP_Jth} for a $0.2\mum$ grain subject to the standard ISRF. For numerical results shown in this paper, we adopt the typical physical parameters of the ISM, including $n_{\H}=30\cm^{-3}$, $T_{\gas}=100\K$, and $T_{d}=20\K$, unless stated otherwise.

The upper panel of Figure \ref{fig:MAP_Jth} shows the typical trajectory map of grain alignment with both low-J and high-J attractors, in the absence of collisional excitation. A fraction of grains is driven to the high-J attractor (marked by a red circle). The characteristic timescale to reach the high-J attractor is less than the gas damping time (see Equation \ref{eq:taugas}). Throughout this paper, we call this {\it fast alignment}, and grains having fast alignment can experience {\it fast} rotational disruption. Moreover, most of grains are driven to the low-J attractor. 

The lower part of Figure \ref{fig:MAP_Jth} illustrates a different setting, namely, the alignment of grains when there is no high-J attractors. It is easy to see that RATs initially accelerate some grains, but the alignment component of torques moves grains toward the low-J attractor, changing $\cos \xi$. The increase of the rate of grain rotation depends on the initial value of $\xi$. For a small portion of initial angles of $\xi$, the increase can be significant. For instance, a single trajectory for $\cos\xi=1$ is shown to move toward the stationary high-J point, which is, however, is a repeller point. Comparing the upper and lower parts of Figure \ref{fig:MAP_Jth}, one can clearly see that the percentage of grains that are spun-up is significantly smaller in the case when the high-J attractor point is absent. 

The dynamics RAT alignment in the presence of gas randomization was studied in \cite{HoangLazarian:2008}, for both cases with high-J attractors and with low-J attractors only, where it is show that grains aligned at the low-J attractor are gradually transported to the high-J attractor by gas collisions. In Section \ref{sec:spinup_down}, we will discuss this issue in more detail.

\subsection{Effects of magnetic dissipation on grain alignment}
\label{sec:mag_diss}

\subsubsection{Enhancement of grain alignment}

In view of the striking differences in the rate of grain rotation at high-J and low-J attractor points, it is obvious that only grains with high-J attractor points are important for the processes of {\it fast alignment} and {\it fast disruption}, as well as for effects of rotation on the grain surface chemistry. The RAT alignment theory has been elaborated in a number of studies that followed LH07 with important effects being added in the subsequent works. For instance, it was shown in \cite{LazarianHoang:2008} that, in the presence of grains with magnetic inclusions, the enhanced magnetic response induces high-J attractor points for the parameter space that only low-J attractor points exist. This makes grains both to be better aligned and rotate fast. In what follows, we discuss how the parameter space for the existence of high-J attractors changes with the increase of magnetic dissipation within grains as well as the change of the fraction of grains that move directly by RATs directly to high-J attractor points.

\begin{figure}
\includegraphics[width=0.5\textwidth]{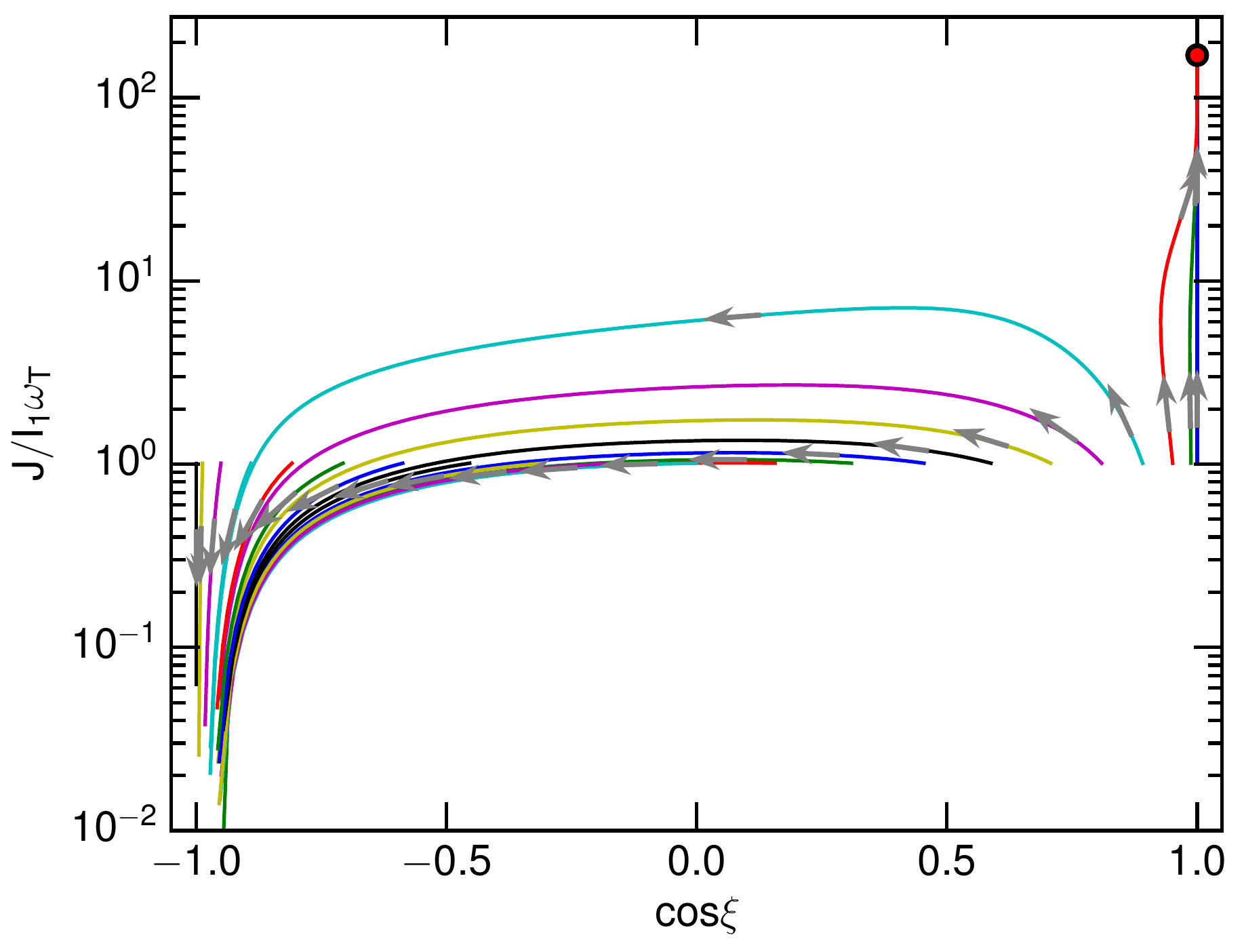}
\includegraphics[width=0.5\textwidth]{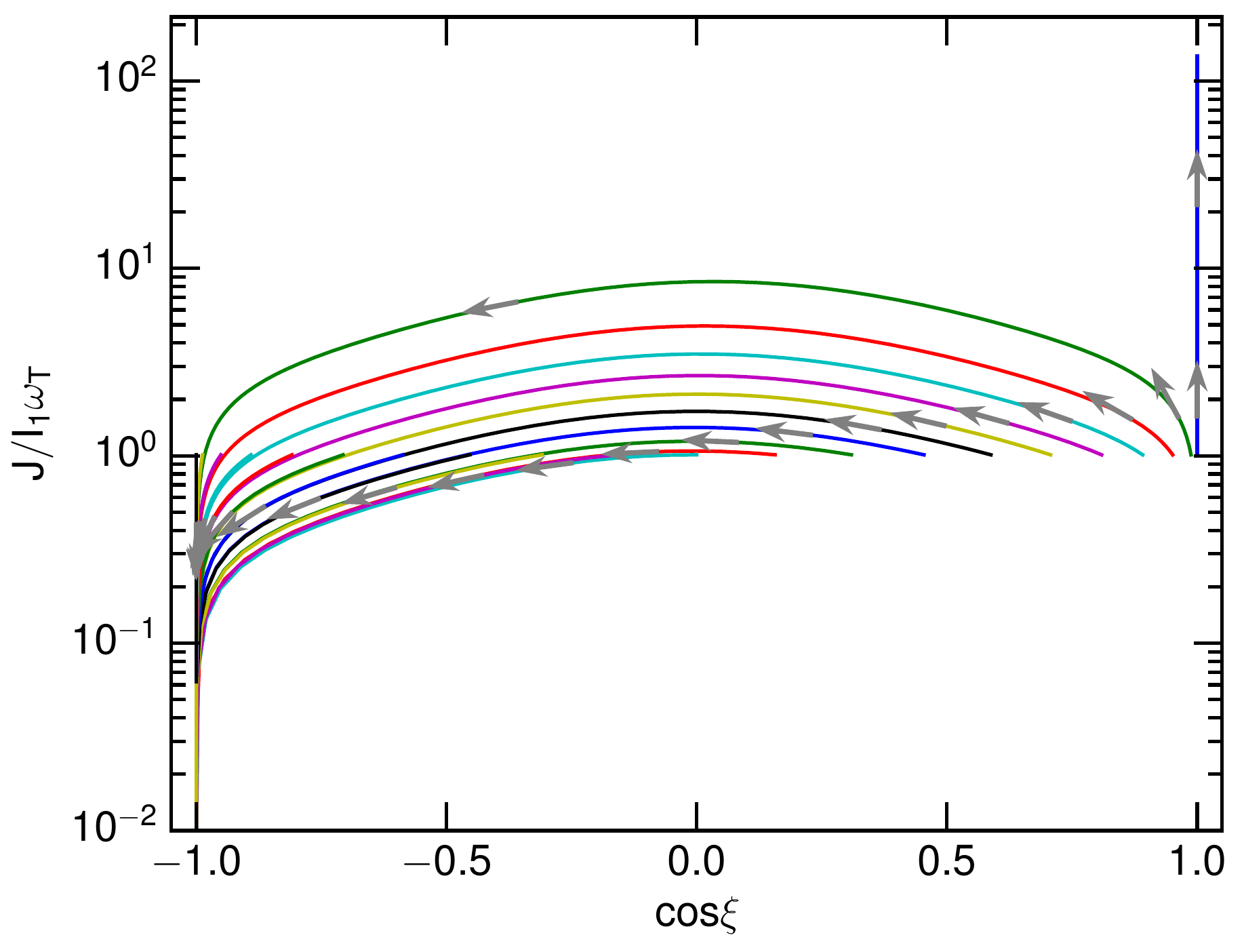}
\caption{Phase trajectory map for alignment of grains starting from $J=J_{\rm th}$. Upper panel: Grain alignment with a high-J attractor ($q^{\max}=2, \delta_{m}=10$). Some grains are rapidly driven to a high-J attractor (red circle), while most grains are driven to the low-J attractor. Lower panel: Grain alignment without a high-J attractor ($q^{\max}=1, \delta_{m}=0.5$). All grains are driven to the low-J attractor. The grain size $a=0.2\mu$m and $\psi=0^{\circ}$, and the standard ISRF are assumed. Collisional excitation by gas are disregarded, and the low-J attractor has $J\rightarrow 0$.}
\label{fig:MAP_Jth}
\end{figure}

The RAT or MET alignment in the presence of magnetic dissipation should not be confused with the classical magnetic alignment in the presence of pinwheel torques (\citealt{Purcell:1979}). The key difference between the two is that RATs and METs can do the alignment with the magnetic effects playing an auxiliary supporting role, which is the case of RAT/MET alignment with high-J attractors. However, for grain alignment by RATs without high-J attractors, the effect of RATs is to first spin-up a fraction of grains (e.g., Figure \ref{fig:MAP_Jth}, lower panel), while magnetic relaxation acts to align grains to the high-J attractor.

The alignment through by torques associated with paramagnetic dissipation was introduced by \cite{DavisGreenstein:1951}. The efficiency of paramagnetic relaxation was shown to be significantly enhanced in for grains with enhanced magnetic response (\citealt{JonesSpitzer:1967}; \citealt{RobergeLazarian:1999}) as well as in the presence of suprathermal rotation induced by Purcell's (1979) torques (see also \citealt{LazarianDraine:1997}). 
For decades, the alignment based on magnetic dissipation was accepted as the dominant mechanism of grain alignment, even though it faced problems with explaining the available set of observational data (see \citealt{Lazarian:2003}). RATs, however, were shown to be stronger and more efficient in terms of grain alignment if the high-J attractors exist (see \citealt{Anderssonetal:2015}). Nevertheless, the grains with enhanced magnetic response can demonstrate a higher degree of RAT alignment \cite{LazarianHoang:2008}. The same is expected to be true for the MET alignment.

The effects of magnetic relaxation on the RAT and MET alignment processes were recently studied in \cite{LazarianHoang:2019}). The characteristic time of the magnetic relaxation is given by 
\bea
\tau_{m} &=& \frac{I_{\|}}{K(\omega)VB^{2}}=\frac{2\rho a^{2}s^{-2/3}}{5K(\omega)B^{2}},\nonumber\\
&\simeq & 6\times 10^{5}\hat{\rho}\hat{s}^{-2/3}a_{-5}^{2}\hat{B}^{-2}\hat{K}^{-1} \yr,\label{eq:tau_DG_sup}
\ena
where $V=4\pi a^{3}/3$ is the grain volume, { $I_{\|}$ is the moment of inertia along the principal axis}, $\hat{B}=B/10\mu$G is the normalized magnetic field strength, and $\hat{K}=K(\omega)/10^{-13}\s$ and
$K(\omega)=\chi_{2}(\omega)/\omega$ with $\chi_{2}(\omega)$ is the imaginary part of complex magnetic susceptibility of the grain material.

To describe the aligning effect of magnetic relaxation relative to the disalignment by gas collisions, we introduce a dimensionless parameter
\bea
\delta_{m}&=&\frac{\tau_{\gas}}{\tau_{m}},\label{eq:delta_mag}
\ena
where 
\bea
\tau_{\rm gas}&=&\frac{3}{4\sqrt{\pi}}\frac{I_{\|}}{n_{\H}m_{\H}
v_{\rm th}a^{4}\Gamma_{\|}},\nonumber\\
&=&7.3\times 10^{4} \hat{\rho}\hat{s}a_{-5}\left(\frac{100\K}{T_{\gas}}\right)^{1/2}\left(\frac{30
\cm^{-3}}{n_{\H}}\right)\left(\frac{1}{\Gamma_{\|}}\right) \yr,~~~~\label{eq:taugas}
\ena
where $v_{\th}=\left(2k_{\B}T_{\gas}/m_{\H}\right)^{1/2}$ is the thermal velocity of a gas atom of mass $m_{\H}$ in a plasma with temperature $T_{\gas}$ and density $n_{\H}$. Above, $\Gamma_{\|}$ is a geometrical parameter, which is equal to unity for spherical grains (see \citealt{RobergeLazarian:1999}). From Equations (\ref{eq:tau_DG_sup}) and (\ref{eq:taugas}), one obtains
\bea
\delta_{m}\simeq 0.1\hat{B}^{2}\hat{K}{a}_{-5}\left(\frac{\hat{\rho}}{\hat{n}_{\H}\hat{T}_{\gas}^{1/2}}\right),\label{eq:deltam}
\ena
where $\hat{n}_{\H}=n_{\H}/(30\cm^{-3})$ and $\hat{T}_{\gas}=T_{\gas}/100\K$, where we used the values of parameters of the typical ISM and paramagnetic grains. In fact, it is easy to see that to contribute to RAT or MET alignment $\delta_m$ should be significantly larger than 1. This can be achieved if grains have magnetic inclusions. The actual $\delta_m$ depend on the size of the inclusions and is difficult to predict. Therefore below we consider a range of possible $\delta_m$ values. 

The enhancement of magnetic response is possible in the presence small superparamagnetic inclusions within grain material similar to those discovered within interstellar grains captured in the Earth atmosphere (\citealt{Bradley:1994}) and measurements by in-situ spacecraft \citep{Altobelli:2016}. We, however, want to stress that it is unclear to what extend the presence of magnetic dissipation is a natural part of the astrophysical grain alignment. Potentially, the observed degrees of polarization may be achieved by RATs with no assistance from the magnetic dissipation because the fraction of grains in their ensemble of shapes with high-J attractors can be larger than $\sim 50\%$ (\citealt{HerranenLazarian:2020}). 

The final answer on the role of magnetic dissipation should be achieved through comparing theoretical predictions with observations. A set of the potential tests is formulated in \cite{LazarianHoang:2019}. Later in this paper, we discuss the possibility of using the processes of grain disruption to test the presence of magnetic inclusions in dust grains.

We note that the magnetic dissipation alignment may dominate, however, for small grains (see \citealt{HoangLazarian:2016a}) including those of the nanoparticle size. The latter can be aligned through the resonance relaxation process that is introduced in \cite{LazarianDraine:2000}. 

\begin{figure}
\includegraphics[width=0.5\textwidth]{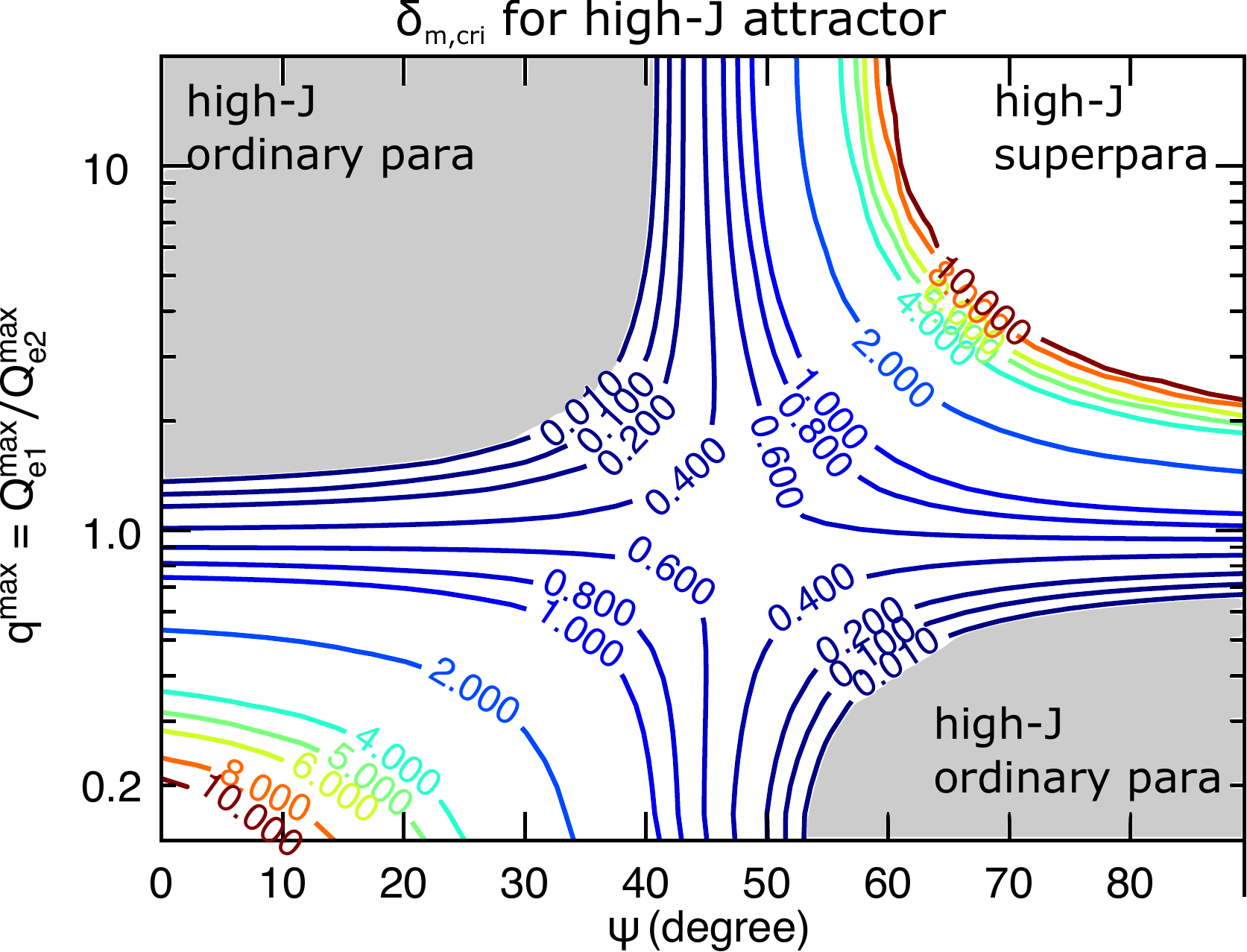}
\caption{The parameter space for grain alignment with low-J and high-J attractors for grains with enhanced magnetic response. The shaded areas correspond to the parameter space defined in LH07 for grains to have high-J attractors for negligible relaxation. The contours are given for borders of the area corresponding to different $\delta_m$. From \cite{HoangLazarian:2016a}.}
\label{fig:lowJ_highJ}
\end{figure}

The torques arising from magnetic dissipation can be synergistic to RATs and increase the degree of alignment arising from RATs. The unified grain alignment theory was introduced in \cite{LazarianHoang:2008} and numerically quantified in \cite{HoangLazarian:2016a}. There, it was shown that the enhanced magnetic dissipation can extend the parameter space in terms of $q_{\rm max}$ (see Equation \ref{qmax}) and $\psi$ for grain alignment with high-J attractors. Figure \ref{fig:lowJ_highJ} shows the contours of $\delta_{m,cri}$, the minimum value of $\delta_{m}$ required for alignment with high-J attractors as function of $q^{\rm max}$ and $\psi$.

\begin{figure}
\includegraphics[width=0.5\textwidth]{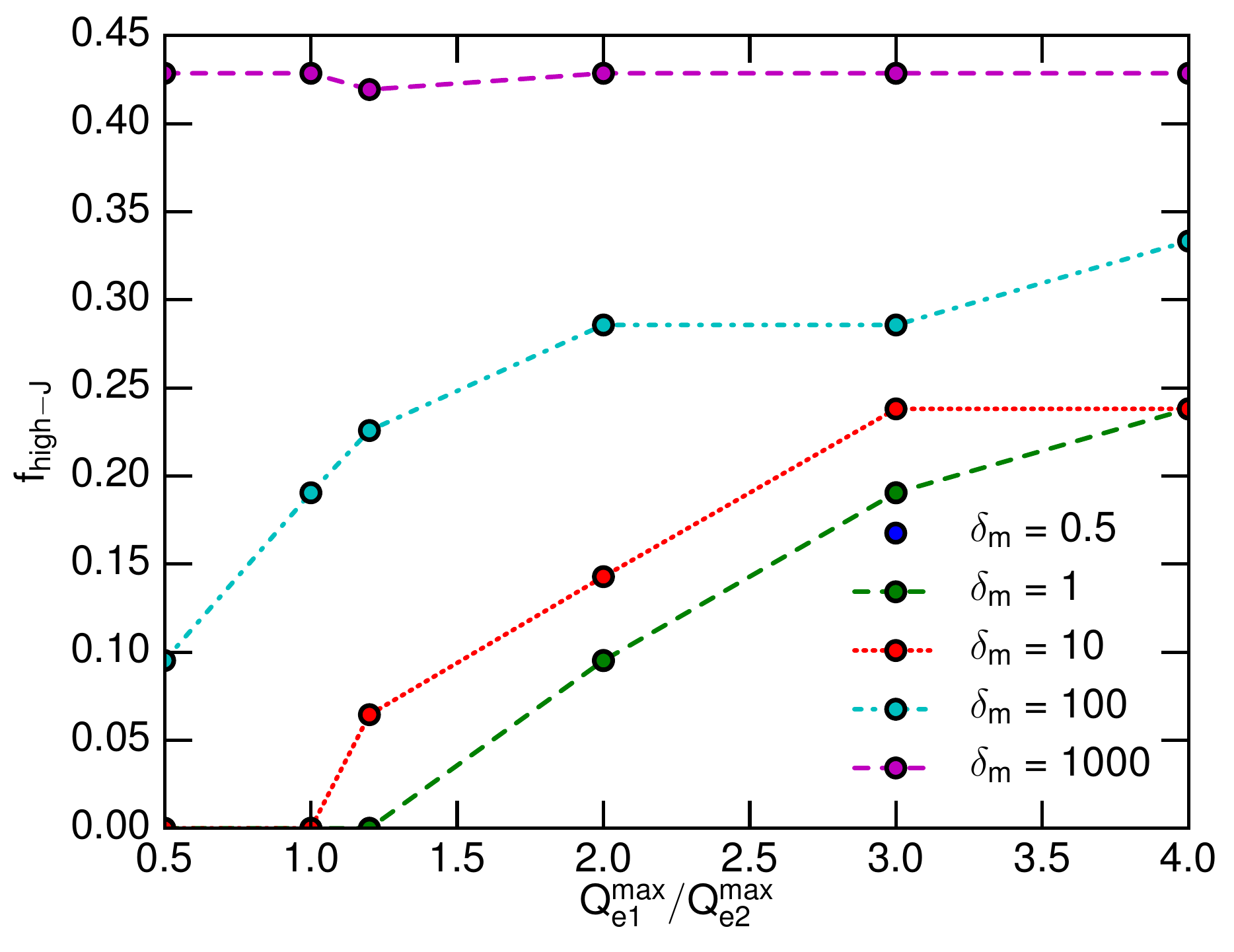}
\includegraphics[width=0.5\textwidth]{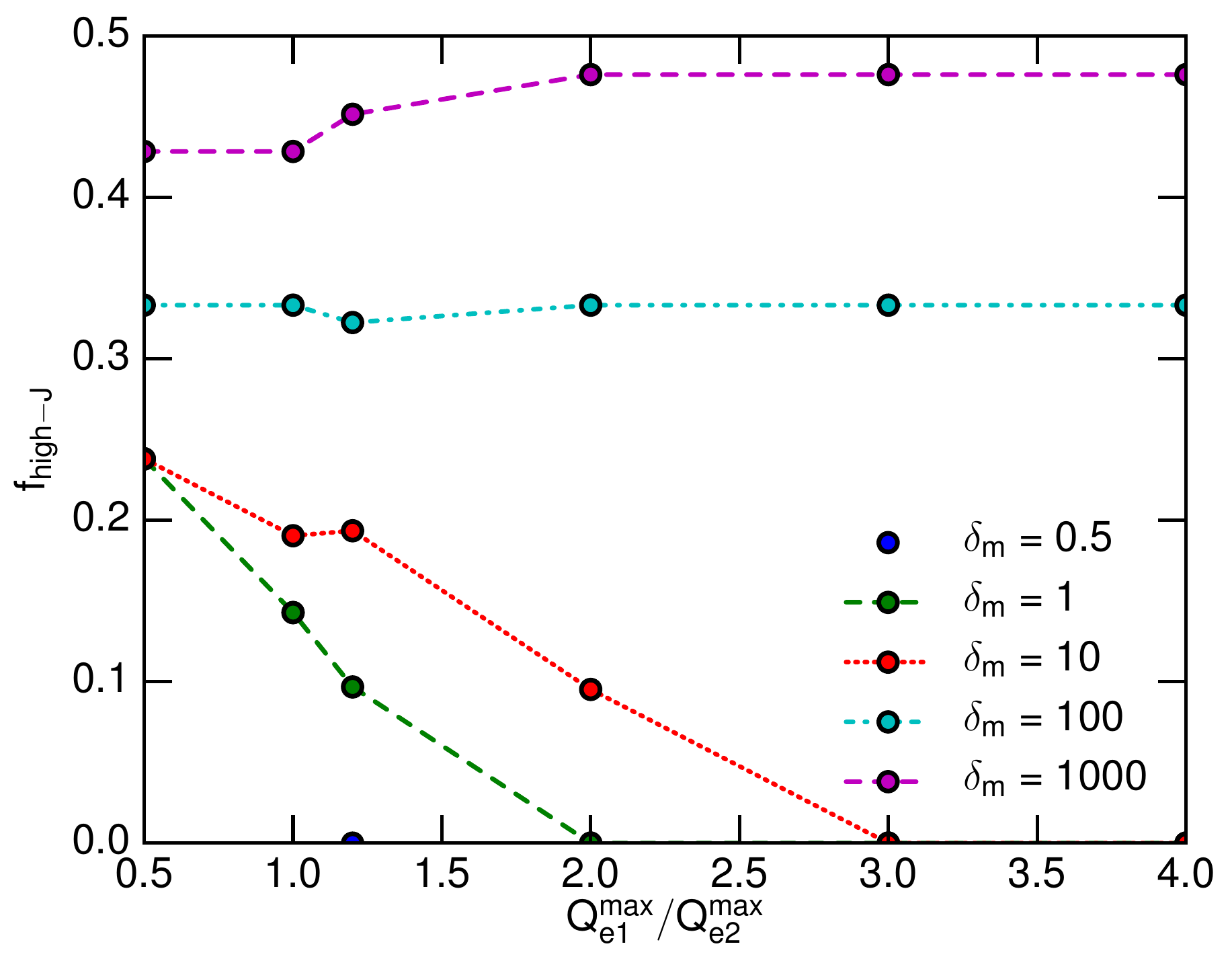}
\caption{The fraction of grains directly driven to the high-J attractor as a function of $q^{\max}=Q_{e1}^{\max}/Q_{e_2}^{\max}$ for different values of the grain magnetic susceptibility, described by $\delta_{m}$, assuming $\psi=0^{\circ}$ (upper panel) and $\psi=60^{\circ}$ (lower panel). The grain size $a=0.2\mu$m and the standard ISRF are assumed. Collisional excitation by gas are disregarded.}
\label{fig:fhighJ}
\end{figure}

\subsubsection{Fraction of grains directly driven to high-J attractors}
Since grain alignment with high-J attractors is important for quantifying the efficiency of grain alignment and rotational disruption, we calculate the fraction of grains in an ensemble that are driven to the high-J attractor, $f_{\rm high-J}$. Here the ensemble refers to grains with the same $q^{\max}$ but initially their angular momenta have different angles with respect to the magnetic field. We first solve the equations of motion for grains subject to RATs, gas damping, and magnetic relaxation, following the approach in \cite{LazarianHoang:2008} where the effect of collisional excitation is disregarded. We then calculate $f_{\rm high-J}$ by counting the fraction of grains that reach the high-J attractor. Thus, the value of $f_{\rm high-J}$ describes the fraction of grains that directly move to the high-J attractor and experience fast alignment and disruption. 

Figure \ref{fig:fhighJ} shows the variation of $f_{\rm high-J}$ vs. $q^{\rm max}$ for ordinary paramagnetic grains and grains with iron inclusions, for two angles of the radiation and magnetic field. The minimal adopted value of $q^{\max}$ is chosen to be 0.5 as suggested by the calculations in \cite{Herranenetal:2019}.  It is evident from Figure \ref{fig:fhighJ} that, the fraction of grains directly moved by RATs to the high-J attractor depends on $q^{\max}$, the angle between the radiation and magnetic field, and magnetic response. For ordinary paramagnetic grains, i.e., grains with $\delta_m<1$. This fraction increases with increasing $\delta_m$ and becomes independent of $\psi$ and $q^{\max}$ for $\delta_m\gtrsim 10^3$ at which $f_{\rm high-J}\sim 0.45$.
 
According to \cite{Herranenetal:2019}, the value $q^{\max}$ mostly lies in the range $0.8$ to $1.5$. The upper panel of Figure \ref{fig:fhighJ} shows $f_{\rm high-J}\lesssim 0.05$ for $\delta_m\lesssim 1$ when the radiation is parallel to the magnetic field, i.e., $\psi=0$. The alignment with respect to the radiation direction corresponds to $\delta_m=0$ and $\psi=0$. This means that the fraction of grains that will have high-J attractors when they are aligned with the radiation direction is expected to be negligible. This suggests that the alignment with respect of the radiation direction is not going to be perfect and the grains aligned with respect to the radiation are not expected to rotate fast. This, as we discuss further, has important consequences in terms of rotational disruption of dust. 

\begin{figure}
\includegraphics[width=0.5\textwidth]{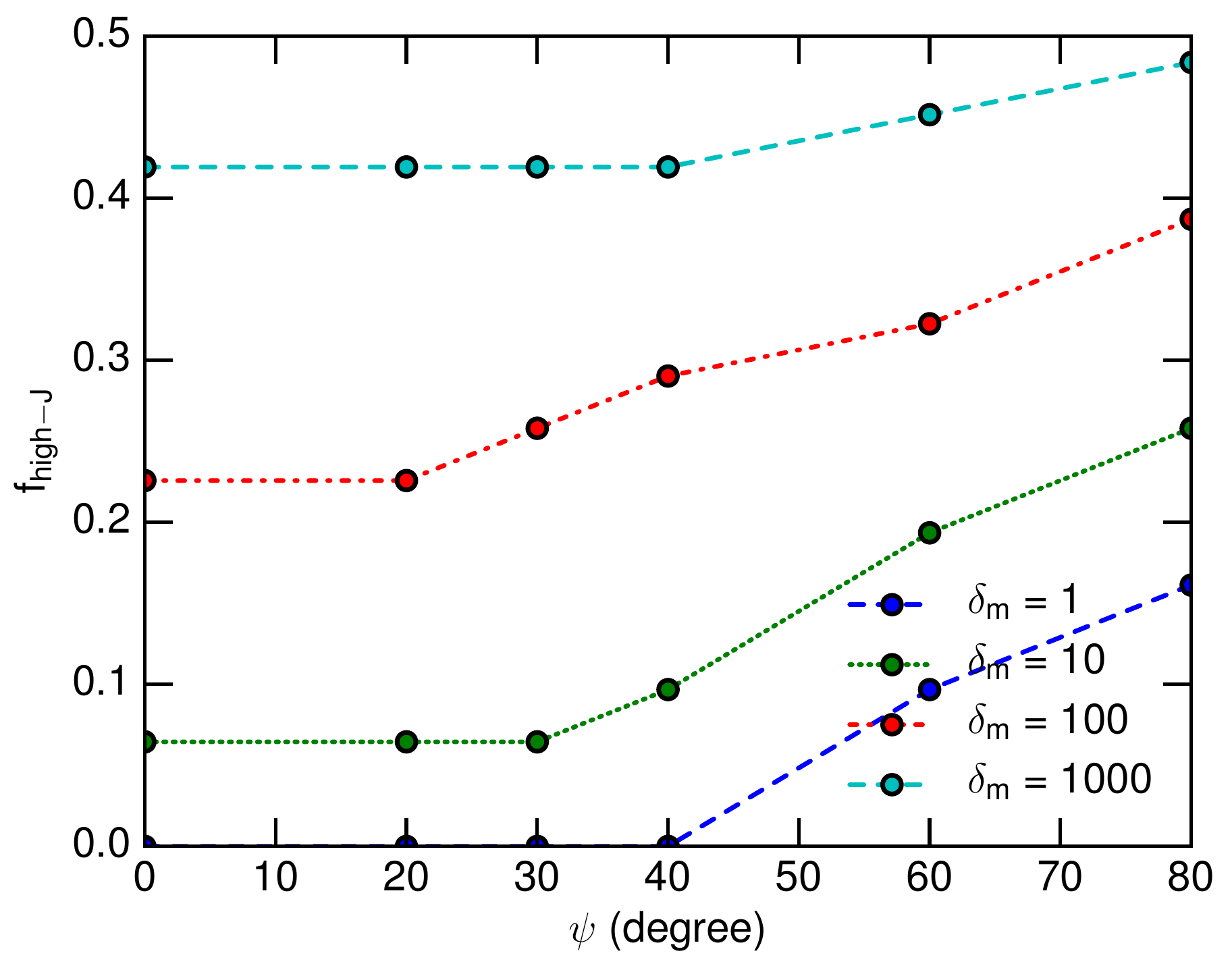}
\caption{Same as Figure \ref{fig:fhighJ}, but for the fraction of grains directly driven to the high-J attractor as a function of the angle $\psi$ between the radiation and the magnetic field for different values of the grain magnetic susceptibility. The typical value of $q^{\max}=1.2$ is assumed.}
\label{fig:fhighJ_deltam}
\end{figure}

Thus, in Figure \ref{fig:fhighJ_deltam}, we adopt $q^{\max}=1.2$ and show $f_{\rm high-J}$ as a function of $\psi$, assuming different values of $\delta_{m}$. We observe a systematic tendency of increasing $f_{\rm high-J}$ with the increase of $\psi$ for low $\delta_{m}$, but it becomes independent of $\delta_{m}$ for grains with high level of iron inclusions.

\subsubsection{Fast alignment at high-J attractors}

{\it Fast alignment} on the timescale of the order of $\Omega_{k}^{-1}$ was introduced in LH07 and was shown to bring the majority of grains to the alignment at low-J attractor points.  On the other hand, a fraction $f_{\rm high-J}$ of grains can be driven directly towards high-J attractor points on the time scale $\Omega_{k}^{-1}$ (see Figure \ref{fig:fhighJ_deltam}). Our present study shows that if grains are strongly magnetic, the value of $f_{\rm high-J}$ could reach $50\%$. This means that for this fraction of grains, the perfect alignment can be achieved fast. Note that grains driven to low-J attractor points are weakly aligned with the net degree of alignment less than $20\%$ \citep{HoangLazarian:2008}. As we will show in the following section, gas randomization can gradually transport grains from the low-J to high-J attractor, providing {\it slow} alignment, and eventually perfect alignment.


\section{Spin-up, spin-down, and randomization of grains}\label{sec:spinup_down}

RAT torques induce grain rotation and its alignment. The rate of grain rotation subject to RATs is controlled by the rotational damping, and the degree of alignment is affected by the processes of grain randomization. 

\subsection{RAT amplitudes}

 Let $u_{\lambda}$ be the spectral energy density of radiation field at wavelength $\lambda$. The~energy density of the radiation field is then $u_{\rm rad}=\int u_{\lambda}d\lambda$. To~describe the strength of a radiation field, let define $U=u_{\rm rad}/u_{\rm ISRF}$ with 
$u_{\rm ISRF}=8.64\times 10^{-13}\erg\cm^{-3}$ being the energy density of the average ISRF in the solar neighborhood (\citealt{Mathisetal:1983}). Thus, the~typical value for the ISRF is $U=1$.

Let $a=(3V/4\pi)^{1/3}$ be the effective size of the dust grain of irregular shape with volume $V$. Such an irregular grain exposed to an anisotropic radiation field experiences radiative torque (RAT) due to differential absorption and scattering of left-handed and right-handed photons. The~magnitude of RATs is defined as
\bea
{\Gamma}_{\lambda}=\pi a^{2}
\gamma u_{\lambda} \left(\frac{\lambda}{2\pi}\right){Q}_{\Gamma},\label{eq:GammaRAT}
\ena
where $\gamma$ is the anisotropy degree of the radiation field, and~${Q}_{\Gamma}$ is the RAT efficiency (\citealt{DraineWeingartner:1996}). Typically, $\gamma\approx 0.1$ for the diffuse ISRF, $\gamma\sim$ 0.3--0.7 for molecular clouds (\citealt{LazarianHoang:2007a}), and~$\gamma=1$ for unidirectional radiation fields (e.g., from~a nearby star). 

The magnitude of RAT efficiency, $Q_{\Gamma}$, can be approximated by a power-law (\citealt{LazarianHoang:2007a}):
\bea
Q_{\Gamma}\sim 0.4\left(\frac{{\lambda}}{1.8a}\right)^{\eta},\label{eq:QAMO}
\ena
where $\eta=0$ for $\lambda \lesssim 1.8a$  and $\eta=-3$ for $\lambda > 1.8a$. Numerical calculations of RATs for several shapes and different optical constants using the DDSCAT code (\citealt{DraineFlatau:1994}) by \citealt{LazarianHoang:2007a} and \cite{Herranenetal:2019} shows good agreement with the approximated torque. The value of $Q_{\Gamma}$ in Equation (\ref{eq:QAMO}) is for the maximum RAT efficiency corresponding to the radiation parallel to the axis of maximum moment of inertia (or $\Theta=0$). These equations are introduced to discuss disruption size and disruption time in Sec 5. The torques averaged over the narrow angles at high-J would be larger than that averaged over low-J attractors. Since we are interested in here the disruption, this equation is justified because at high-J both internal alignment is perfect, producing $\Theta=0$.

Let $\overline{\lambda}=\int \lambda u_{\lambda}d\lambda/u_{\rm rad}$ be the mean wavelength of the radiation spectrum. For~the ISRF, $\overline{\lambda}=1.2\mum$. The average radiative torque efficiency over the radiation spectrum is defined as
\bea
\overline{Q}_{\Gamma} = \frac{\int \lambda Q_{\Gamma}u_{\lambda} d\lambda}{\int \lambda u_{\lambda} d\lambda}.
\ena

For interstellar grains with $a\lesssim \overline{\lambda}/1.8$, $\overline{Q}_{\Gamma}$ can be approximated to (\citealt{HoangLazarian:2014})
\bea
\overline{Q}_{\Gamma}\simeq 2\left(\frac{\overline{\lambda}}{a}\right)^{-2.7}\simeq 2.6\times 10^{-2}\left(\frac{\overline{\lambda}}{0.5\mum}\right)^{-2.7}a_{-5}^{2.7},\label{eq:QRAT_avg}
\ena
and $\overline{Q_{\Gamma}}\sim 0.4$ for $a>\overline{\lambda}/1.8$. For convenience, let $a_{\rm trans}=\bar{\lambda}/1.8$ be the transition size of grains from a flat to the power-law stage of RATs. A rigorous derivation of the average RAT efficiency over a radiation spectrum is presented in \cite{Hoangetal:2020}, where a slightly different scaling is obtained.

Plugging $\overline{Q}_{\Gamma}$ into Equation~(\ref{eq:GammaRAT}) yields the radiative torque averaged over the radiation~spectrum,
\bea
\Gamma_{\rm RAT}&=&\pi a^{2}
\gamma u_{\rm rad} \left(\frac{\overline{\lambda}}{2\pi}\right)\overline{Q}_{\Gamma}\nonumber\\
&\simeq&  5.8\times 10^{-29}a_{-5}^{4.7}\gamma U\overline{\lambda}_{0.5}^{-1.7}\erg,\label{eq:Gamma_avg1}
\ena
for $a\lesssim a_{\rm trans}$, and
\bea
\Gamma_{\rm RAT}\simeq & 8.6\times 10^{-28}a_{-5}^{2}\gamma U\overline{\lambda}_{0.5}\erg,\label{eq:Gamma_avg2}
\ena
for $a> a_{\rm trans}$, where $\overline{\lambda}_{0.5}=\overline{\lambda}/(0.5\mum)$.

\subsection{Classical randomization}
\label{classical}

Thermally rotating grains experience randomization of their orientation on the time scale of the damping of their rotation. Different processes, including gas collisions, infrared emission, plasma drag (see \citealt{DraineLazarian:1998}), are responsible for the grain rotational damping. The total rate of rotational damping is
\begin{equation}
\tau_{\rm damp}^{-1}=\tau_{\rm gas}^{-1} + \tau_{\rm IR}^{-1} + \tau_{\rm other}^{-1},\label{eq:tdamp}
\end{equation}
where the first term denotes the gaseous damping as given by Equation (\ref{eq:taugas}). The second term describe the damping by IR emission, which can be written as
\bea
\tau_{\rm IR}^{-1}=F_{\rm IR} \tau_{\gas}^{-1},\label{eq:tauIR}
\ena
where $F_{\rm IR}$ is a damping coefficient for a grain having an equilibrium temperature $T_{\rm d}$, which is given by
\bea
F_{\rm IR}\simeq \left(\frac{0.4}{a_{-5}}\right)\left(\frac{u_{\rm rad}}{u_{\rm ISRF}}\right)^{2/3}
\left(\frac{30 \cm^{-3}}{n_{\H}}\right)\left(\frac{100 \K}{T_{\gas}}\right)^{1/2}.\label{eq:FIR}
\ena

Additional processes, e.g., related to plasma-dust interactions, emissions of microwave emission (see \citealt{DraineLazarian:1998}; \citealt{Hoangetal:2010}) etc., can be important for rotational damping. Those correspond to the rate $\tau_{\rm other}^{-1}$ in Equation (\ref{eq:tdamp}). For large grains of $a\gtrsim 0.01\mum$, these processes are subdominant, and we disregard them in the current paper. 

\subsection{Grain randomization and alignment cycles at the low-$J$ attractor point} 

Grain randomization is most significant when the value of angular momentum $J$ is minimal. Therefore the randomization is very significant at the low-$J$ attractor points. The effects of the randomization by the gaseous bombardment in the context of RAT alignment was considered in \cite{HoangLazarian:2008}. We note that the randomization can happen due to pure radiation interaction with a thermally wobbling grain (see Section \ref{wobble}). Indeed, according to Equation (\ref{therm_dist}) grains with $E_{kin}$ of the order $kT_{d}$ or lower exhibit significant wobbling with respect to the alignment axes. For some of these angles they experience RAT component $F$ that both increase $\theta$ and torques $H$ that accelerate the grain (see Figure \ref{torques}). The increase of the angular momentum perpendicular to the original direction of grain angular momentum at the low-$J$ attractor point can be $\sim H \tau_{\rm int}$ and can significantly change the direction of the alignment.

In addition, toward the low-J attractor point, the grains get randomized both through the gaseous bombardment, emission of IR photons, and internal thermal fluctuations related through the Fluctuation Dissipation Theorem (\citealt{Weber:1956}) to the processes of internal energy dissipation within a wobbling body. The gaseous randomization time is $\tau_{\gas}\approx (T_{d}/T_{\gas}) \tau_{\rm damp}$, where $\tau_{\rm damp}$ is the gas damping time given by Equation (\ref{eq:taugas}). The factor $T_{d}/T_{\gas}$ arises here due to the fact that the angular momentum of a grain in a low-J attractor is reduced by the factor $(T_{d}/T_{\gas})^{1/2}$ and the randomization proceeds as a random walk process. We expect the same factor to be present for other types of classical randomization in Equation (\ref{eq:tdamp}), but keep cautious about the anomalous randomization that requires further studies.

During the low-J rotation stage, the grains are experiencing fluctuations of the angle between $\bf J$ and the grain axes. The characteristic time of these fluctuations is given by Equation (\ref{NR}) for the nuclear relaxation process. This sort of randomization does not involve the change of the direction ${\bf J}$. It does complicate the analysis, but our estimates show that this process does not appreciably accelerate the transport of grains over the phase space. Therefore we do not expect this process to increase the grain transport to the high-J points. As a result, the grains can be moved to alignment over the time scale of the order of grain precession time given by Equation (\ref{rad_p}). This process moves was termed in LH07 "fast alignment"\footnote{In LH07 the fast alignment was explored numerically and the initial angular momentum was taken to be $ 60 I_\| \omega_T$. Thus the time of the fast alignment there was $\sim 60$ precession times of the grain with thermal value of angular momentum $I_\| \omega_T$.} and this process was predicted to drive alignment near the variable radiation sources, as was confirmed by the subsequent research (see \citealt{Hoang:2017}). 


\begin{figure}
    \includegraphics[width=0.5\textwidth]{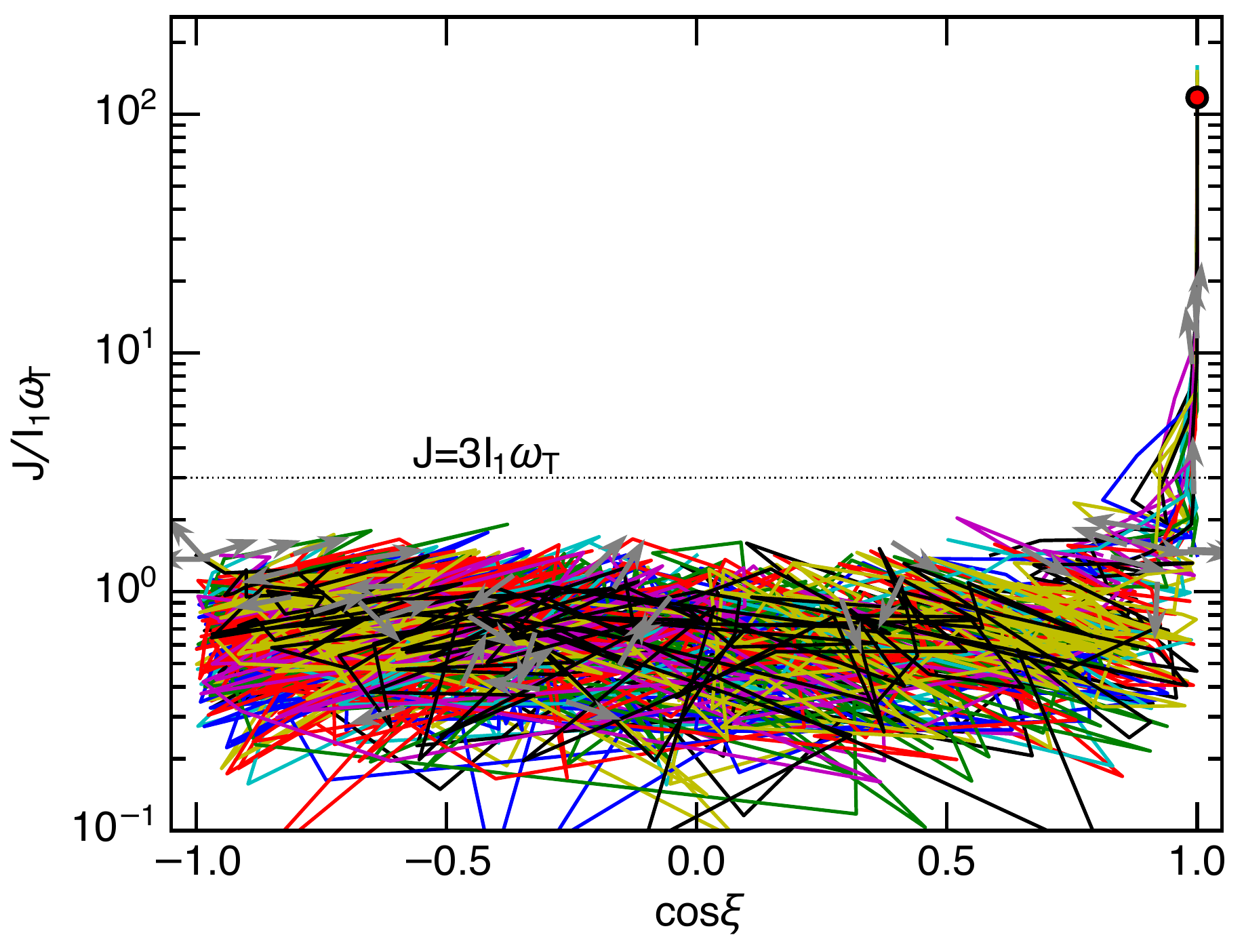}
    \includegraphics[width=0.5\textwidth]{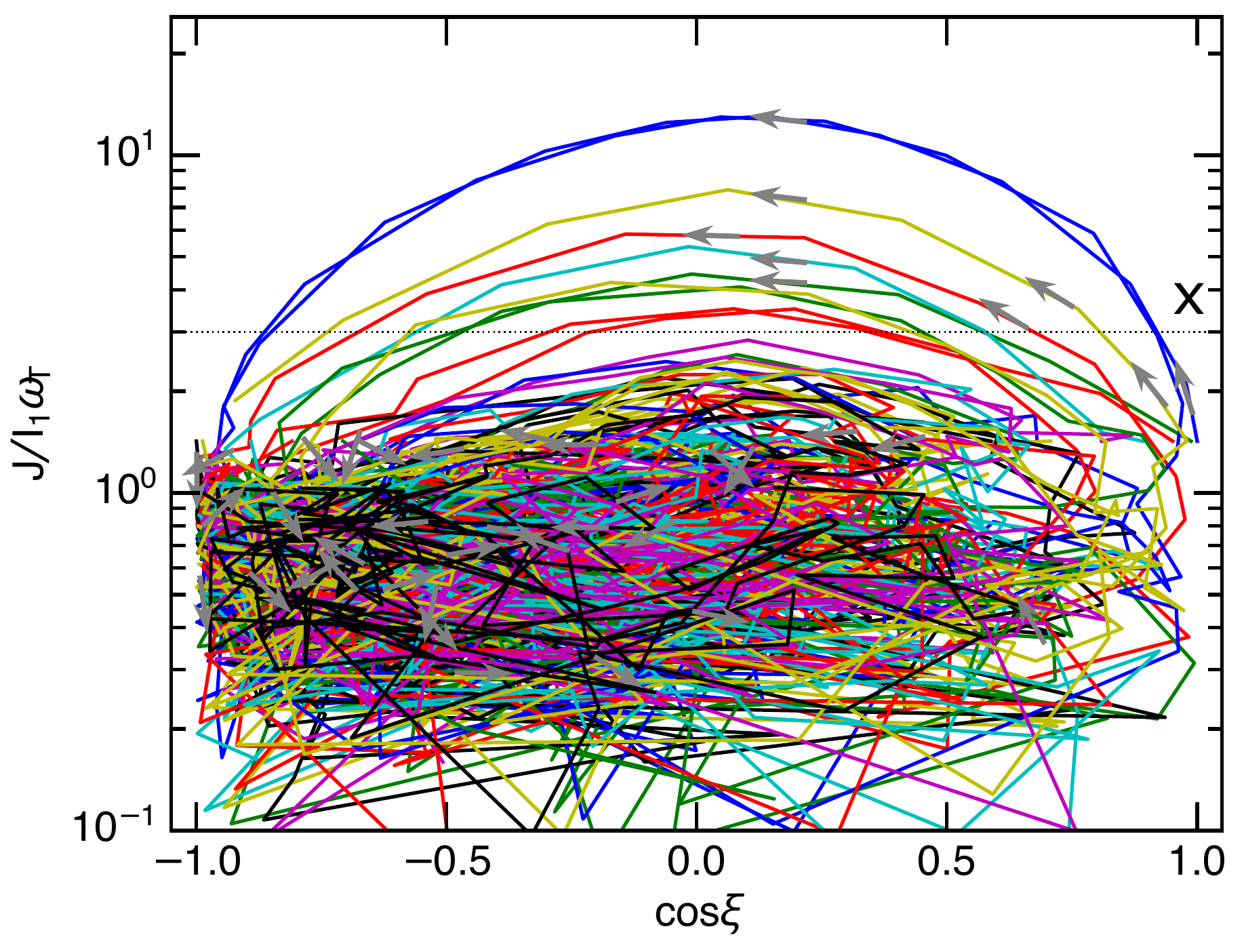}
    \caption{Upper panel: The phase trajectory map for grain alignment in the presence of gas randomization for the case with a high-J attractor (marked by a red filled circle). Lower panel: the phase map for alignment without a high-J attractor, i.e., with a high-J repellor point (marked by X). The arrows indicate the time-evolution of the grain angular momentum in the phase space. The grain size $a=0.2\mum$ and the standard ISM are assumed. The horizontal dotted line mark the angular momentum of $J=3I_{1}\omega_{T}$ above which the grain trajectory is deterministic. For $J<3I_{1}\omega_{T}$, the grain's trajectory is random due to gas collisions.}
    \label{fig:map_lowJ_random}
\end{figure}

To illustrate the effect of gas randomization on the low-J attractor, we follow the numerical approach in \cite{HoangLazarian:2016a} to obtain the phase trajectory map for grains that initially start at $J=J_{\rm th}=I_{1}\omega_{T}$. The upper panel of Figure \ref{fig:map_lowJ_random} shows the phase map for the alignment with high-J attractors. As shown, the trajectory of grains at low-J ($J\lesssim J_{\rm th})$ is randomized strongly. Eventually, grains are driven to the high-J attractor due to transport by gas collisions. The lower panel of Figure \ref{fig:map_lowJ_random} shows the phase trajectory map for grain alignment without high-J attractors. Grains are strongly randomized during the slow rotation stage of $J< 3J_{\rm th}$, but their trajectory is more deterministic for $J\gtrsim 3J_{\rm th}$. When grains approach the high-J repellor (marked by X), they are repelled back and follow their trajectory back to the low-rotation stage (see arrows). This process continues and grains undergo many cycles back and forth between the low-J and high-J repeller points. It is evident that in the absence of high-J attractor, the probability of grain getting high angular momentum significantly is reduced, and grains do not get as high angular velocities compared to the case when the high-J attractor is present.

As we discussed in Section \ref{sec:pinwheel}, the pinwheel torques can stabilize the grain rotation in the low-J attractor point. In this situation, the time scale of grain at low-J alignment and therefore of relatively slow rotation can be significantly enhanced. This is the process that decreases the rotational disruption of grains (see Section \ref{sec:rot_disr}). However, due to thermal fluctuations in grain material (see Appendix \ref{app:D}), grains may experience thermal flipping, which for some grains can significantly decrease the efficiency of pinwheel torques (see Section \ref{purcell}).  

\subsection{Effects of grain electric moment variations}

Grain alignment and the rate of grain rotation can be significantly affected by the randomization processes. Indeed, the dynamics of grains that rotate at low-$J$ or high-$J$ attractors is very different from the dynamics of grains that randomly sample the entire parameter space. 

A process of strong anomalous randomization was proposed by \cite{Weingartner:2006}, (henceforth W06). This process questioned the applicability of all earlier studies to the grain alignment in realistically turbulent ISM.

Astrophysical fluids, in particular, in the ISM, are turbulent (see \citealt{Larson:1981}; \citealt{Armstrong:1995}; \citealt{ChepurnovLazarian:2009}, see \citealt{McKeeOstriker:2007} for a review). The turbulent motions induce relative motions of dust with respect to the magnetic field. 
Grains moving with respect to the magnetized gas due to turbulent motions (\citealt{LazarianYan:2002}; \citealt{YanLazarian:2003}; \citealt{Hoangetal:2012}) experience the electric field $E_{\rm ind}=V_{\rm grain}/c\times B$ while moving with velocity $V_{\rm grain}$ in the magnetic field $B$. This is the field that is experienced by the dust grain of velocity $V_{\rm grain}$ as it, for instance perform Larmor rotation around magnetic field.  

The electric field interacting with the grain electric dipole moment induces an additional precession. It was noted  by W06 that if the electric dipole moment of a grain changes, this causes an additional randomization. This process that can be termed "anomalous randomization" was further elaborated in \cite{Weingartner:2009}(henceforth JW09).

We address the problem in Appendix B and our calculations demonstrate that the effect of the variations of grain electric moment is suppressed and not important for astrophysical settings.

\subsection{Significance of damping rate}

The most trivial consequence of the damping torques is that the rotational rate of grains in high-$J$ attractors is determined by the balance of RATs or METs and the damping torques acting on a grain. From the point of view of alignment the comparison of the times of the alignment and damping is of more value. The process of fast alignment was introduced in LH07. This fast alignment drives grains to the low-$J$ attractor and it takes the time of the order of the period of grain precession $\sim \Omega_{\rm rad}^{-1}$ (see Equation (\ref{rad_p}). Similarly, for METs the corresponding time is of the order of $t_{\rm mech}$ given by Equation (\ref{t_mex}). If the $\tau_{\rm damp}$ given by Equation (\ref{eq:tdamp}) is less than any of these time scales, the corresponding process of alignment is completely suppressed. 

However, in the presence of gas randomization, alignment in the low-J attractor is unstable (see Figure \ref{fig:map_lowJ_random}; also \citealt{{HoangLazarian:2008},{HoangLazarian:2016a}}). Perfect alignment with the high-$J$ attractor points takes place over a few $\tau_{\rm damp}$ (see Figure \ref{fig:RQJ_CHI}, upper panel). Thus, rather counterintuitively, the faster damping may correspond to faster perfect alignment of grains. However, this can be physically understood because the perfect alignment is achieved by the transport from the low-J to high-J attractors. In this picture, fast damping corresponds to faster randomization of grains at the low-J attractor and faster transport of the grains to the high-J attractor.

The process of grain alignment depends on the efficiency of internal alignment, i.e., the alignment of ${\bf J}$ with respect to the axis of the maximal moment of inertia. If the rate of internal relaxation $\tau_{\rm int}^{-1}$ given by Equation (\ref{intern}) is less than the rate of $\tau_{\rm damp}$, the alignment takes place without effects of internal relaxation. It was shown in \cite{HoangLazarian:2009a} that its properties are different from the alignment in the presence of internal relaxation. In particular, the alignment at low-$J$ attractors can happen with grain long axes parallel to the alignment axis. In case of $\Omega_B$ being the largest precession frequency, this means the alignment with grain long axes parallel to ${\bf B}$. This may be the case of sufficiently large grains. At the same time, depending on the radiation intensity,  $\tau_{\rm int}^{-1}$ can be longer than the rate of fast alignment $\sim \Omega_{\rm rad}$. This means that, on short scales, the transient RAT alignment can happen with long grain axis parallel to magnetic field. This is an important effect to keep in mind for the grain alignment and disruption in the vicinity of the time-dependent radiation sources. Naturally, the same conclusions are applicable to METs with shocks aligning long grain axes parallel to the direction of shock propagation. Note, that on the time scales $\sim \tau_{\rm int}$ the alignment can changes its direction.

\section{Rotational disruption of grains by RATs}
\label{sec:rot_disr}

Fast rotating grains can be disrupted due their centrifugal stress. We show below that the rotational disruption is very closely associated with grain alignment, and its efficiency depends on different parameters in analogy to grain alignment. 
\subsection{Disruption size and time for fast rotating grains}

Astrophysical grains cannot withstand high centrifugal stress with the critical velocity of rotation that depends on the grain composition, grain structure and shape (\citealt{Hoangetal:2019}). Rotational disruption of grains by RATs (namely RATD) was evaluated in \cite{Hoang:2019} for different structures, assuming the maximal efficiency of RATs arising from the typical ISRF and the enhanced values of the radiation field. Composite grains were shown to disrupt starting at the size $\sim 0.2\mum$ for the standard ISRF and gas properties (\citealt{Mathisetal:1983}). For grains closer to a star, for instance, with the 10 times the ISRF, even compact grains of $\sim 0.5\mum$ are subject to disruption. 

Let $S_{\rm max}$ be the tensile strength of the grain material. A spherical grain of radius $a$ spinning at angular velocity $\omega$ experiences an average centrifugal stress $S=\rho a^{2}\omega^{2}/4$ acting on the plane through the grain center. The critical angular speed above which the grain is disrupted is determined by setting $S=S_{\rm max}$, yielding,
\begin{eqnarray}
\omega_{\rm disr}=\frac{2}{a}\left(\frac{S_{\rm max}}{\rho}\right)^{1/2}\simeq \frac{3.65\times 10^{8}}{a_{-5}}\left(\frac{S_{\rm max,7}}{\hat{\rho}}\right)^{1/2}~\rm rad s^{-1},\label{eq:w_disr}~~~~
\end{eqnarray}
where $S_{max,7}=S_{\rm max}/(10^{7}\rm erg \cm^{-3})$.

If the grain is being spun-up to the high-J attractor point by the radiation source with stable luminosity, the radiative torque $\Gamma_{\rm RAT}$ is constant, and the grain velocity is steadily increased over time. The maximum angular velocity of grains spun-up by RATs is given by
\bea
\omega_{\rm RAT}=\frac{\Gamma_{\rm RAT}\tau_{\rm damp}}{I_{1}},~~~~~\label{eq:omega_RAT0}
\ena
where $\Gamma_{\rm RAT}$ is given by Equation (\ref{eq:GammaRAT}), and the factor $\alpha_{1}$ in the inertia moment $I_{1}$ is taken to be unity. The situation is different for grains being moved to the low-J attractor point. The angular momentum of such grains decreases. 

For a general radiation field, the maximum rotation rate induced by RATs is given by
\bea
\omega_{\rm RAT}&\simeq & 9.22\times 10^{7}\gamma_{-1} a_{-5}^{0.7}\bar{\lambda}_{0.5}^{-1.7}
\left(\frac{U}{n_{1}T_{2}^{1/2}}\right)\nonumber\\
&&\times \left(\frac{1}{1+F_{\rm IR}}\right)\rad\s^{-1},\label{eq:omega_RAT}
\ena
for grains with $a\lesssim a_{\rm trans}$, and~\bea
\omega_{\rm RAT}&\simeq& 1.42\times 10^{9}\gamma_{-1}a_{-5}^{-2}\bar{\lambda}_{0.5}
\left(\frac{U}{n_{1}T_{2}^{1/2}}\right)\nonumber\\
&&\times \left(\frac{1}{1+F_{\rm IR}}\right)\rad\s^{-1},
\ena
for grains with $a> a_{\rm trans}$. Here $\gamma_{-1}=\gamma/0.1$ is the anisotropy of the radiation field relative to the typical anisotropy of the ISRF. 

The minimum size of grain disruption is obtained by setting $\omega_{\rm RAT}=\omega_{\rm disr}$, yielding
\bea
a_{\rm disr}\simeq 0.22\bar{\lambda}_
{0.5}S_{\max,7}^{1/3.4}
 (1+F_{\rm IR})^{1/1.7}\left(\frac{n_{1}T_{2}^{1/2}}{\gamma_{-1}U}\right)^{1/1.7}\mum.~~~\label{eq:adisr}
\ena

The disruption size depends on the tensile strength, gas property, and radiation field. For composite grains of $S_{\rm max}\sim 10^{7}\erg\cm^{-3}$, large grains of $a>0.2 \mum$ are already disrupted even with the standard ISRF of $U=1$. The minimum size of rotational disruption decreases with increasing the radiation intensity, but increases with increasing the gas density. The maximum rotation rate increases rapidly with $a$, implying that largest grains are easier to be disrupted than small ones, but very large grains (e.g., $a\gtrsim 10\mum)$ may not be disrupted due to the decrease of $\omega_{\rm RAT}$ with $a$ for $a>a_{\rm trans}$ (\citealt{Hoang:2019}; \citealt{Hoangetal:2019}; see \citealt{Hoang:2020} for a review).


The characteristic timescale for grain disruption is defined as
\bea
t_{\rm disr,min}=\frac{I\omega_{\rm disr}}{\Gamma_{\rm RAT}} \simeq 10^{5} (\gamma U)^{-1}\bar{\lambda}_{0.5}^{1.7}\hat{\rho}^{1/2}S_{\max,7}^{1/2}a_{-5}^{-0.7} \yr~~~~\label{eq:tdisr1}
\ena
for $a_{\rm disr}<a \lesssim a_{\rm trans}$, and~
\bea
t_{\rm disr,min}\simeq 7.4(\gamma U)^{-1}\bar{\lambda}_{0.5}^{-1}\hat{\rho}^{1/2}S_{\max,7}^{1/2}a_{-5}^{2}{~\yr}\label{eq:tdisr2}
\label{eq:tmin}
\ena
for $a_{\rm trans}<a<a_{\rm disr,max}$ (\citealt{Hoang:2019}). 

The minimum disruption time is the disruption time for {\it fast disruption} case, or for grains already aligned at high-J attractors prior the enhancement of the radiation field by transients (\citealt{Hoangetal:2019}; \citealt{Giangetal:2020}). The disruption time also depends on the angle between the radiation and magnetic field because of the dependence of the angular momentum spun-up by RATs on $\psi$ as shown in Figure \ref{fig:Jmax_angle}. The disruption time varies within a factor of 3 for $\psi<45^{\circ}$, but it can increase significantly for $\psi \rightarrow 90^{\circ}$.

\subsection{Dependence of disruption on angle $\psi$ between alignment axis and radiation}
If grains are aligned with respect to the magnetic field, as shown in Figure \ref{fig:fhighJ_deltam}, the fraction of grains aligned with high-J attractors increases with increasing the angle $\psi$ between the radiation and the magnetic field, assuming fixed $\delta_{m}$ and $q^{\rm max}$. 

\begin{figure}
\includegraphics[width=0.5\textwidth]{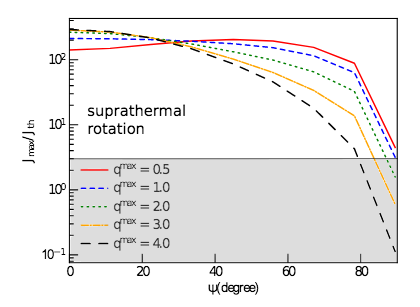}
\caption{Dependence of the angular momentum at the high-J stationary point as a function of the angle $\psi$ between the radiation beam and the magnetic field, assuming different values of $q^{\rm max}$ for RATs from the AMO. We assume a grain size $a=0.2\mum$, physical parameters of the ISM and ISRF. The darkened area corresponds to the region where effects of grain randomization become important. Modified from \cite{LazarianHoang:2019}.}
\label{fig:Jmax_angle}
\end{figure}

Moreover, the value of $J$ at the high-J stationary point, i.e., the maximum angular momentum, depends on the angle between the magnetic field and the radiation direction. Figure \ref{fig:Jmax_angle} illustrates this dependence. For $\psi <45^{\circ}$, the maximum rotation rate essentially decreases by a factor of 3, depending on the value of $q^{\max}$. When the radiation is nearly perpendicular to the magnetic field, $\psi\rightarrow 90^{\circ}$ and grains are aligned with respect to the magnetic field, then the rate of grain rotation decreases significantly. As a result, the disruption size and time of rotational disruption change slowly for $\psi<45^{\circ}$ due to weak dependence of $a_{disr}$ on $\omega_{\rm RAT}$ (see Eq. \ref{eq:adisr}). The disruption is inefficient when the radiation is perpendicular to the magnetic field, such that grains can withstand higher intensity radiation exposure. Note that sufficiently strong radiation beam can change the direction of grain alignment, imposing the dominant precession with respect to the radiation direction.


\subsection{Alignment of Grain Fragments}

The fragments of the disrupted grains will have the angular momentum of the initial grain. This does not necessarily guarantee that the fragments are going to have their angular momenta in the same direction as the disrupted grain, but, nevertheless, makes their alignment with the initial angular momentum rather likely. Therefore, if the fragments have the high-J attractor point, the fragments are also likely to find themselves in the high-J attractor points. Such fragments are likely to be both aligned and, if RATs are strong enough, may be subject to further disruption.

If the fragments have only low-J attractors, they are bound to slow down and lose a high degree of alignment. Such fragments are unlikely to experience further disruption. If grains had magnetic inclusions, the fragments with the inclusions will be aligned and may be further disrupted. Within sufficiently strong radiation sources, this disruption process can result in a population of grains that only consist of the pure magnetic material with high tensile strength, provided that these resulting inclusions are sufficiently large that they can be spun up by RATs.

\section{Time dependence of grain alignment and rotational disruption}\label{sec:timedepend}
In this section, we discuss time-dependence alignment and disruption, which is particularly important for grains subject to transients such as supernovae (SNe), novae and kilonovae, and Gamma-ray bursts (GRBs).

\subsection{Role of grain randomization and transport of grains from low-J to high-J attractors}

If RATs/METs align grains with high-J attractors, a fraction $f_{\rm high-J}$ of grains are rapidly driven to the high-J attractor (see Figure \ref{fig:MAP_Jth}). For grains with strongly magnetic inclusions, the values $f_{\rm high-J}$ increases to $\sim 50\%$ (see Figures \ref{fig:fhighJ} and \ref{fig:fhighJ_deltam}). Thus, a fraction $f_{\rm high-J}$ grains experience {\it fast alignment} and {\it fast disruption}. In this section, we discuss the effect of {\it slow alignment} and disruption that occurs with $1-f_{\rm high-J}$ fraction of grains that are first driven to the low-J attractor by RATs (see Figure \ref{fig:MAP_Jth}).
 
As shown in the upper panel of Figure \ref{fig:map_lowJ_random}, randomization by gas collisions occasionally drive grains from low-J into high-J. Eventually, all grains establish their stable alignment at high-J attractors (see also \citealt{{HoangLazarian:2008},{HoangLazarian:2016a}}). This corresponds to time-dependent grain alignment and disruption. The terminal timescale is several gas damping time $\tau_{\rm damp}$ for ordinary paramagnetic grains.  

 To understand the effect of dust in vicinity of transients (SNe and GRBs) on their signal, it is important to know how the time-scale for the alignment of an ensemble of grains toward the high-J attractor. To understand how this time-scale is changing with the intensity of radiation we performed numerical calculations as in \cite{HoangLazarian:2008} of grain dynamics for an extended period of time for two different intensities of incoming radiation. Specifically, we solved the equation of motion for an ensemble of $N_{\rm gr}$ grains with initial orientation uniformly distributed, subject to RATs with thermal fluctuations, gas damping and randomization, and magnetic relaxation.

\begin{figure}
\includegraphics[width=0.5\textwidth]{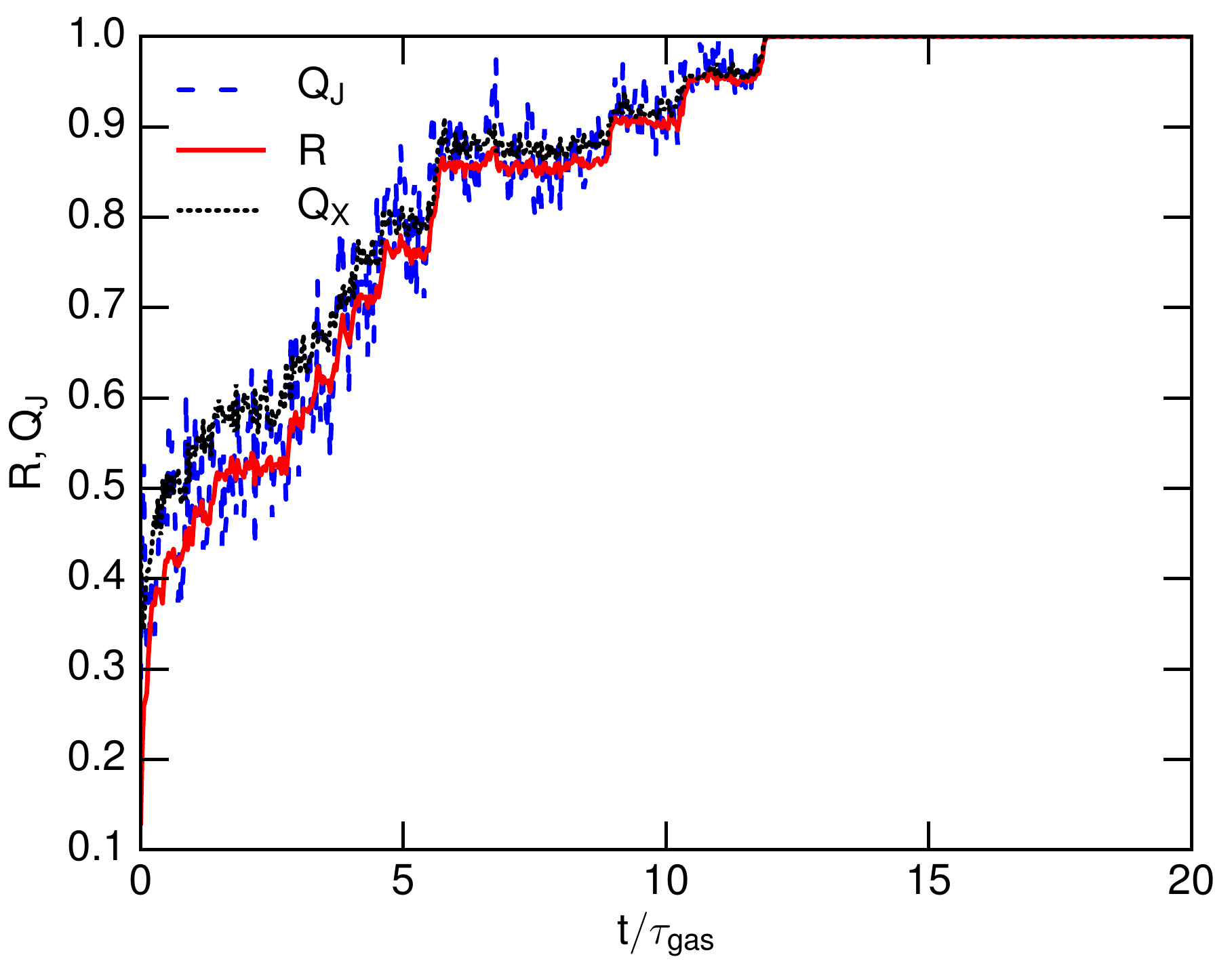}
\includegraphics[width=0.5\textwidth]{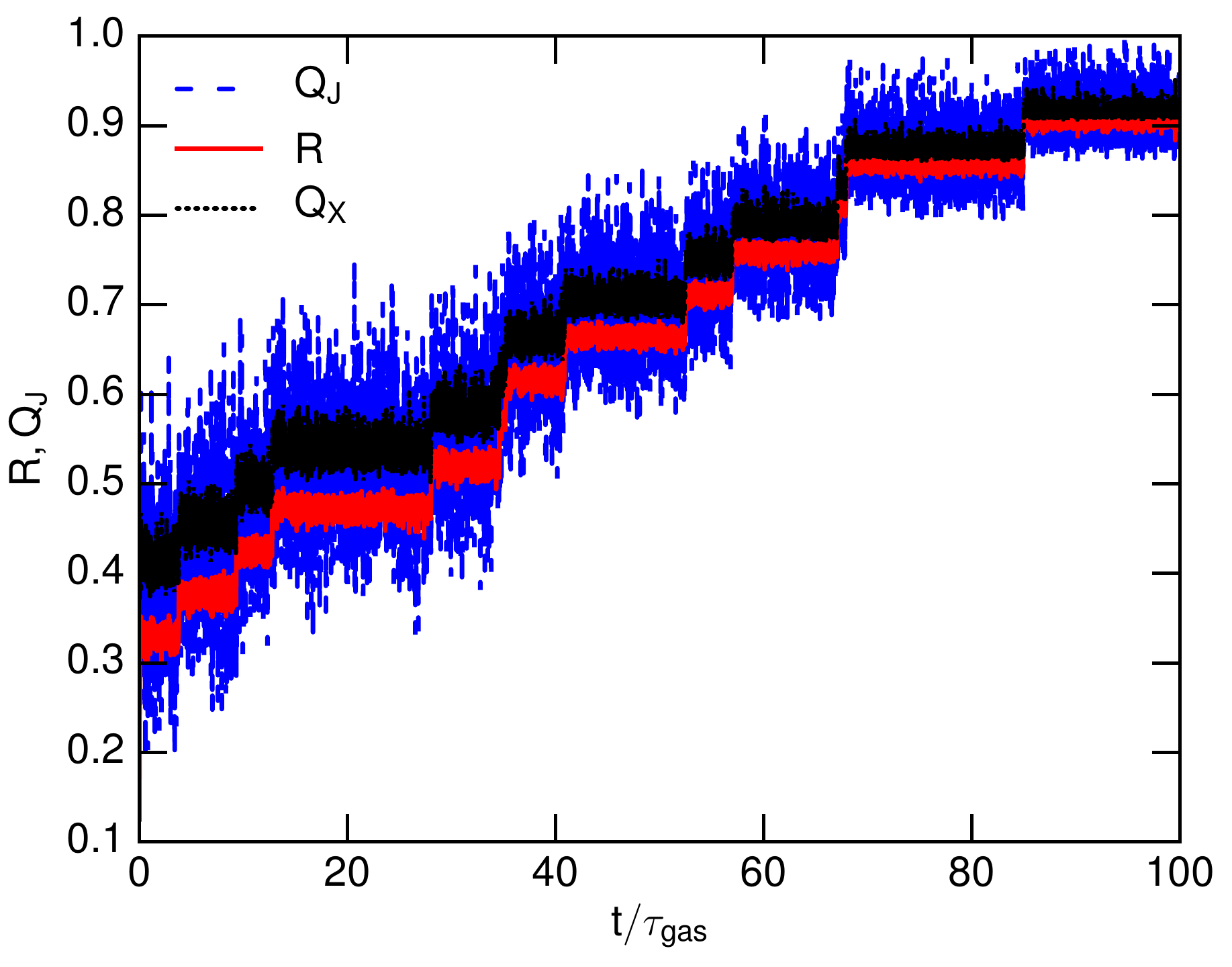}
\caption{Temporal evolution of the alignment degrees for different radiation strengths, $U=1$ (upper panel) and $U=5$ (lower panel). The grain size $a=0.2\mum$, $q^{\max}=1$, $\delta_{m}=100$, and the standard ISM are considered. Perfect alignment achieves at a later time for a stronger radiation field.}
\label{fig:RQJ_CHI}
\end{figure}

Figure \ref{fig:RQJ_CHI} shows the temporal evolution of different alignment measures for two radiation strengths, $U=1$ and $U=5$. Here,
$R$ is the Rayleigh reduction factor which describes the alignment of the grain principal axis with the magnetic field and calculated as
\begin{eqnarray}
 R=\frac{1}{N_{\rm gr}}\sum_{i=1}^{N_{\rm gr}} \frac{(3\cos^{2}\xi_{i}-1)Q_{X}(J_{i})}{2},
\end{eqnarray}
$Q_J$ is the measure of the alignment of ${\bf J}$ with the magnetic field:
\begin{eqnarray}
    Q_{J}=\frac{1}{N_{\rm gr}}\sum_{i=1}^{N_{\rm gr}} \frac{(3\cos^{2}\xi_{i}-1)}{2},\label{eq:QJ}
\end{eqnarray}
and $Q_X$ characterizes the alignment of grain axes with respect to ${\bf J}$. For an oblate grain,
\bea
Q_{X}(J_{i})\propto \int_{0}^{\pi} q_{X}\exp\left({-J_{i}^{2}[1+(h-1)\sin^{2}\theta]}\right)\sin\theta d\theta,~~~
\ena
where $h=I_{\|}/I_{\perp}$, and $q_{X}=(3\cos^{2}\theta-1)/2$ with $\theta$ the angle between $\ba_{1}$ and ${\bf J}$.

In both cases, the alignment degrees reach $\sim 50-100\%$ for $t\sim 5-10t_{\gas}$.  To achieve perfect alignment, it requires 10 and 100 $t_{\gas}$ for the radiation strength $U=1$ and $5$, respectively. For the typical ISM, the corresponding time is $\lesssim 1 Myr$ for $t\lesssim 10t_{\gas}$ using Equation (\ref{eq:taugas}). This time is short compared to the typical time scales of the ISM evolution or grains near stable sources like stars, but it is very long compared to the time of the evolution of the evolution of the radiation flux from a transient like SNe and GRBs. 

The increase of the time-scale of alignment at high-J with the increase of the radiation flux happens as stronger RATs force grains to the low-J attractors. Thus, for the population of grains the increase of the radiation intensity can {\it decrease} the disruption of grains. Our calculations show that the rotational disruption is expected only to a fraction of the grains in the vicinity of the transient radiation source. The fraction of disrupted grains increases if grains have magnetic inclusions as the consequence of the increase of the percentage of grain trajectories leading to the high-J attractors (see Figure \ref{fig:fhighJ}).

If RATs align grains without high-J attractors, i.e., only low-J attractors exist, in the state of low-J alignment, grains are subject to randomization. Gaseous bombardment presents such a source of randomization. It was shown in Figure \ref{fig:map_lowJ_random} (lower panel) that if a grain leaves the stationary point of low-J, it wanders in the phase space for a while before it returns to the low-J point. Potentially, during these wandering the grain angular momentum can exceed $\omega_{disr}$ and, as a result, be disrupted. However, this is an inefficient process, as RATs force the grain towards low-J attractors.

Due to thermal fluctuations in grain material, it is difficult to decrease the rotational temperature of a grain less than $T_{d}$. As a result of collisions with the mass of hydrogen atoms equal to the fraction of the grain mass $m_{\rm grain} T_{d}/T_{\gas}$, the grain gets randomized at the low-J alignment point. During this time RATs push the grain towards the low-J alignment. Therefore, very strong RATs may not induce any appreciable RATD if only low-J attractors exist. 

If the radiation is strong enough, the alignment axis can change to be the radiation axis. This change is more likely to paramagnetic grains compared to the grains with strong magnetic response.

\subsection{Fast and slow disruption}
In Section \ref{sec:AMO}, we have shown that a fraction of grains aligned by RATs are directly moved to the high-J attractor and experience {\it fast alignment}. Assuming that the radiation field is strong enough, such {\it fast alignment} grains can be disrupted by centrifugal stress on a timescale less than the gas damping, which is called {\it fast disruption}. The timescale of the fast disruption is given by Equation (\ref{eq:tmin}). 

The parameter space for the high-J attractor points for the RAT alignment, i.e., fast disruption, is shown in Figure \ref{fig:lowJ_highJ}. There, the angle $\psi$ is the angle between the radiation beam direction and the axis of alignment, which is the direction of the ambient magnetic field ${\bf B}$, provided that the Larmor precession is faster than that induced by the radiation. In the opposite case, the $\psi=0$, i.e., the direction of the beam coincide with the axis of alignment. 

The fraction of grains that experience fast disruption, $f_{\rm high-J}$, is shown in Figures \ref{fig:fhighJ} and \ref{fig:fhighJ_deltam} for different parameters of grain shapes ($q^{\max})$, the radiation direction ($\psi)$, and the magnetic susceptibilities ($\delta_{m}$). As shown, for strongly magnetic grains with $\delta_{m}\gtrsim 10^{3}$ (i.e., superparamagnetic material), the existence of high-J attractors is guaranteed for any angle between the radiation direction and the magnetic field $\psi$, and a substantial fraction of grains can experience {\it fast disruption}. 

In Section \ref{sec:spinup_down}, we have shown that grains aligned at the low-J attractor point are gradually transported to the high-J attractor due to gas collisions, and eventually all grains reach the high-J attractor point. Such grains can experience rotational disruption in more than a gas damping time, which we term {\it slow disruption}. Thus, a fraction of $1-f_{\rm high-J}$ of grains experience {\it slow disruption} for the case of alignment with high-J attractors.

The characteristic timescale for the transport from the low-J to high-J attractor, which determines the time of {\it slow disruption} can be described by
\begin{equation}
    t_{\rm slow}\approx t_{\rm disr, min} +\alpha \frac{T_{d}}{T_{\gas}} \tau_{\rm damp},
    \label{eq:tint}
\end{equation}
where $\alpha$ is a factor that depends on the parameter space of the $\xi$ corresponding to the trajectories towards to high-J attractor point (e.g., $q^{\max}$ and $\delta_{m}$) as well as the radiation intensity. 

Figure \ref{fig:RQJ_CHI} shows that it takes $\alpha$ is increasing with the radiation strength $U$. For instance, it can be of the order of $10^3$ for $U=5$ compared to $\sim 10^2$ for $U=1$. This is an important effect of "capturing" of grains at the low-J attractor points by strong radiation. Therefore we expect that near the very strong sources of radiation, i.e. novae or supernovae, the large grains which were at low-J attractors before the flash of light, will not be rotationally disrupted. The actual dependence of $\alpha(U)$ requires further numerical studies.  


If the grains do not have high-J attractor point, all the phase trajectories, follow towards the low-J attractor. However, some of the grains can get sufficient angular velocities as their phase trajectories approach the high-J repeller point. The probability of grains reaching the velocities comparable to the case of the grain with high-J attractor point is low and we expect the grains with only low-J attractor points to take significantly more time for the RAT disruption. From the observational perspective the grain without high-J attractor points are expected to be significantly more stable in terms of rotational disruption. 




\subsection{Effect of variations of radiation flux on alignment and disruption}

With new observational facilities at optical-NIR wavelengths, including GMT, SPHEREx, LSST, TMT, E-ELT, JWST and WFIRST millions supernovae explosions and Gamma-Ray bursts will be detected. The signals from these transients will be affected by the properties of dust. Those are being changed both to grain alignment and disruption.  

Variable radiation sources, including variable stars, novae, and SNe, make the grain dynamics even more complex. However, for local interstellar grains, the situation is simplified because prior transient explosions, they are already being aligned by the average ISRF. In the case of alignment with high-J attractors, most grains are driven to the high-J attractor (see the upper panel of Figure \ref{fig:map_lowJ_random}) because there is enough time for gas randomization to transport grains from the low-J attractor to the high-J attractor. As a result, the enhanced radiation by transients rapidly spin-up grains on high-J attractors to the disruption limit. In this case, rotational disruption is efficient. On the other hand, for the alignment with low-J attractors, grains mostly spend their time experiencing random motion during the rotation stage with low-angular momentum (see the lower panel of Figure \ref{fig:map_lowJ_random}). Therefore, for grains being formed in the supernova ejecta, only a fraction of grains with the high-J attractor, $f_{\rm high-J}$, can be fast disrupted by RATs, whereas most of grains survive because those grains experience slower disruption that may be longer than the timescale of SNe flash (see Equation \ref{eq:tint}).

The increase of the radiation flux can change the axis of grain alignment if the radiation precession rate is larger than the Larmor precession. Since the precession rate $\Omega_k$ is inversely proportional to the magnitude of the angular momentum $J$ as given by Equation (\ref{rad_p}), it changes in the process of alignment. Therefore, when the grain that is aligned with respect to the magnetic field is subject to the increased radiation flux, if $\Omega_{k}< \Omega_{B}$, the averaging over the fast Larmor precession still induces $F(\xi=0)=0$ i.e., RATs cannot change the direction of ${\bf J}$ with the magnetic field because $d\xi/dt\propto F(\xi)=0$. As a result, grains are directly spun-up to larger $J$ by the spin-up torque component $H$. On the other hand, if $\Omega_{k}> \Omega_{B}$, RATs may start realigning grains with respect to the radiation direction.  Depending on the angle between the radiation direction and the magnetic field, grains can first be spun-up before spun-down to low-J attractors or are directly driven to a new low-$J$ corresponding to $k$-alignment (see Figure \ref{fig:MAP_Jth}). However, if the new position corresponds to the parameter space with the high-$J$ attractor, on the time scale of the order of a damping time the grain phase trajectory will be directed to the that attractor. The corresponding increase of the angular momentum can make $\Omega_B>\Omega_k$, which will induce the realignment of the grain with its phase trajectory following to the initial low-$J$ attractor with respect to ${\bf B}$ till the angular momentum decreases to the degree that $\Omega_k>\Omega_B$. After that the grain will follow again to the low-$J$ attractor point with respect to ${\bf k}$. 

Similarly, the alignment with respect to the electric field can be changed to the alignment with respect to the magnetic field if the grain angular momentum increases. The latter can happen due to the grain trajectory moving towards the high-$J$ attractor or due to the increase of the amplitude of RATs or METs due to the change of the external driving.

\subsection{Grains near stable intense radiation sources}\label{sec:stable}
As shown in Figure \ref{fig:RQJ_CHI}, grains subject to intense radiation sources take longer time to move from the low-J to high-J attractors under gas randomization. To describe the rate of {\it slow alignment}, i.e., rate of driving from low-J to high-J attractors via gas randomization, we introduce a dimensionless parameter $U/(n_{1}T_{2}^{1/2})$, which is proportional to the ratio of the radiation flux to the gas flux. The timescale of slow alignment and disruption is expected to increase with increasing $U/(n_{1}T_{2}^{1/2})$. Thus, the time-dependence alignment shown in Figure \ref{fig:RQJ_CHI}, which are calculated for the typical $n_{\H}$ and $T_{\rm gas}$, can be applicable for arbitrary environments with the same $U/(n_{1}T_{2}^{1/2})$. 

For example, in environments surrounding young stars and photodissociation regions (PDRs) where both $n_{\H}$ and $U$ are substantially elevated from the diffuse ISM (see \citealt{HoangTram:2020}), the characteristic timescale of slow alignment/disruption is not different from the diffuse ISM. As a result, if high-J attractor exists, grains can be driven to its in less than $10t_{\rm damp}$, which is $\lesssim 10^{2}$ yr for $n_{\H}>10^{4}\cm^{-3}$ (see Equation (\ref{eq:taugas})). This is short compared to the dynamical timescale, thus, rotational disruption and grain chemistry can be efficient in star-forming regions and PDRs.

\section{Application to METs}\label{sec:METs}

MEchanical torques (METs) acting on irregular grains were introduced in \cite{LazarianHoang:2007b} in the direct analogy with RATs. They are not explored as well as RATs. Observations (see Andersson et al. 2015) testify that RATs provide the dominant mode for grain alignment and, in the general ISM, the alignment of grains with sizes smaller than those that are expected to be aligned by RATs is marginal. METs, however, may be dominant in particular astrophysical settings.  

\subsection{Alignment and rotation induced by METs}

 Indeed, the AMO is a model of a helical grain and its alignment and rotation does not depend whether the grains are exposed to the flux of photons or the flow of atoms bombarding the mirror (see Fig. \ref{AMO}). The details of the interaction of atoms with the mirror, e.g. reflection or absorption with the subsequent evaporation, change the magnitude of METs by a factor of unity and do not change the nature of alignment for the model. The numerical simulations in \cite{Hoangetal:2018} confirmed that the effect of the MET alignment is present and strong. However, they testified that while AMO provides good accuracy for representing RATs, its ability to reproduce the properties of METs is limited.     

As a result, the MET alignment theory is less developed compared to the RAT theory. In particular, it has been established that the properties of uncompensated MET torques vary significantly from those predicted within the AMO. This is the consequence of the fact that while radiation samples the entire grains and therefore feels the overall grain helicity, the flow of impinging atoms samples only the effective helicity of the given rotational-averaged facet of a grain. As the grain gets aligned and shows 
another facet, the effective helicity can change, making the process of both alignment and corresponding spin-up significantly more complicated. This explanation is consistent with the properties of METs revealed both by analytical calculations in \cite{DasWeingartner:2016} and numerical calculations in  \cite{Hoangetal:2018}.  Therefore, in our present study, we mostly focus on the effects of RATs and use the analogy between RATs and METs to obtain the corresponding conclusions about METs.

\subsection{Calculation of METs}

\begin{figure}
\includegraphics[width=0.5\textwidth]{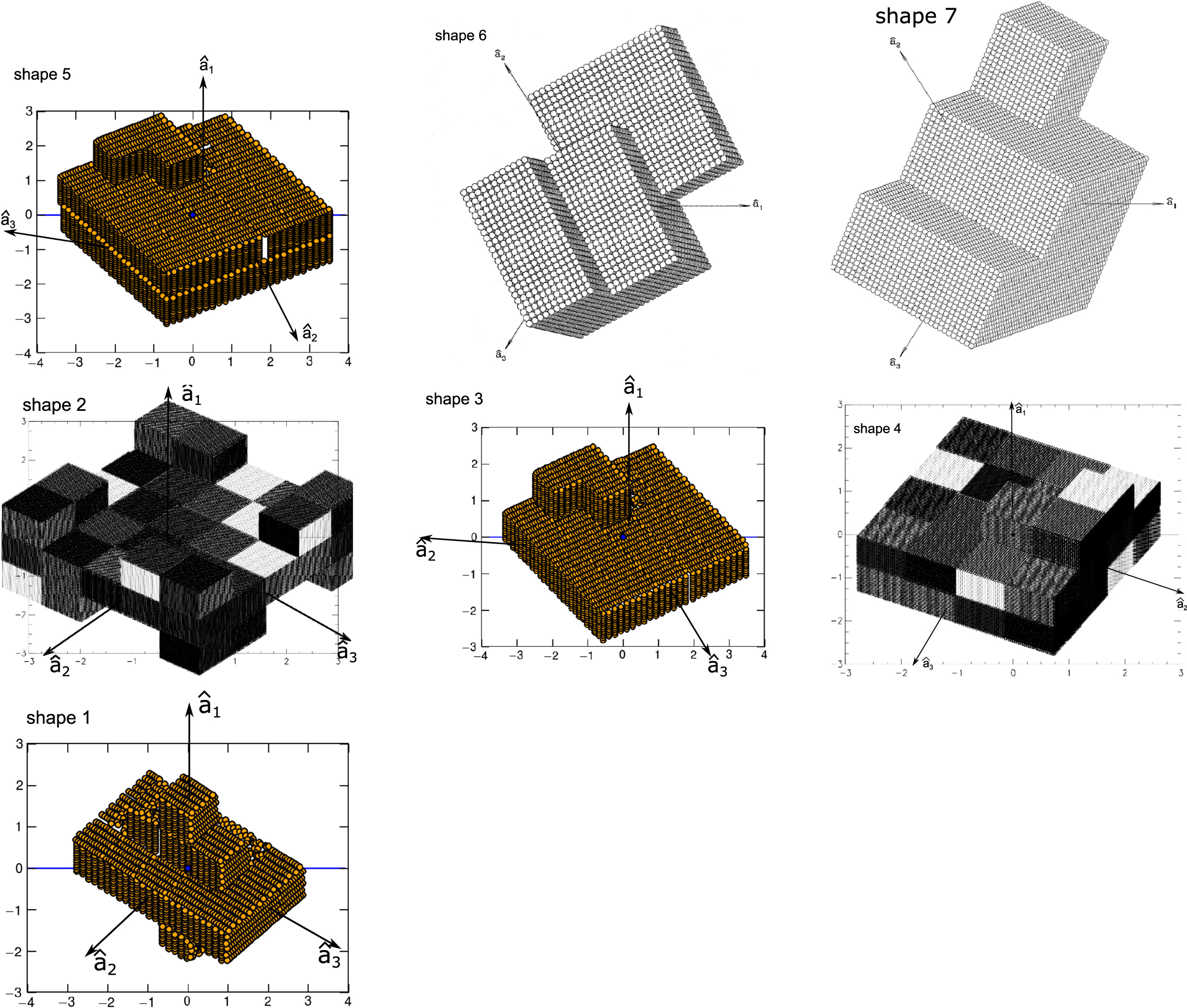}
\includegraphics[width=0.5\textwidth]{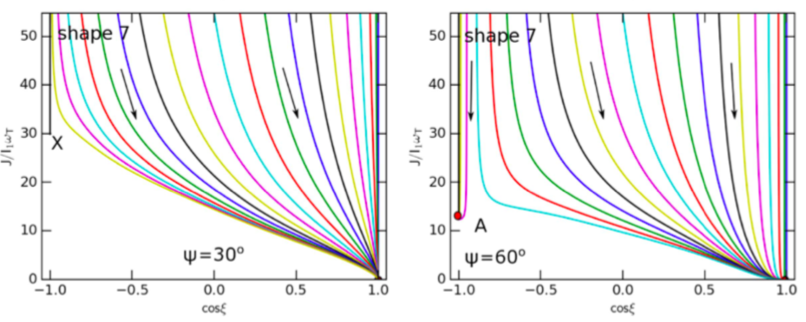}
\caption{Upper panel: Grain shapes used for study METs. Lower panel: The phase trajectories of number 6 grain shape for different $\psi$. For $\psi=30$ degrees the stationary point is is high-J repeller. This changes to the high-J attractor for $\psi=60$ degrees. From \cite{Hoangetal:2018}. }
\label{MET}
\end{figure}

The analytical calculations of METs were performed for the toy model of grain modeled by AMO in \cite{LazarianHoang:2007b}. AMO is a too simple model quantitative calculations of METs and corresponding calculations suggesting that the efficient alignment is possible for very subsonic flows interacting with the grain are likely to over-estimate the efficiency of METs. \cite{DasWeingartner:2016} used 13 irregular grain shapes modeled by the Gaussian Random Spheres and provided elaborate calculations supporting the idea of MET alignment. Both similarities and differences from the AMO calculations were revealed there. Finally, \cite{Hoangetal:2018} performed numerical simulations bombarding grains of selected random shapes with a flow of atoms. The upper panel of Figure \ref{MET} illustrates the shapes studied, and the lower panel shows the trajectory map of grain alignment by METs.

METs calculated in \cite{Hoangetal:2018} exhibit a strong dependence on grain shape. Nevertheless all the calculations supported the core idea in \cite{LazarianHoang:2007b} that irregular grains generically exhibit helicity while interacting with gaseous flows and the corresponding regular torques dominate the stochastic \cite{Gold:1952} torques. The strength of METs systematically increase with the grain drift velocity.   

The difference between RATs and METs in terms of the numerical studies stems from the available tools, the complexity of the problem and the importance of the problem as it is perceived by the community. The progress in understanding of the strength of RATs was possible due to the DDSCAT code in \citealt{DraineFlatau:1994}. The analog of such powerful code does not exist for the mechanical torque studies. In terms of the complexity, METs do not show a similar degree of the universality that is observed for RATs. This is reflected by the fact that while AMO provides a good quantitative model for RATs, for METs, it gives only the qualitative guidance. In addition, while RATs have provided the main explanation for the interstellar grain alignment (see \cite{Anderssonetal:2015}), METs are viewed as an auxiliary mechanism for particular astrophysical settings. Nevertheless, as the numerical simulations in \cite{Hoangetal:2018} testify the dynamical importance of METs, these torques deserve more detailed study.

\subsection{New types of METs: Photoemission Torques and torques from variations of grain properties}

As we discuss in Section \ref{purcell} photoemission, the variations of the accommodation coefficient and the formation of H$_2$ molecules over grain surface can produce pinwheel torques \cite{Purcell:1979}. This torques assume the isotropic arrival of atoms of photons to the grain surface. This property corresponds to the isotropic RATs (see Appendix \ref{app:F}).  However, the fact that there should exist the anisotropic Purcell's torques that have properties of METs was missed in earlier studies. 

Similarly, in the presence of the directed flux of atoms that induced METs, the variations of grain accommodation coefficient also induce helical grain response. In other words, such grains can be not only spin up, but can be aligned by the radiation and the gaseous flux, respectively, even if the grains are regular, e.g., perfect spheroids. For grains with the variations of the accommodation coefficient, the alignment is one of the versions of the MET alignment that was introduced and tested for grains with irregular shapes. Similarly, we note that grains with irregular shapes are expected to experience isotropic pinwheel torques being subject to isotropic bombardment. The corresponding torques are expected to be long-lived, i.e. survive for more than a rotational damping time $\tau_{\rm damp}$.

Consider the photoemission from grains. If radiation is anisotropic, the photoemission of electrons can render effective helicity to the grain. At the same time, the photoemission torques represent a completely new Photoemission Torque (PT) alignment mechanism that shares significant similarities with the MET mechanism. Naturally, grain shape irregularities can increase further the efficiency of the PTs. The PTs are subdominant compared to RATs when $\lambda \sim a$, but may dominate for $\lambda \ll a$. In these settings, we can expect the PT alignment to take place. We shall provide the quantitative study of the PT process elsewhere. 

The directed flux of hydrogen atoms can also produce the MET torques, provided that the diffusivity of atoms over grain surface is limited. In this situation, the directed ejection of $H_2$ atoms formed on catalytic sites is expected to induce the METs. If, however, atoms are free to diffuse over grain surface, the torques can induce only grain spin-up rather than the alignment.

We feel that PTs may be one of the most promising incarnations of the METs. 
While the suppression of traditional pinwheel torques can take place due to thermal flipping (see Section \ref{purcell}), the PT alignment and other types of alignment induced by the mechanical flows are not suppressed by the aforementioned effect. Similar to the RAT alignment the alignment arising from Purcell torques from a directed flow proceeds irrespectively whether grains perform thermal thermal flipping or not.

\subsection{METs for grains with magnetic inclusions}

METs were introduced in \cite{LazarianHoang:2007b} using the same model of AMO as RATs employed. Later in \cite{LazarianHoang:2008} the combination of AMO and enhanced magnetic relaxation was introduced for the RAT alignment. The potential importance of this model for explaining the high value of the observed interstellar polarization was demonstrated in \cite{HoangLazarian:2016a}. The MET alignment was in the shadow of more promising RAT alignment and therefore there were no similar developments for the MET alignment. Here we attempt to compensate this deficiency by explaining how enhanced magnetic relaxation changes the MET alignment as well as the MET rotational disruption.

Figure \ref{MET} shows that, similar to the case of RATs, the phase portraits of MET may or may hot have high-J attractors. Magnetic dissipation is expected to stabilize high-J stationary points, creating the high-J attractors. For instance, the we expect the high-J repeller for $\psi=30^{\circ}$ to transfer into an attractor point if $\delta_m$ gets sufficiently large. 

Similar to RATs, METs, in the presence of randomization, are expected to bring the a significant part of grains with strong magnetic response into the high-J attractor points over a grain randomization time $\tau_{\rm rand} T_{d}/T_{\gas}$. This is expected to make grains perfectly aligned with the magnetic field. Similar to the case of RATs the increase of magnetic susceptibility is going to increase $\Omega_B$, making the alignment with the radiation direction less probable. 

In terms of disruption, the presence of magnetic inclusions will increase the percentage of grains that are being disrupted by the shocks that are not perpendicular to magnetic field. For the shocks perpendicular to magnetic field one may expect, similar to the case of RATs, a reduction of the torque amplitude.

For the fragments of disrupted grains, we, similar to the case of RATD, expect them to be aligned and be subject to further disruption if the METs are sufficiently strong. Unlike RATs, we do not expect the decrease of the grain size to affect the METs efficiency. As the fragments with magnetic inclusions are expected have high-J attractors, the disruption of grains with magnetic inclusions in shocks is likely to separate magnetic inclusions from the rest of the grain material. This can result in the appearance of isolated magnetic nanoparticles in the ISM. Such nanoparticles can be an important ingredient of the ISM responsible for the microwave foreground emission (\citealt{HoangLazarian:2016b}).

\subsection{METs and other types of torques}

In \cite{HoangTram:2019} the disruption of small grains in shocks was considered. Such grains passing through the shock are subject to stochastic bombardment of high velocity atoms and get significant rotational velocities that are sufficient to induce grain disruption. 

It was mentioned in \cite{HoangTram:2019} that METs can provide higher degrees of rotation and increase the efficiency of grain disruption. Indeed, being regular, METs dominate the stochastic torques. Therefore, the dynamics of realistic irregular grains should be mostly dominated them.

Assuming that METs dominate RATs in shocks, we can discuss the response of grains to the MET alignment. This response depends on the orientation of magnetic field with respect to the shock as well as to the magnetic properties of grains. 

For grains with high magnetic response, e.g. grains with inclusions, the alignment happens with respect to the magnetic field direction. If grains are already aligned with magnetic field at high-J, the shock will increase their rate of rotation and can result in the grain disruption over the time $\sim I\omega_{\rm disr}/\Gamma_{\rm MET}$, where $\Gamma_{\rm MET}$ is the amplitude of MET torques, provided that this time is less than the time of the shock passage, which is 
\begin{equation}
\tau_{\rm sh}=\frac{L}{V_{\rm sh}}
\end{equation}
where $L$ is the shock thickness that depends on the physical parameters including gas density, magnetic field, and shock velocity (see \citealt{HoangTram:2019}). For slow shocks considered in \cite{HoangTram:2019}), the grain stopping time is longer than the shock crossing time. The stochastic torques increase the rate of grain rotation making grain rotational disruption more probable. The magnitudes of METs are expected to decrease when shocks are perpendicular to the magnetic field. 

If grains do not have an enhanced magnetic response, they can be aligned both at high-J and at low-J attractor points initially. If the grains are aligned with magnetic field at high-J attractors prior to the passage of the shock, the shock is expected to spin-up such grains further, provided that $\Omega_B>\Omega_k$.

If grains are aligned with magnetic field at low-J attractor points, their realignment on the time scale $\sim \Omega_k^{-1}$ with the shock direction will take place provided that $\Omega_k$ is larger than $\Omega_B$ of a slowly rotating grain. During this realignment most irregular grains will be moved to the low-J attractors and therefore will be slowed down by METs. In other words, METs can force grains in the states of low-$J$ and this way METs can decrease the rate of grain rotation compared to the rotation induced by pure stochastic torques. This presents an interesting example of {\it rotational cooling} induced by METs.

\section{Alignment and selective disruption of different types of grains}\label{sec:graincomp}

 For both RATs and METs their action on grains depend on whether the grains are getting aligned with the magnetic field, or the flow or the electric field.

\subsection{Silicate grains}

In the absence of magnetic inclusions the silicate grains are expected to exhibit paramagnetic properties. The corresponding precession rate of silicate grains in typical interstellar magnetic fields are significantly larger to the precession rates of grains with respect to the anisotropy direction of the ISRF. According to \cite{HerranenLazarian:2020} a significant part of grains in their ensemble of shapes is expected to be aligned at high-J attractors, which makes $\Omega_B\gg \Omega_k$ for the typical radiation field. In the case of METs, only highly supersonic shocks can realign grains with respect to ${\bf k}$.

\subsection{Carbonaceous grains}

The alignment of carbonaceous grains by RATs was recently described in \cite{Lazarian:2020}. The difference between the pure carbonaceous grains and the silicate ones arises from the rate of grain precession in the magnetic field, which is related to their diamagnetic properties vs. paramagnetic properties of silicates. In this situation, other effects that induce grain precession become more important. For instance, this increases the relative role of RATs in defining the grain alignment axis, making the k-RAT alignment more important for carbonaceous grains compared to their silicate counterparts. 

\cite{Lazarian:2020} identified the alignment of grains with respect to the electric field an important mode of carbonaceous grain alignment (Lazarian 2020). The physical reason of this is that grain precesses fast in the electric field as discussed in Section \ref{precession}. The RAT and MET alignment happen with respect to electric field, and the electric field is perpendicular to the magnetic field. Therefore the RAT and MET alignment for the carbonaceous grains with fast internal relaxation happens with long grain axes with respect to electric field. This means that in the process of such E-alignment the grain get aligned with long axes parallel to magnetic field. 

If gyrofrequency $\omega_{\rm gyro}$ is much slower than $\Omega_k$ the alignment happens instantaneously. Different grains will have different $\psi_E$ and some may have high-J attractor points. At the same time for some $\psi_E$ there may be no high-J attractor points. As a result, the dynamics of grain alignment is getting much more complicated compared to the alignment of silicate grains. The details of this dynamics will be presented elsewhere.

Due to lower magnetic susceptibilities of carbonaceous grains, it is more likely to get situations with $\Omega_k$ being the highest frequency, i.e., the alignment with respect to ${\bf k}$ is more likely. According to Figure \ref{fig:lowJ_highJ} the probability of having $q^{\max}$ appropriate for having high-J attractor points is low. Therefore, the perfect alignment of carbonaceous grains aligned in ${\bf k}$ direction by RATs is unlikely. Similarly, we expect to have a significantly reduced disruption of such grains by RATs. Keeping in mind the analogy of the action of RATs and METs, we conclude that similar reduced alignment and reduced disruption by METs is expected for grains aligned in ${\bf k}$-direction.

If the randomization of carbonaceous grains happens faster than the alignment, the disruption of such grains is going to take several grain damping times. If, however, the alignment happens in relation to the radiation or magnetic field, the RAT disruption of carbonaceous grains will depend on whether the grains are aligned at the low-J or high-J attractor points and what is the angle between the grain axis of alignment and the direction to the radiation source. We expect the MET disruption to follow qualitatively to the same dependence, the predictions for this case are less certain due to the difficulties of applying AMO to METs. 

\subsection{Composite silicate-carbonaceous grains and grains with magnetic inclusions}

Composite grains containing both silicate and carbonaceous fragments are expected to get aligned similar to the silicate grains. Indeed, the precession rate of composite grains is expected to be comparable with that of silicate grains and therefore the E-alignment is going to be subdominant, unless grains have extremely large velocities or magnetic field experience really dramatically fast compression, e.g., as in supernovae shocks. 

Grains with iron inclusions have both significant additional paramagnetic relaxation (\citealt{JonesSpitzer:1967}) and significantly larger magnetic moment. Both effects result in much more efficient B-RAT and B-MET alignment. Indeed, it was shown in \cite{LazarianHoang:2008} that the enhanced magnetic dissipation stabilizes high-J stationary points, making them attractor points. Further studies in \cite{HoangLazarian:2016a} revealed that even a moderate increase of grain magnetic susceptibility, e.g., an increase by a factor of 10, can ensure that grains for the entire parameter space have high-J attractor points (see Figure \ref{fig:fhighJ_deltam}) and can be perfectly aligned with long axes perpendicular to the magnetic field (see Figure \ref{fig:RQJ_CHI}).\footnote{Incidentally, the presence of magnetic inclusions makes the internal relaxation more efficient (\citealt{LazarianHoang:2008}), making sure that the parameter space for which the grain alignment happens in the absence of internal relaxation. Therefore, the parameter space the low-J alignment that happens with long grain axes parallel to axis of the fastest precession introduced in \cite{HoangLazarian:2009a} shrinks.} The higher magnetic moment of the grains increases the range over which the magnetic field provides the fastest precession and therefore acts as the alignment axis. 

It was argued in \cite{HoangLazarian:2016a} that a significant part of interstellar grains may have magnetic inclusions and thus exhibit an enhanced magnetic response. In this situation, over a few damping times such grains will find themselves at high-J attractor points. The lower rate of grain rotation for $\psi$ close to $90^{\circ}$ would allow the grains in such situations to withstand the centrifugal stress (see Figure \ref{fig:Jmax_angle}). Therefore, we predict that large grains are expected to exist for $\psi$ of the order $90^{\circ}$, and detection of such grains may determine the magnetic field direction provided that the location of the radiation source is known. 

We note, however, that even in the presence of magnetic inclusions, a fraction $f_{\rm high-J}$ are rapidly disrupted, whereas $1-f_{\rm high-J}$ fraction of grains first move to the low-J attractor points. Therefore, we expect the {\it perfect} grain disruption to take several damping times, i.e., the time that it takes for grains at low-J to move to high-J attractor points by gas collisions.

The previous considerations are relevant to B-RAT or B-MET alignment. In the case $\Omega_k>\Omega_B$, the k-RAT alignment takes place. In this case, the role of magnetic inclusions is limited to the enhancement of internal relaxation (\citealt{LazarianHoang:2008}) which decreases the parameter space for the "wrong" alignment with grain axes parallel to ${\bf k}$.

\section{Effects of Pinwheel Torques}\label{sec:pinwheel}

For decades the studies of the grain alignment field were limited by stochastic torques. The situation changed when \cite{Purcell:1979} introduced pinwheel torques. These torques can be important both for the RAT and MET alignment. In this section, we discuss how Purcell's torques can modify the alignment and disruption. We also outline the uncertainties associated with these torques.

\subsection{Pinwheel torques}
\label{purcell}

The regular uncompensated torques proposed by \cite{Purcell:1979} act in the local frame of grain reference. They can arise, for instance, from the variations of the accommodation coefficient over grain surface. This coefficient controls whether the impinging atom sticks to the surface or rebounds. It is natural to assume that this coefficient is not uniform but varies over the surface. Provided that these variations are large enough, the grain can be spun up to the velocities significantly larger than grain thermal velocity. Similarly, interstellar grains are likely to form H$_2$ not over their entire surface, but on the particular catalytic sites which are randomly distributed. The recoil from the nascent $H_2$ was shown by \cite{Purcell:1979} as a potent way of spinning up the grains. Finally, the photoemission of electrons from grain surface was the third \cite{Purcell:1979} process of grain suprathermal spin-up. This process is caused also by the expected inhomogeneity of grain surface, this time the inhomogeneity in terms of the variations of the photoelectric yield. 

It is usually assumed that the strongest among the Purcell's torques are the torques arising from H$_2$ formation over catalytic sites distributed over the grain surface. The comparison of these torques with the typical RATs arising from the ISRF is shown in Figure \ref{H2}.  We provide more details about pinwheel torques in Appendix \ref{sec:prop_purcell}.

\begin{figure}
\includegraphics[width=0.5\textwidth]{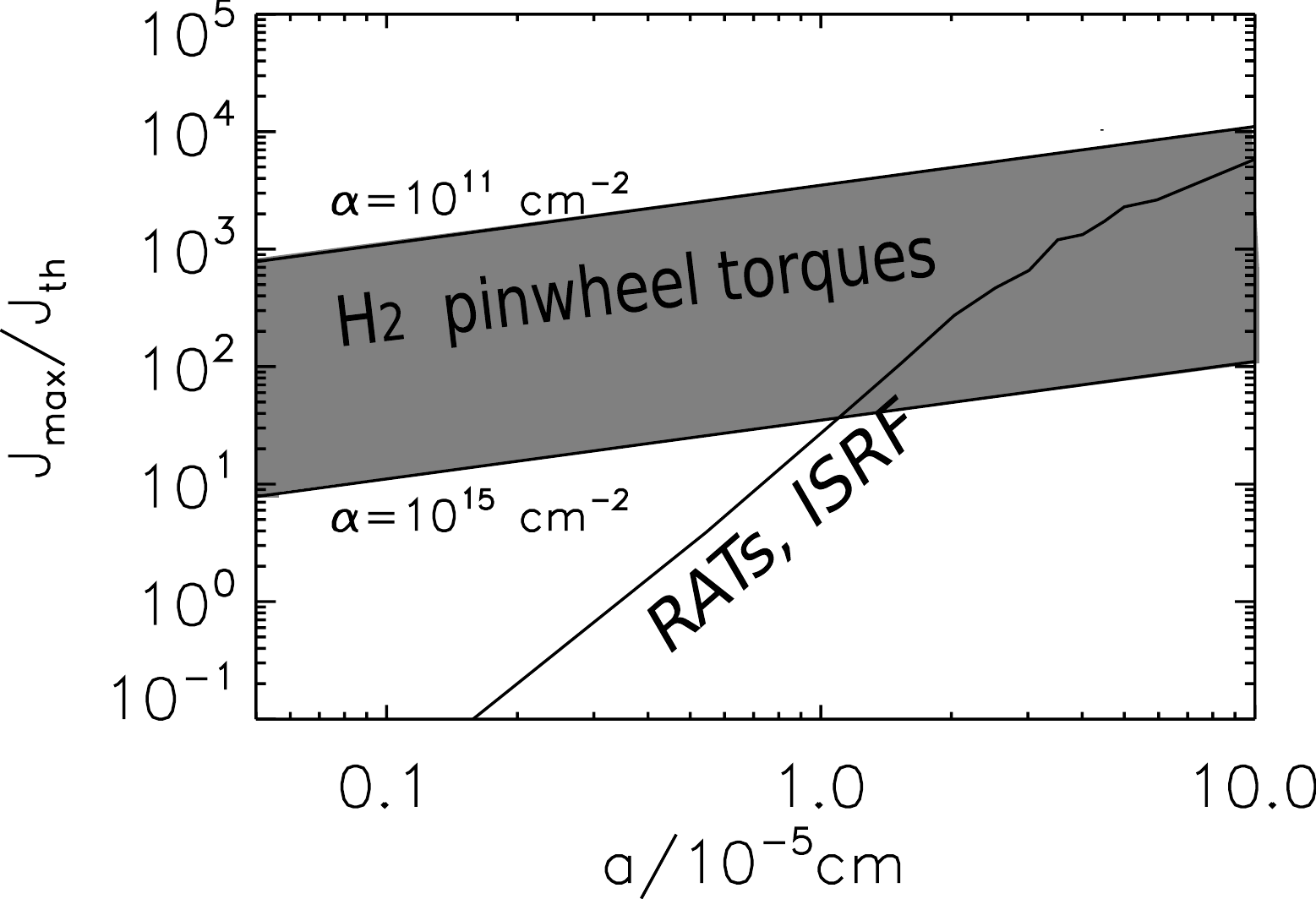}
\caption{Comparison of H$_2$ pinwheel torques and RATs as a function of grain size for the ISM. The shaded area represents the range of normalized to the thermal value pinwheel torques arising the H2 formation with a range of the active site density $\alpha$, and the solid line shows RATs for an irregular grain (shape 1 in LH07) and the ISRF. For H$_2$ pinwheel torques, the upper boundary corresponds to one site per a grain of $a = 10^{-6}$ cm. For RATs, the radiation direction is parallel to the magnetic field. From \cite{HoangLazarian:2009a}.}
\label{H2}
\end{figure}

\subsection{Effects for high-J attractors and fast disruption}
The interaction of the Purcell torques with the RAT alignment was explored in \cite{HoangLazarian:2009b} for the case of classical grains with fast internal relaxation. The effects of Purcell torques for grains aligned with high-J attractor points is rather straightforward. Provided that the RAT or MET torques dominate, the increase of rotational rate is expected for 1/2 of the grains with the corresponding decrease of the rotational rate for the other half of the grains. As a result, pinwheel torques can help to increase the angular velocity of grains at high-J attractors established by RATs, increasing the efficiency of grain alignment and dust polarization.
Therefore, pinwheel torques extend the range of grain sizes that can experience {\it fast disruption}. The minimum size of grain disruption given by Equation (\ref{eq:adisr}) is decreased as a result of pinwheel torques. Pinwheel torques can also decrease the timescale of fast disruption. Observational evidence for enhancement of grain alignment by pinwheel torques was presented in \cite{Anderssonetal:2013}, and numerical modeling of grain alignment by RATs and pinwheel torques in support of this observation was shown in \cite{Hoangetal:2015}.

\subsection{Effects for low-J attractor points and slow disruption}
\label{effect}

We now discuss the effect of pinwheel torques for the alignment at low-J attractor point, which is important for {\it slow disruption} process. In \cite{HoangLazarian:2009b}, it was shown that the Purcell torques increase the angular momentum for the grain low-J attractor points. They also make phase portraits of grain motion asymmetric. What is important for our discussion of the grain disruption is that RATs and the Purcell torques act in opposite directions when the grain is at the low-J attractor point. Indeed, the effect of RATs when they move the grain towards low-$J$ is to decrease the grain angular momentum. Thus RATs and the Purcell torques act against each other when the grain is at the low-J attractor point. 

For typical ISM radiation field RATs dominate for $\gtrsim 0.1$ $\mu$m grains. In spite of that, the role of Purcell's torques may not be negligible. The strength of RATs decreases in the phase space in the vicinity of the low-J attractor (see Figure \ref{torques}). Thus Purcell's torques can increase the angular momentum that the grain has at the low-$J$ attractor point significantly, even if the amplitude value of RATs is larger than that of the Purcell's torques. It is discussed in \cite{HoangLazarian:2009b} that the grain with the increased value of $J$ at the low-$J$ attractor demonstrates higher degree of grain alignment. 

This is a challenge to the close connection between the disruption and grain alignment that we discuss in this paper. Indeed, grains stabilized at low-J attractors by pinwheel torques may show the high degree of alignment, but not be subject to rotational disruption. This is a challenging issue that can be addressed by simultaneous measurement of the polarization and the rate of disruption in the vicinity of transients.   

Another effect of the increased value of angular momentum is that the grain rotation at the position of low-J alignment gets much more stable. Thus, the efficient randomization of grains at low-J attractors described \cite{HoangLazarian:2008} gets suppressed. Therefore, the grain can stay at low-J attractor point for long time. This has significant implications that we list below.

First of all, in the presence of high-J attractor points and in the absence of the stabilization of low-J attractor points by Purcell's torques, the grains are expected to diffuse to the state of high-J alignment in several damping times. If pinwheel torques are present, the fraction of grains directly to the high-J attractor increases. At the same time, the corresponding residence time at the low-J attractor may be significantly enhanced, making the state of low-J alignment long lived. Thus, pinwheel torques can both increase the efficiency of fast alignment/disruption and decreases the efficiency of slow alignment/disruption. The time scale enhancement depends on the lifetime of the Purcell torques as well as on what happens as the torques change their direction, such as thermal flipping as discussed below.

A very non-trivial effect is related to the alignment of grains with only low-J attractor points in the presence of both RATs and pinwheel torques. While, in this settings, the pinwheel torques are increasing the degree of alignment, the increase of the amplitude of RATs can induce a {\it decrease} of the alignment degree.

Another implication of the stabilization of low-J attractors is related to the alignment of grains with slow internal relaxation. The latter drops with the increase of the grain size and is also weaker for pure carbonaceous grains (see Appendix D). The dynamics of grains changes if the grains do not experience internal relaxation on the time scale of their alignment. \cite{HoangLazarian:2009a} have shown that grains in the low-J attractor point will be aligned in a "wrong way", i.e. with long grain axes {\it parallel} to the axis of alignment, e.g., with long axes parallel to magnetic field (see Appendix E). However, the wobbling of grain direction at low-J attractor is expected to decrease the effect of such "wrong" alignment. Subject to gaseous bombardment, the grain is expected to spend a significant time rotating with long axes {\it perpendicular} to the axis of alignment. If, however, the rotation at low-J alignment is stabilized by the pinwheel torques, a prominent "wrong" alignment is expected. Thus the high degree of "wrong" alignment can indicate the action of pinwheel torques. 
 
\subsection{Thermal fluctuations and suppression of pinwheel torques}
\label{wobble}

The internal relaxation within a rotating grain tends to align the grain axis corresponding to the grain maximal moment of inertia with the direction of ${\bf J}$. However, the high rate of internal relaxation taking place in grain material does not ensure that the internal alignment is perfect. It was demonstrated in \cite{Lazarian:1994} that the grain axes directions experience thermal fluctuations with respect to ${\bf J}$ with the rate $\tau_{\rm int}^{-1}$. In agreement with the Fluctuation-Dissipation Theorem, the amplitude of fluctuations was shown to increase as the value of $J$ becomes comparable of smaller than the grain angular momentum at thermal rotation. The analytical representation of grain internal alignment in the presence of thermal fluctuations was obtained in \cite{LazarianRoberge:1997}. 

The rotational energy of the grain is minimal when it rotates around the axis of the maximal moment of inertia ${\bf a}_{1}$. If $h=I_{\|}/I_{\perp}>1$, the deviations of ${\bf a}_{1}$ from the direction of ${\bf J}$ are associated with the change of the grain kinetic energy:
\begin{equation}
    E(\theta)=\frac{J^2}{2 I_{\|}}\left[1+(h-1)\sin^2\theta\right].
    \label{energy}
\end{equation}

The distribution of angle fluctuations obeys the Boltzmann distribution 
\begin{equation}
    f(\theta) \sim \exp\left(-\frac{E(\theta)}{kT_{d}}\right),
    \label{therm_dist}
\end{equation}
which testifies that for $J\gg J_{\rm th}$ the fluctuations in $\theta$ are negligible, while they dominate the grain dynamics for $J\lesssim J_{\rm th}$. For instance, for the plot of the alignment torques $F$ corresponding to $J=10^{-3} J_{\rm th}$ in Figure \ref{torques}, we observe the change of the curvature in the vicinity of $\cos\xi$ equal to plus and minus one. This is the result of thermal wobbling and it shifts the position of the low-$J$ attractor points.  

The thermal fluctuations change the dynamics of alignment, i.e., they change the dynamics of crossovers that were introduced by \cite{SpitzerMcGlynn:1979} as it discussed in  \citealt{LazarianDraine:1997}). In addition, thermal fluctuations can induce the flipping of grains on the scales much shorter than the time of grain resurfacing. This effect termed {\it thermal flipping} was introduced in \cite{LazarianDraine:1999b}. The nature of this effect is simple: as amplitude of thermal fluctuation increases, the grain can flip reversing the direction of Purcell's torques. Naturally, if the thermal flipping is sufficiently fast, the Purcell's torques may not be able to accelerate grains giving rise to a new element of grain dynamics termed {\it thermal trapping}. Thermally trapped grains rotate thermally in spite of the presence of the pinwheel torques (\citealt{LazarianDraine:1999b}). 

The rate of thermal flipping is uncertain, however.
The nature of thermal flipping was questioned in \cite{Weingartner:2006}, where it was noted that for oblate grains there thermal fluctuations cannot alone flip the grain and reverse the Purcell's torques. However, considering a more realistic model of flipping which includes not only thermal, but also external random torques, \cite{HoangLazarian:2009b} demonstrated that the process of the "thermal + bombardment" torques is valid. In a more recent paper, \cite{KolasiWeingartner:2017} found that even without external torques the process of flipping is possible provided that the grain is not oblate, but triaxial. The quantitative study of flipping similar to that in \cite{LazarianDraine:1999b} for the triaxial grains and accounting for grain stochastic randomizing torques from gaseous bombardment should be addressed in future research.

Pinwheel torques definitely a part of grain dynamics. These torques {\it definitely} act on grains rotating at high-J attractor points. 
The current uncertainty is at what grain sizes these torques are important for grains at low-J attractor points. There they can stabilize low-J attractors and significantly extend the time scale necessary for the transfer of grains from low-J to high-J attractor points. This can allow grains survival in high intensity radiation fields.

Potentially, pinwheel torques can be strong enough to provide grain disruption even without RATs.\footnote{In fact, \cite{Purcell:1979} mentioned the possibility of grain disruption due to centrifugal stress, but did not elaborate the process and its consequences.} For instance, Figure \ref{H2} shows that for the typical interstellar field the torques potentially can be comparable to RATs for grains larger than $0.1\mum$.
Interestingly enough, the RATs weak for the grain disruption may stabilize the grains from flipping and the pinwheel torques can disrupt the grains. This shows the the synergy of the pinwheel torques with METs and RATs. With the decreased efficiency of RATs acting on grains with $a\gg \lambda$ and the Purcell torques may become the dominant torques for the grain rotational disruption, e.g., grains larger than 1 $\mu$m, in the circumstellar accretion disks.

\section{Discussion}\label{sec:discuss}

\subsection{Deep connection of grain alignment and rotational disruption}

Rotational disruption is different from other known processes of dust destruction (see \citealt{Hoang:2020} for a review). The selection of grains to be destroyed includes the ability to withstand the centrifugal stress, which acts against fluffy fractal grains as well as well grains composed from the fragments with weak chemical bonds. Observed far-IR spectrum of interstellar dust testifies against the existence of fractal grains in the interstellar environments \citep{1990ARA&A..28...37M}, which can be seen as the evidence in favor of rotational disruption controlling the grain structure.

Our present study reveals an additional feature of rotational disruption, i.e., its close connection to efficient grain alignment. This alignment is present for grains with high-J attractor points, and these grains constitute a population of grains that can be rapidly and efficiently disrupted by RATs and METs. The existence of grain alignment with high-J attractor points and the fraction of grains with high-J attractor, $f_{\rm high-J}$, depends on grain shapes, magnetic susceptibility, and the angle between the flux and the magnetic field (see Figures \ref{fig:fhighJ} and \ref{fig:fhighJ_deltam}). This presents a possibility of observational testing grain properties and grain environments using rotational disruption as well as dust polarization. 

Observations of interstellar polarization report of high polarization degree, and modeling works usually require a high degree of grain alignment to reproduce observed data. For instance, modeling of starlight polarization by \cite{Draine2009} reports the alignment degree between $20-100\%$, where the model with only silicate grains aligned requires higher degree of alignment than the model with both silicate and carbonaceous grains aligned. Modeling of {\it Planck} polarization data by \cite{Guilletetal:2018} requires the alignment degree between $60-100\%$. The alignment at only low-J attractors provides the degree of alignment less than $\sim 20\%$ (\citealt{HoangLazarian:2008}). In particular, the alignment with only low-J attractors is expected not to correlate with or even decrease with increasing radiation flux, which is inconsistent with numerous observations (see \citealt{Anderssonetal:2015}). Therefore, grain alignment with high-J attractors is most consistent with observations of interstellar polarization.

On the other hand, observational evidence for a high abundance of small grains near an enhanced radiation sources implies an efficient rotational disruption and grain alignment. For example, early-time observations toward SNe Ia report unusual properties of dust extinction with $R_{V}<2$ (e.g., \citealt{Burnsetal:2014}) and dust polarization of $\lambda_{\max}<0.4\mum$ (\citealt{Patatetal:2015}). Rotational disruption is found to be the most plausible mechanism to reproduce small grains \citep{Hoang:2017}; and numerical modeling could reproduce such anomalous observational properties (\citealt{Giangetal:2020}; \citealt{Hoangetal:2020}). 

In particular, various observations (\citealt{Guilletetal:2018}; \citealt{Santosetal:2019}) show a non-monotonic increase of the polarization degree with increasing grain temperature (i.e., radiation intensity) that is implied by the RAT alignment theory (\citealt{Leeetal:2020}). Numerical modeling with including rotational disruption could reproduce this feature (\citealt{Leeetal:2020}; \citealt{Trametal:2020}). These observed features suggest a significant fraction of grain ensemble to be aligned with high-J attractors. The tentative evidence of the grain rotational disruption may indicate that grains have enhanced magnetic response. The comparison of the observations with models of the dust properties evolution arising from dust disruption can provide the answer whether astrophysical dust has strong magnetic response. Very recently, \cite{HerranenLazarian:2020} found that the fraction of grains aligned with a high-J attractor by RATs acting on ordinary paramagnetic grains might be large of $f_{\rm high-J}\gtrsim 50\%$. This implies efficient alignment and fast disruption of ordinary paramagnetic grains.

\subsection{Fast and slow rate of disruption}
As studied in Section \ref{sec:spinup_down}, rotational disruption by RATs and METs can occur for grain alignment with high-J attractors. The rotational disruption can occur at different rates, {\it fast and slow}.

Assuming grains initially at thermal rotation, our results show that, under the effect of RATs, a fraction, $f_{\rm high-J}$, of grains can be directly driven to a high-J attractor point in less than a damping time (see Figure \ref{fig:MAP_Jth}, upper panel). Such grains potentially experience {\it fast disruption}. This fast process is crucial for dust disruption by transient sources such as SNe, novae, GRBs (\citealt{2020ApJ...895...16H}), and cometary nuclei (\citealt{2020ApJ...901...59H}). The fastest rotational disruption occur with interstellar grains already being aligned at the high-J attractor by the ISRF and then further spun-up by enhanced radiation flux (e.g., by transients), which has the minimum disruption time given by Equation (\ref{eq:tdisr2}). 

A fraction $1-f_{\rm high-J}$ of grains are first driven to the low-J attractors, and subsequently transported to the high-J attractor by grain randomization (e.g., gas collisions). This process takes several gas damping times, and we termed {\it slow disruption}. 
The process of slow rotational disruption is important for interstellar and circumstellar dust, dust surrounding Active Galactic Nuclei (AGN) where the lifetime of radiation sources is much longer than the gas damping time.

As we have shown in Figures \ref{fig:fhighJ} and \ref{fig:fhighJ_deltam}, the value $f_{\rm high-J}$ increases with the increase of the magnetic dissipation in grains. If grains have strong magnetic response but not originally aligned, we find $f_{\rm high-J}\sim 50\%$ (see Figure \ref{fig:fhighJ_deltam}). 


\subsection{Observational predictions of RAT Disruption}

We found in the paper that the nature of RAT disruption (RATD) is {\it selective} with only grains in high-J attractors subject to efficient disruption. Since grains in high-J attractors correspond to perfect alignment, this implies a correspondence between perfect alignment and efficient disruption. For such grains, on the top of this, the angle between the direction of radiation and the axis of alignment matters because the fraction of grains with the high-J, $f_{\rm high-J}$, increases (see Figure \ref{fig:fhighJ_deltam}) but the maximum rotation rate induced by RATs decreases with increasing that angle (see Figure \ref{fig:Jmax_angle}). For instance, the rotational disruption is efficient if the radiation makes an angle of $\psi\lesssim 45^{\circ}$ with respect to the magnetic field, but the disruption is significantly reduced if the radiation is nearly perpendicular to the magnetic field. Thus, unlike other disruption mechanisms, the RATD destroys grains selectively, i.e., fast disrupting some grains, slowly disrupting other grains, and not disrupting some other grains. The conditions for the selective disruption depend on the grains shape, chemical composition, magnetic properties, angle between magnetic field and the radiation source, etc. This provides a way to get into the basic properties of dust by studying simultaneously the polarization induced by grain alignment and effects of rotational disruption from dust grains. Studying these effects is possible observing the emission and polarization from the media around strong radiation fields, including stars, transients, and Active Galactic Nuclei (AGN). 

Consider the aforementioned effect of the RAT amplitude decrease when the radiation is coming at angles $\psi\approx 90$ degrees. For grains to be well aligned their rotational speed should be a factor of 2 or 3 larger than the speed of thermal rotation. At the same time, for the rotational disruption the required rotational speeds may be a factor of $\sim 100$ larger (see Eq. \ref{eq:w_disr}), depending on the tensile strength. Therefore, one can expect to observe regions of aligned grains that of $\psi$ is close to 90 degrees, while grains subject for RATD for other $\psi$. While predicting the disruption of grains accounting for the angle $\psi$ is important. For instance, $\psi\approx 90^{\circ}$ is the configuration expected for the circumstellar accretion disks with toroidal magnetic field. For grains aligned with the radiation direction, i.e., with $\psi=0$ the disruption may be also decreased as the grains have low probability of having high-J attractors for this typical angle.


A more complicated situation happens when grains are initially aligned with the magnetic field, but a transient changes the angle $\psi$ of the anisotropic radiation. For the aligned grains that for a new $\psi$ have high-J attractor point, the RATD can be fast. A fraction of aligned grains that does not have a high-J attractor point for a new $\psi$ can still be disrupted being spun up towards the high-J repeller. Others will go to the low-J attractor points and will not be disrupted. However, some grains which did not have high-J attractor point for the earlier $\psi$ can get high-J attractor points for a new $\psi$. Such grains can be disrupted, but via a slow disruption process. 

Our calculations show that the increase of the RAT's amplitude decreases the transfer and therefore the process of {\it slow} rotational disruption, whereas the increase of gas density can increase the transfer and the rotational disruption.

Grain magnetic properties significantly affect the process of grain transfer to high-J attractor points. Our study shows that for grains with strong magnetic response, a significant fraction of grains (i.e., upto $50\%$) can move directly to high-J attractor points (see Figures \ref{fig:fhighJ} and \ref{fig:fhighJ_deltam}). This can significantly decrease the timescale of grain disruption. This timescale is directly measurable for the time variable sources of radiation, i.e., SNe explosions, gamma ray bursts. This presents a unique way of testing magnetic properties of dust grains. 

Grains with weak magnetic response, e.g., carbonaceous grains, may, on the contrary, be disrupted with lower efficiency. Such grains are more likely to be aligned in the direction of the radiation with $\psi=0$. For this setting, the fraction of grains aligned with high-J attractors is of $f_{\rm high-J}\lesssim 20\%$ (see Figure \ref{fig:fhighJ}, upper panel). At the same time, the disruption of grains with only low-J attractors is a very inefficient process.

Observations are necessary to test grain alignment and rotational disruption. Grain alignment is only constrained by polarimetric observations, whereas rotational disruption can be constrained by photometric (starlight extinction, thermal and spinning dust emission), polarimetric, and chemical observations because rotational disruption affects the grain size distribution and grain surface chemistry (see \citealt{Hoang:2020} for a review).

In the presence of the depolarization related to the turbulent magnetic field and the variations of grain shape and composition, it is difficult to distinguish the ordinary RAT alignment and the RAT alignment enhanced through magnetic inclusions. Thus, the variations of grain alignment at different distances from the radiation sources as well as in the presence of variations of radiation flux were considered in \cite{LazarianHoang:2019} as a way of probing whether the magnetic inclusions were present. It can be complementary to the approach based on disruption of grains by RATs.

\subsection{Grain alignment and grain surface chemistry}

Our present study is focused on revealing the connection of grain alignment and grain rotational disruption. In terms of physics, this connection arises from the close relation between the RAT and MET grain alignment and the rate of grain rotation. The latter rate, however, is affecting not only the rotational disruption, but also physical and chemical processes taking place over the grain surface.

\cite{HoangTram:2020} identified grain rotation arising from RATs as an important factor of removing grain ice mantles, while \cite{HoangTung:2019} demonstrated that chemical processes on the grain surfaces can be significantly modified due to the centrifugal forces that act on the molecules adsorbed on the grain surface. \cite{Hoang:2019b} studied the effect of rotation on hopping and segregation of ice species in the ice mantle. Naturally, METs can induce similar effects and this can be a reason of removing ice mantles and change of the chemical processes that are induced by shocks. The effect of METs seems promising particularly in cold environments where the numerous complex organic molecules are detected, which is unexpected due to the lack of radiation to warm up the ice mantles (see \citealt{Herbst:2009} for a review). 

Similar to the grain disruption that we described in the present paper, the processes above can be significantly different depending whether the grains are aligned and whether they are aligned with high-$J$ attractor points or only low-J attractors. For grains aligned with high-J attractors, the effect on grain rotation on surface chemistry in star-forming regions and PDRs is efficient because grains at the low-J attractor are moved to the high-J attractor in a short timescale, much shorter than the dynamical timescale (see Section \ref{sec:stable}). However, for grains aligned with only low-J attractor points, its effect on surface chemistry is expected to decrease, depending on the time fraction that grains spend on the rapid rotation between the low-J attractor and high-J repellor points (see Figure \ref{fig:map_lowJ_random}).

\subsection{Consequences for astronomical "standard candle" research}

Measuring distances in astronomical research is frequently based on using the  "standard candle" approach, i.e., using the astrophysical sources with robust well known properties. A well known example of "standard candle" that is used for measuring the Hubble constant is SNe Ia. 

Naturally, the variations of dust properties on the line of sight from the source to the observer introduce uncertainties that are being accounted for by using dust models. The present study testifies that accounting for the dynamics of dust alignment and dust disruption can be an important factor that must be considered for high precision measurements involving the "standard candle" approach. 

The alignment changes not only the polarization from the signal, but also the extinction.  Indeed, grains aligned with the magnetic field perpendicular to the line of sight have less extinction compared to the case of randomized grains. The effects of grain disruption have even more serious consequences for the extinction. The grain fragments have both different total extinction and differentially change the extinction for different wavelengths. 
Accounting for the grain disruption and the alignment of grains fragments by SNe Ia was performed in \cite{Giangetal:2020} and provided the important predictions on the evolution of the properties of dust in the vicinity of SNe. Our present study adds to this an additional parameter to account for, namely the angle $\psi$ between the direction of the magnetic field and the radiation direction. Indeed, our study shows that the disruption of grains can be significantly affected the angle $\psi$ as well as the grain composition (see Section \ref{sec:graincomp}). These two parameters are expected to introduce the additional dispersion of the measurements obtained with SNe Ia and accounting for them can contribute to resolving the ongoing controversy on the value of the Hubble constant. Indeed, analysis of SNe observational data by \cite{Brout:2020} and \cite{Uddin:2020} suggest dust in the host galaxies as a potential factor for the large dispersion in the Hubble constant. 

\subsection{New physical processes and resolved problems}

Our research revealed a process of reduced grain rotational disruption as the intensity of radiation increase. This process is intrinsically related to the properties of RATs that tend to move the majority of grains from an isotropic distribution to the low-J attractor point. In addition, we revealed that the number of grains that are moved to high-J attractor point increases for the alignment with respect to ${\bf B}$ as the magnetic susceptibility of grains increases. 

Another interesting effect described in our study is the action of the torques of different nature against each other and thus limiting the grain disruption. For instance, in the case of METs aligning grains in the low-$J$ attractor points can act against the spin-up induced by strong random torques. Such an alignment can extend the time-scale of the grain disruption up to a several $\tau_{\rm rand}$. Another example is the action of Purcell's torques on the grains aligned at high-$J$ attractor points. If the Purcell's torques are weaker than the value of RATs at the high-$J$ attractor point, the increase of $J$ is expected for a half of the grains at high-$J$ alignment, while for the other half the value of $J$ decreases. This can prevent the disruption of 1/2 of the grains over the time of resurfacing of the grain, the latter can be comparable to $\tau_{\rm damp}$ (\citealt{SpitzerMcGlynn:1979}; \citealt{LazarianDraine:1997}). 

We also noticed that the evolution of $q^{\rm max}$ is another effect that can affect the disruption processes. Indeed, if for a given setting the grains for a range of $q^{\rm max}$ get aligned at high-J attractors and be destroyed, this would result in the preferential selection of grains with only low-J attractors. Potentially, this process can act to destroy grains with high magnetic response in general ISM. At the same time, in accretion disks, grains aligned with toroidal magnetic field experience reduced torques due to $\psi\sim 90^{\circ}$ (see Figure \ref{fig:Jmax_angle}). For this setting the disruption of grains with magnetic inclusions may be reduced compared to the grains for which $\Omega_k>\Omega_B$ and which can be aligned at low-J attractors.

The process of randomization of grains arising from the variations of the electric dipole moment was a serious puzzle for the RAT theory that was  developed disregarding the process of anomalous randomization. Nevertheless the theory provided predictions that well corresponded to observations (see \citealt{Anderssonetal:2015}). In the present paper we have resolved this puzzle, namely, we have shown that the anomalous randomization is a subdominant process (see Section \ref{anomalous}).

\subsection{Synergy between MET and RAT studies}

METs are the major source of grain alignment for grains in mechanical flows. Our present study identified new mechanisms of MET alignment, photoemission torques (PT) and H$_2$ torques (H2T), which arise when the grain are subject to radiation and atomic hydrogen anisotropic flows. Unlike RATs, METs do not have the requirement of grain size to be of the order of the radiation wavelength, which ensures that METs have their niche for inducing both alignment and disruption. 

There have been much less research of METs compared to RATs and the MET research relies on the analogy between the MET and RAT torques established on the basis of AMO and supported by the subsequent studies.
We expect the properties of RATs and METs to be similar with METs both aligning and disrupting grains whenever there is a significant relative motion of dust and grains. Shocks provide the most natural setting for METs to act. 

The difference between METs and RATs arise from the fact that the gas collisions sample not the helicity of the entire grain, but only the helicity of the facet of the grain that is exposed to the directed gaseous flux. As the grain size increases, we, however, expect to see more analogy between the properties of METs and RATs. Indeed, in the limit of the grain size $a$ much larger than the characteristic wavelength of radiation $\lambda$, the interaction of photons with the grain surface is expected to be more similar to the interaction of atoms. In this limit, the results obtained with numerical studies of METs become applicable to RATs acting on large $a\gg \lambda$ grains

Although the functional dependence of METs deviates much more from AMO than that for RATs, the physical essence and the main global features of the two types of alignment are similar. Both RATs and METs can align grains in low-J and high-J attractor points. 
METs, similar to RATs can disrupt grains moving them to high-J points and decrease grain disruption by moving grains into low-J attractors. For instance, one may expect that some grains that can be spun-up and rotationally disrupted by stochastic torques, be aligned by METs at low-J attractors and survive. 

When grains are disrupted, for both processes the grain fragments are likely to be brought initially at a high-J alignment state. The further evolution of the fragments depends on whether this state corresponds to a repeller or an attractor. For both RATs and METs, the presence of magnetic inclusions increases the rate of alignment and the disruption efficiency. As a result, we expect that in shocks METs may preferentially further disrupt grain fragments with magnetic inclusions, producing bare magnetic grains eventually.

\begin{table}[]
    \centering
    \caption{Summary of alignment and disruption by RATs and METs}
    \begin{tabular}{l l l}\toprule
    {\it Feature} &   {\it Alignment}  & {\it Disruption} \cr
    \hline\cr
    Superparamagnetic Grains & & \cr
    ~~~Fast & Imperfect & Imperfect \cr
    ~~~Slow & Perfect & Perfect \cr
    \hline\cr
    Paramagnetic Grains &  &  \cr
    ~~~High-J &  \cr
    ~~~~~~Fast & Imperfect & Imperfect\cr
    ~~~~~~Slow & Perfect & Perfect \cr
    ~~~Low-J & Inefficient & Inefficient\cr
    \hline\cr
    Diamagnetic Grains & & \cr
    ~~~High-J & &  \cr
    ~~~~~~Fast & Imperfect & Imperfect\cr
    ~~~~~~Slow & Perfect & Perfect \cr
    ~~~Low-J & Inefficient & Inefficient\cr
    \hline
    \multicolumn{3}{l}{{\it Notes}: Fast (slow) process: timescale shorter than}\cr
   \multicolumn{3}{l} {(greater than) a damping time.}\cr
	\hline\hline
    \end{tabular}
    \label{tab:summary}
\end{table}

Table \ref{tab:summary} summarizes our results on the MET and RAT grain alignment and rotational disruption for different grain properties. The alignment and disruption depends closely on whether grain alignment implied by RATs and METs exist high-J and low-J attractors. Note that low-J attractors are not stable under gas randomization, and its notion here is inferred from only RATs and the gas damping, excluding gas randomization. For the case of alignment with high-J attractor, both alignment and disruption are efficient. For the case grains aligned only on low-J, rotational disruption does not occur, but the degree of alignment is rather low, $\sim 10-20\%$.

\subsection{Progress in understanding grain dynamics and existing uncertainties}

The grain alignment research has a long history (see \citealt{Lazarian:2003}) and the corresponding studies resulted in many unexpected discoveries. Some of them include the new physical effects, e.g.  Barnett relaxation (\citealt{Purcell:1979}), nuclear relaxation (\citealt{LazarianDraine:1999b}, see also Appendix \ref{app:D}), resonance magnetic relaxation (\citealt{LazarianDraine:2000}). More recent advances in understanding of grain dynamics include the analytical description of RATs in LH07, understanding of the role of high-J and low-J attractors, as well as stabilization of high-J attractors by magnetic dissipation (\citealt{LazarianHoang:2008}; \citealt{HoangLazarian:2016a}), the transport from low-J to high-J attractor induced by gaseous bombardment (\citealt{{HoangLazarian:2008},{HoangLazarian:2016a}})) as well as the stabilization of low-J attractor point by pinwheel torques (\citealt{HoangLazarian:2009b}). These and earlier classical studies form the foundations of modern understanding of grain alignment and grain disruption.

It was based on the understanding of RATs, that the process of RAT disruption (RATD) was introduced in \cite{Hoangetal:2019}. RATD has provided a completely new insight into the evolution of astrophysical dust with its numerous consequences being evaluated actively (see \citealt{Hoang:2020} for a review). Additional effects of RATs include a very exciting new avenue of the research is the changes of grain surface chemistry induced by fast grain rotation. 

Through appreciating the role of grain helicity for grain alignment, it was possible to introduce also mechanical torques acting on irregular dust grains, i.e., METs (\citealt{LazarianHoang:2007b}). Although less well studied compared with RATs (although see \citealt{DasWeingartner:2016}; \citealt{Hoangetal:2018}), METs have proven to dominate the stochastic mechanical torques (see \citealt{Gold:1952}) and are expected to be stronger than RATs for special astrophysical settings, e.g., small grains $a\ll \lambda$ in shocks. Our current study reveals additional types of METs, e.g., METs arising from the photoemission. In analogy with RATD we introduce in this paper METD and expect this process to be of major significance for interstellar dust evolution. In analogy to the effects of RATs on interstellar grain surface chemistry, we may also expect similar changes induced by METs.  

This work capitalizes on the earlier research on the physics of RAT and MET alignment. The present paper demonstrates a very strong connection between grain alignment and rate of grain rotation induced by RATs and METs. An important advance achieved in the present study is our understanding that the variations of grain electric dipole moment do not induce any efficient randomization (see Appendix \ref{anomalous}). This effect could potentially change the entire picture of RAT and MET alignment as well as RATD and METD processes. We also demonstrated a new effect of suppression by intensive radiation of the transport of grains from low-J to high-J attractors and described the importance of this effect for decreasing both grain alignment and RATD/METD. In addition, our study reveals significant differences in alignment and rotational disruption of grains with different magnetic responses. For instance, we found more efficient disruption of grains with magnetic inclusions and much less efficient disruption of carbonaceous grains that demonstrate only weak magnetic response. 

We, however, have to accept the existing uncertainties. For instance, we accepted in Section \ref{purcell} that the existing ambiguities in the process of thermal flipping of irregular grains prevents us from reaching solid quantitative conclusions related to the role of Purcell pinwheel torques for the stabilization of grain rotation at the low-J attractor point. In particular, at this point, it is difficult to evaluate at which grain sizes the pinwheel torques are getting important. The corresponding problem requires further studies. The progress in pinwheel torque understanding will help to evaluate better both the efficiencies of grain alignment and grain disruption. It also should help to our understanding of the alignment of grains with low degree of internal relaxation (see Appendix \ref{app:D}). Such grains according to \cite{HoangLazarian:2009b} can align with long axes parallel to magnetic field, i.e., 90 degrees to the classical alignment with the grain axis of maximal moment of inertia being along magnetic field (see Appendix \ref{app:E}). The pinwheel torques are important for such "wrong" alignment to stabilize it.

Last, but not the least, one should not overestimate the accuracy of AMO while evaluating the alignment and the corresponding processes of disruption for realistic irregular grains. One should keep in mind that AMO was used in LH07 to successfully explain the physical process of RAT alignment, i.e. the process that was mysterious before its introduction. The fact that a simple mechanical model happened to be able also to provide a good quantitative description of RATs for a limited sample of the irregular grains at a range of studied wavelengths was a pleasant surprise. At the same time, LH07 already reported that AMO requires modifications in order to reproduce the properties of RATs e.g. at the UV wavelengths. These modifications could be achieved by the change of the angle of the mirror employed in AMO, but this direction was not pursued as it was depriving AMO of its simplicity. More recent research with a large ensemble of irregular grains in Herranen et al. (2019, 2020) shows both the remarkable utility of AMO for explaining the alignment, but also demonstrates the limitations of this simple model. The application of AMO for METs (Lazarian \& Hoang 2007b) is limited mostly to identifying the grain helicity as the major driver for MET alignment. The structure of torques obtained numerically in Hoang et al. (2018) is much more complicated compared to simple dependence of AMO (see Appendix A). Therefore, for studies of RATs and RATD, as well as METs and MeTD, the physical picture of the processes obtained by AMO should be augmented by the corresponding numerical studies of torques acting on ensembles of irregular grains.  

As it stands, our present study provides a number of predictions related in relation to the rate of grain disruption, as well as peculiarities of grain surface chemistry with the grain alignment. Testing these predictions can provide the insight into grain chemical composition and grain structure.  

\subsection{Selective grain disruption and its effect on grain alignment}

The efficient rotational disruption of grains aligned at high-$J$ attractors discussed in the paper induces a new important effect of grain selection. It follows from our study that the rotational disruption is forcefully destroying the grains with $q^{max}$ corresponding for the given setting to high-$J$ attractors. This leads to the selection of the grain shapes have for the given $\psi$ have only low-$J$ attractor points (see Figure \ref{fig:lowJ_highJ}.
It is also obvious that if grains have enhanced magnetic susceptibility, they are more likely to be aligned with high-$J$ and therefore are more likely to be disrupted. As a result, grains with magnetic inclusions, e.g., with small superparamagnetic clusters (\citealt{Morrish:2001}), are more likely to be disrupted. This process tends to move inclusions to small grains that are not rotationally disrupted in the given settings. In larger grains, the inclusions will tend to survive in more solid compact grains. 

We can predict that this selection can {\it decrease} the degree of alignment in the vicinity of bright sources as well as to decrease the magnetic susceptibility of grains corresponding to larger sizes.

\section{Summary}\label{sec:summary}

The problem of grain alignment is of longest standing, more than 70 years, while the mechanism of rotational disruption of dust grains has been identified in \cite{Hoangetal:2019}. The former problem still has unclear features, but a number of theoretical prediction has been successfully tested with the existing observational data. We demonstrate in the present paper that the problems of grain alignment and grain rotational disruption are intrinsically connected and the present-day theory of grain alignment can guide our quantitative understanding of the disruption. Similarly, as the modification of the chemical processes induced by rapid grain rotation (see \citealt{HoangTung:2019}) also depends on the grain rotational rate, the theory of grain alignment can be used to understand catalytic reactions on grain surfaces. 

In the paper above, we considered the alignment and disruption of astrophysical dust grains by RAdiative Torques (RATs) from AMO and MEchanical Torques (METs). In addition, we discuss the pinwheel torques introduced in \cite{Purcell:1979} and their effect on the RAT and MET alignment.

While RATs have proven to be a powerful agent for controlling the alignment and spin-up of astrophysical dust, their mechanical analogs, i.e., METs, require more studies. Nevertheless, METs also look promising in terms of spinning up grains in astrophysical conditions. Rotational disruption induced by both RATs and METs is controlled by the same physics that govern the alignment. In the paper above we have shown that the processes of RAT/MET alignment and RAT/MET disruption are intrinsically connected. Our principal findings are summarized as follows: 

\begin{itemize}

\item The rotational disruption of grains induced by RATs/METs occurs for grains aligned with high-$J$ attractor points and does not occur for low-$J$ attractor points. This makes the grain disruption selective and opens ways of testing the predictions of the grain alignment theory by studying grain disruption.  

\item We describe the {\it fast disruption} process for grains that are directly driven to the high-J attractor by RATs, and {\it slow disruption} for grains that are transported from the low-J to high-J attractor by gas collisions. We calculate $f_{\rm high-J}$ for different grain shapes and magnetic properties using the AMO. In case of RATs the fast disruption is applicable for dust surrounding transient sources, for METs this process can take place in shocks. The slow disruption is applicable for interstellar and circumstellar dust. 

\item The increase of magnetic relaxation (e.g., via iron inclusions) extends the parameter space corresponding to the alignment with high-$J$ attractors (see Figure \ref{fig:lowJ_highJ}) and increases the fraction of grains moving directly to the high-J attractors (see Figures \ref{fig:fhighJ} and \ref{fig:fhighJ_deltam}). This results in more efficient {\it fast disruption} and increases the rate of {\it slow disruption}.

\item Numerical calculations for an ensemble of grain shapes and compositions by \cite{HerranenLazarian:2020} show that RATs can induce grain alignment with high-J attractor of $f_{\rm high-J}\gtrsim 50\%$ for ordinary paramagnetic grains. This implies efficient fast alignment and rotational disruption of a significant part of ordinary paramagnetic grains by RATs. 

\item If grains are aligned with magnetic field, the efficiency of rotational disruption of grains depends on the angle $\psi$ between the magnetic field an the direction of the radiation flux or gas flow. For instance, it decreases $\psi$ close to $90^{\circ}$ as the consequence of the decrease of RATs/METs in amplitude.

\item The sudden increase of the radiation flux or of the inflow of bombarding atoms can induce the realignment of grains from the direction given by magnetic field to the direction of the flux or the inflow, i.e. induce the transition from B-alignment to k-alignment. In this situation only the grains that for k-alignment have high-$J$ attractor points will be rotationally disrupted and their disruption will take the time of the order or longer than the damping time. This is the consequence of the initial moving of the absolute majority of grains into low-$J$ attractor points. This process can mitigate the rotational disruption of grains in shocks. 

\item Pinwheel torques can increase the efficiency of {\it fast} rotational disruption by RATs through increasing the angular momentum of high-J attractors as well as increasing the grain angular momentum on their way to low-J attractors in the absence of high-J attractors. At the same time, pinwheel torques may reduce the efficiency of {\it slow disruption} by stabilizing the low-J attractors, although it is uncertain due to efficient flipping rate at low-J.

\item The timescale of fast disruption decreases with increasing the radiation intensity, but the timescale of slow alignment/disruption increases. Thus, counterintuitively, the time-scale of {\it perfect} grain disruption can be longer for a fraction of grains closer to the time-dependent source of radiation. This is the consequence of  stronger RATs delaying the transport of grains from the low-J to high-J attractor. At the same time, increasing the local gas density can increase the slow disruption.

\item Turbulence can accelerate grains and this changes grain dynamics, e.g. changes the alignment of carbonaceous grains by 90 degrees, but it does not introduce anomalous randomization of grains that would change the physics of the rotational disruption. The disruption of carbonaceous grains may be reduced due to lower probability of them being at high-J attractor points.

\item{} The selective rotational disruption of grains aligned at high-$J$ attractor points reduces percentage of perfectly aligned large grains, especially those with high magnetic response near the bright sources. In some cases, counterintuitively, this even can decrease the observed alignment efficiency near bright sources.

\item{} We provide predictions for the observational testing for grain rotational disruption and believe that the corresponding tests can shed light on the key processes of grain alignment, i.e., high-J or low-J attractors, as well as grain properties (geometry and magnetic susceptibility).

\end{itemize}

\acknowledgments
We thank the anonymous referee for helpful comments that improved the paper, in particular, the presentation of our results. A.L. acknowledges the support of the NSF grants AST 1715754, 1816234 and NASA ATP AAH7546. A.L. also thanks David Spergel for his pointing out to the connection of the dispersion of the Hubble constant measurements with the effects discussed in the paper. T.H acknowledges the support from the National Research Foundation of Korea (NRF) grant funded by the Korea government (MSIT) through Mid-career Research Program (2019R1A2C1087045). The Flatiron Institute is supported by the Simons Foundation.

\appendix

\section{A: Physical understanding of RAT alignment}
\label{sec:physical}
The RAT alignment is easy to understand assuming that the angular momentum ${\bf J}$ is aligned with the grain axis of the maximal moment of inertia $\ma_1$. This is the natural arrangement in the situation when the rate of internal dissipation within grains is faster than the rate of randomization of angular momentum in grain axes. The former is induced by a number of processes that include inelastic relaxation (\citealt{Purcell:1979}; \citealt{LazarianEfroimsky:1999}), Barnett relaxation (\citealt{Purcell:1979}; \citealt{HoangLazarian:2009b}) and nuclear relaxation (\citealt{LazarianDraine:1999a}). It is easy to see that if ${\bf J}$ that is parallel to $\ma_1$ makes an angle $\xi$ with the axis of fast precession, the precession-averaged component of RAT torque that changes the angle $\xi$ is $\sim \sin\xi$, i.e., it gets to zero when the grain axis $\ma_1$ is perfectly aligned with the precession axis. Therefore, for grains angular momentum along the precession axis, the aligning torque component vanishes. Therefore, the positions of $\xi$ equal to zero or $\pi$ correspond to the stationary points. These points can be stable, corresponding to attractor points or unstable, corresponding to repeller points. The alignment can happen with long grain axes parallel to the magnetic field direction if the grain precession about the magnetic field is the fastest. It was shown in LH07 that the radiation direction constitutes the axis of alignment if the precession induced by the RAT component $Q_{e3}$ is the fastest.

\begin{figure}
\includegraphics[width=0.5\textwidth]{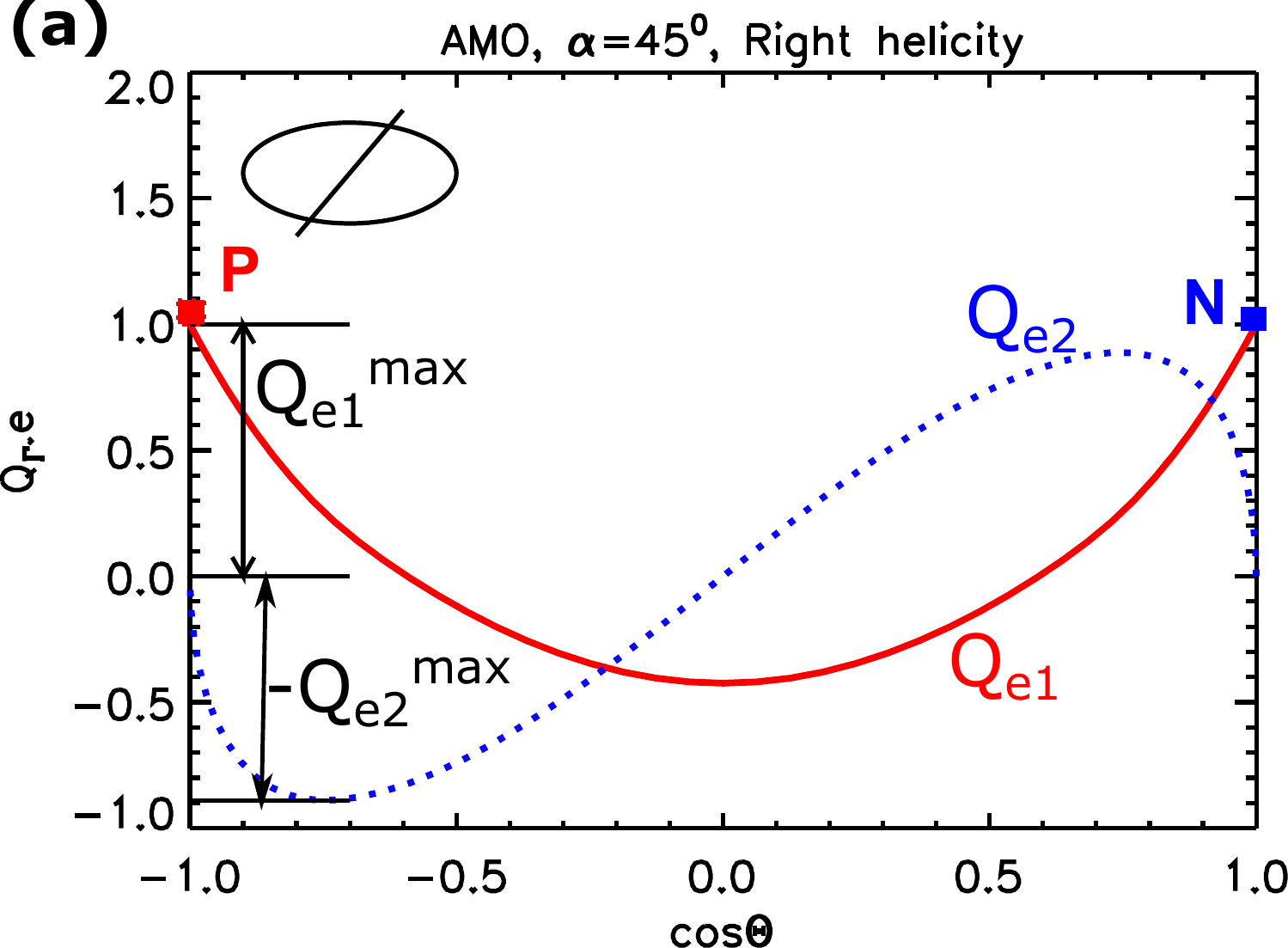}
\includegraphics[width=0.48\textwidth]{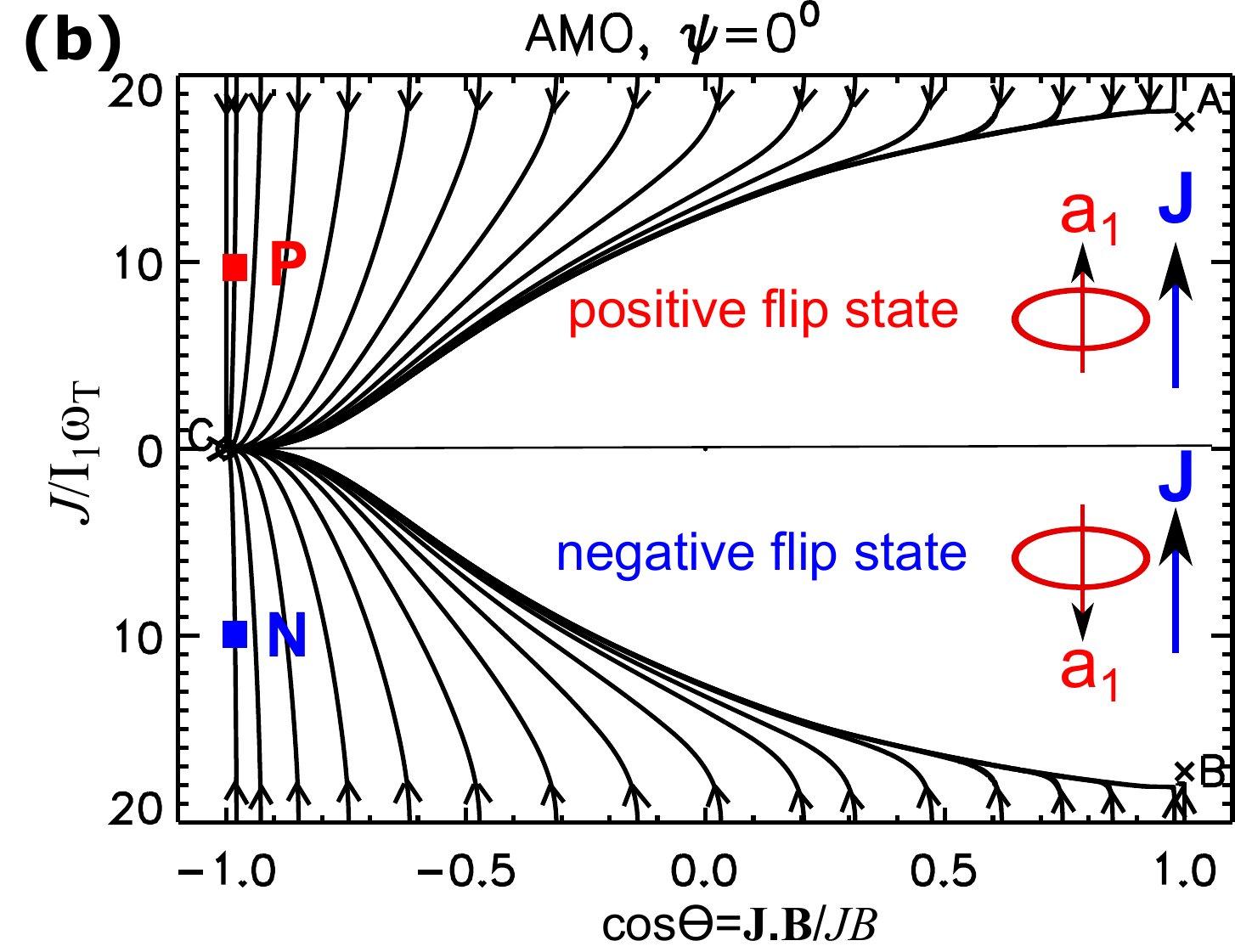}
\caption{Panel (a): Normalized RAT components, $Q_{e1}$ and $Q_{e2}$, as function of $\Theta$ from the AMO. Panel (b): Phase trajectory map for alignment of grains starting from fast rotation at $J=20I_{1}\omega_{\rm T}$. The phase trajectory map of grain alignment without a high-J attractor and only low-J attractors (C) where arrows indicate the time-evolution of grain orientation under the effect of RATs. Points A and B are repellor points. Grains initially have a positive and negative flipping states. Gas randomization is disregarded. From \cite{LazarianHoang:2007a}.}
\label{phase}
\end{figure}

A phase trajectory of the AMO is illustrated in the lower panel of Figure \ref{phase} (Panel (b)), where both low angular momentum, i.e., low-$J$, stationary points, as well as high angular momentum, i.e., high-$J$ stationary points are shown. For the given setting the high-$J$ stationary points A and B are the repeller points, and therefore the trajectories of all grains that we assumed to start at $J/I_1 \omega_{T}=20$, where $\omega_{T}$ is the grain angular momentum corresponding to the grain thermal rotation corresponding the temperature of ambient gas, converge at the point low-J attractor point C. Formally, the angular momentum at this point is zero, but in reality the value of angular momentum at this point gets a bit less than $I_1 \omega_{T, grain}$, where $\omega_{T, grain}$ is the angular velocity corresponding to the grain temperature. This happens due to the fact that dissipation processes get inefficient for the thermal equilibrium conditions and the direction of ${\bf J}$ get randomized with respect to $\ma_1$. The upper and lower parts of the phase map correspond to ${\bf J}$ being parallel or anti-parallel to $\ma_1$. The grain can transfer from the parallel to anti-parallel state by thermal flipping (\citealt{LazarianDraine:1999b}; \citealt{Weingartner:2009}; \citealt{HoangLazarian:2009b}; \citealt{KolasiWeingartner:2017}). However, the phase trajectories are symmetric for the parallel and antiparallel states, which is an intrinsic property of grain helicity.

The requirement of the trajectories to be symmetric with respect to the horizontal axis are obvious when AMO is considered. Indeed, the flip corresponds to the rotating the grain over $\pi$ around the axis $a_3$ in Figure \ref{AMO}. As a result of this turn, the mirror also turns over $\pi$. The plane of the mirror preserves its direction in space with the only change being that the front side of the mirror becoming the back side and vise versa. If the reflection of light by the front and back side of the mirror is the same, the flip does not change the value and direction of RATs. Numerical calculations using DDSCAT (LH07) and T-matrix (\citealt{Herranenetal:2019}) for many grains shapes confirm that RATs do not change with the flip of the grain, confirming the conclusions obtained using AMO. 

\section{B: Stochastic alignment by mechanical torques}
\label{app:A}

Stochastic process of mechanical alignment of thermally rotating grains proposed by \cite{Gold:1952} who considered a flow of atoms bombarding a grain, which shape was approximated by a stick. A more elaborate studies (\citealt{Purcell:1969}; \citealt{RobergeHanany:1995}; \citealt{Lazarian:1994}; \citealt{Lazarian:1995a}) provided qualitative description of the process. In particular, in the latter papers, ubiquitous magnetohydorodynamic turbulence (see \citealt{BeresnyakLazarian:2019}) was identified as a source of relative grain-gas motion. The later studies, indeed, showed that turbulence can accelerate grains to supersonic speeds with respect to the ambient gas. Such speeds are required for the Gold process to be efficient.  

The \cite{Purcell:1979} torques that spin up grains to suprathermal rates (see Section \ref{purcell}) significantly diminish the efficiency of the Gold process. However, the mechanisms of {\it cross-sectional} or {\it cross-over} alignment proposed for such fast rotating grains (\citealt{Lazarian:1995a}; \citealt{LazarianEfroimsky:1999}).

The expected niche for importance of the mechanical stochastic torques (MST) shrunk after it became clear that for typical interstellar conditions they cannot compete with RATs. After the introduction of METs it became clear that the latter regular torques are bound to be more important for the alignment compared to the MSTs. The cross-sectional alignment found its narrow niche for the alignment of nano-particles subjected to UV radiation (\citealt{HoangLazarian:2018}). 

\section{C: Calculation of anomalous randomization}
\label{anomalous}

W06 considered both thermal flipping of grains and variations of the grain charge as the reasons for the causes of grain electric moment change $p_{el}$. These changes induce the fluctuations torque $\Gamma_{el}=p_{el}\times E_{\rm ind}$ that act to randomize charged grains. The effect of the these randomizing torques depend on whether the time-scale of their action is longer or shorter than the period of grain precession. The precession frequency is
$\delta \theta \approx \Gamma_{el}/J\tau_{fluct}$. 

Unlike W06 and JW09, we distinguish the alignment with respect to electric field and the alignment arising from the grain motion with respect to magnetic field and the precession with respect to magnetic field. As we discussed in Section \ref{precession} these are two very different types of alignment resulting in orthogonal directions of polarization. Therefore we disagree with W06 and JW09 who applied their process to carbonaceous grains assuming that these grains are also aligned with respect to the ambient magnetic field. Instead, we limiting our discussion to the silicate grains providing our simple derivation of the expressions in W6 and then providing our arguments why we believe that the anomalous randomization is, in fact, inefficient. 

For silicate grains the factor $\aleph$ given by Equation (\ref{aleph0}) is approximately $0.05$ and therefore the precession is taking place with respect to magnetic field with electric field just slightly modifying grain precession about magnetic field. W06 noticed, however, that the component of the electric moment $p_{el}$ can vary both in the grains stochastically losing and gaining individual electrons. W06 considered the flip of the direction of $p_{el}$ as a result of this process. Another possibility for grain to flip the direction of $p_{el}$ is thermal flipping. The latter process introduced in \cite{LazarianDraine:1999a} happens for grains with lower angular value momentum as a result of coupling of rotational and vibrational degrees of freedom (see more in \citealt{Weingartner:2009}, \citealt{HoangLazarian:2009b}, \citealt{KolasiWeingartner:2017}). As a result of grain flipping, the change of the $p_{el}$ is also taking place. These two processes were discussed in W06 as a source of the randomization grain precession around magnetic field. 

W06 considered two regimes of randomization. In the first regime, the significant variation/flipping of the electric moment $p_{el}$ happens over time $\tau$ which is much shorter than the grain Larmor precession time $\Omega_B^{-1}$. {\it For the stochastic}\footnote{Later we explain why the process is not a random walk in reality.} change of $p_{el}$ the change perturbation in the angle between the grain angular momentum ${\bf J}$ and the magnetic field direction can be estimated as 
\begin{equation}
\delta \theta_{\rm fast} \approx \aleph \Omega_B \tau.
\label{theta_fast}
\end{equation}
W06 assumed that the fluctuations of $\delta \theta_{\rm fast}$ are accumulated as a random walk. Then for the fast fluctuation regime that we consider the number of steps over one grain precession is $\sim (\tau \Omega_B)^{-1}$ and during one precession time the direction of ${\bf J}$ deviates by 
\begin{equation}
\theta_{\rm period}\sim \Omega^{-1}  \tau \aleph^2 \Omega_B^2.
\end{equation}
To randomize the direction of ${\bf J}$ the required number of steps can be found from $N_{steps, fast} \theta_{\rm period}^2 \sim 1$, which gives 
$N_{steps, fast}\sim 1/\theta_{\rm period}^2$. These steps takes time 
\begin{equation}
\tau_{anom, fast}\sim N_{steps, fast} \Omega_B^{-1}\sim \tau^{-1} (\aleph \Omega_B)^{-2},
\end{equation}
which coincides with the expression obtained in a more complicated way in W06.

In the opposite regime when $\tau$ is significantly longer than $\Omega^{-1}$ within the approach in W06 it is assumed that the randomization happens at the last precession of the grain during which the flipping of $p_{el}$ takes place. Therefore the time of the action of perturbing torque $\aleph\Omega$ on the grain is $\sim \Omega^{-1}$ and the corresponding variation in ${\bf J}$ direction is
\begin{equation}
\delta \theta_{\rm slow}\sim \aleph \Omega \Omega^{-1}.
\end{equation} 
Similarly to the case of fast flipping the number of steps $N_{total, slow}$ can be estimated from the relation $N_{total, slow}(\delta \theta)^2\sim 1$, i.e.
$N_{total, slow}\sim \aleph^{-2}$. The randomization time is
\begin{equation}
\tau_{anom, slow}\sim N\tau \sim \aleph^{-2}\tau,
\end{equation}
which transfers into the expression in W06 for $\aleph^{-2}\gg 1$. 

The most significant assumption that we, following W06, adopted in the derivations above was that the process of flipping of $p_{el}$ can be described as a random walk process. It is clear, however, that this is an incorrect assumption. Indeed, consider thermal flipping. The direction of the electric moment
 $p_{el}$ during the flip changes to the opposite. This constitutes one step in $\delta \theta_{\rm fast}$, for instance (see Equation (\ref{theta_fast}), which we can assume to be in the positive direction, i.e. increasing the precession angle of ${\bf J}$. For the random walk the next step should be equally probable in both directions, i.e. next $\delta \theta_{\rm fast}$ should be possible to be positive and negative. This, however, cannot happen as the next flip of $p_{el}$ can only get it into its original orientation, thus determining the direction of torques and the direction of change of ${\bf J}$. In other words, the next change of $\delta \theta_{\rm fast}$ is necessarily negative, compensating the deviation at an 

The nature of the randomization due to flips may be generalized for more gradual variations of $p_{el}$ due to changes arising from attaching of individual electrons to the grain. For instance,  the arguments in W06 may be modified if one presents $p_{el}$ as a sum of the mean $p_{el, mean}$ and the fluctuating $\delta p_{el}$.  Instead of considering the regular precession in the electric plus magnetic field for $\aleph\ll 1$ one use the original equation in W06 for change of the precession angle:
\begin{equation}
\label{dtheta_dt}
\frac{d\theta}{dt} = - \Omega_B \aleph \sin (\phi + \omega_{\rm gyro} \, t)~~~,
\end{equation}
where $\omega_{\rm gyro}$ is a slow gyration frequency given by Equation (\ref{gyro}). It is evident from Equation(\ref{dtheta_dt}) that the changes in $\aleph$ 
do not present a random walk as of $\delta p_{el}$ cannot change in the random walk fashion. 

The variations of the magnitude of angular momentum $J$ that happen simultaneously with the variations of $p_{el}$ can bring the true stochasticity
in the variations of $\delta \theta$. However, these variations happen over the time scale of grain damping or acceleration and this makes this
process subdominant in most settings. 

Another anomalous randomization process is discussed in \cite{LazarianHoang:2019} and it is related to the thermally induced magnetic fluctuations in grain material as the grain is being aligned by RATs with respect to the radiation direction (see Section \ref{precession}). In this process it is the grain magnetic 
moment that fluctuates and its interactions with the magnetic field induces the fluctuating torques. However, this process of anomalous magnetic 
randomization is also inefficient due to the same arguments that we discussed above.

\section{D: Internal relaxation and internal randomization}
\label{app:D}

The processes that we described earlier are the processes of the alignment of grains with respect to the ambient magnetic field or radiation direction.
A wobbling grain is subject to internal relaxation. The inelastic relaxation being the simplest relaxation process (\citealt{Purcell:1979}). For instance, for an oblate graphite grain \cite{LazarianEfroimsky:1999} obtained:
\begin{equation}
\tau_{inelast}\approx 2.3\times 10^{-1} a_{-5}^{5.5} X_{10}^{3/2} \left( \frac{\rho}{3 \g\cm^{-3}} \right)^{1/2} {\rm yr}, 
\end{equation}
where $X_{10}$ is the grain axis ratio normalized by 10.

A rotating paramagnetic grain induces magnetization of its material due to the Barnett effect and the this magnetization
 is aligned with the grain angular momentum (\citealt{Dolginov:1976}). It was discovered by \cite{Purcell:1979} that as the grain wobbles around
its angular momentum ${\bf J}$, the dissipation takes place due to the change of the direction.

The Barnett relaxation time for an oblate grain with of dimensions $2a\times 2a \times a$ was estimated in \cite{Purcell:1979}
\begin{equation}
\tau_{\rm BR}=\frac{\rho a^2 \gamma^2}{K(\omega) \omega^2}\approx 4\times 10^{7} \left(\frac{10^5~\s^{-1}}{\omega}\right)^2 {\rm s},
\end{equation}
which is a short time compared to the characteristic timescales involved in the alignment of interstellar grains.\footnote{It is wrong to assume that grains must have ${\bf J}$ and ${\bf a}$ {\it perfectly} aligned. It was noted by \cite{Lazarian:1994} that for thermally rotating grains, the alignment is far from being perfect due to the thermal fluctuations within grain body (see more in \citealt{LazarianRoberge:1997}).} 

\cite{LazarianDraine:1999a} identified another mechanism of internal relaxation related to nuclear spins. The time scale for the nuclear relaxation for a "brick" with dimensions $a\times a\sqrt{3}\times a\sqrt{3}$ is:
\begin{equation}
\tau_{\rm NR}= G_{\rm NR}\hat{\rho}^2 a_{-5}^{7},
\label{NR}
\end{equation}
where
\begin{equation}
G_{\rm NR}=610 \left(\frac{n_e}{n_n}\right) \left(\frac{\omega_d}{\omega}\right)^2 \left(\frac{g_n}{3.1}\right) \left(\frac{2.7 \mu_N}{\mu_n}\right) \left[1+\left(\omega \tau_n \right)^2 \right] \s, 
\label{Gn}
\end{equation}
where $\hat{\rho}\equiv \rho/(2 {\rm g~cm}^{-3})$, $\omega_d$ is the angular velocity of a grain rotating thermally at $T_d=10$K, i.e. $\approx 3\times 10^4$ s$^{-1}$. The nuclear spins with magnetic moments $\mu_n=g_{n}\mu_{N}$ are normalized by the magnetic moment of the proton $\mu_N\equiv 5.05 \times e\hbar/2m_pc=10^{-24}$ erg G$^{-1}$, $\tau_n$ is the time of spin-spin relaxation within the system of nuclear spins.  Compared to \cite{LazarianDraine:1999a} we, following arguments in \cite{LazarianHoang:2019}, use a different asymptotic behavior of the relaxation.

In general, the rate of internal relaxation is the sum of the individual internal relaxation rates, i.e.
\begin{equation}
\tau_{\rm int}^{-1}=\tau_{NR}^{-1} + \tau_{BR}^{-1} + \tau_{inelast}^{-1}, 
\label{intern}
\end{equation}

All these processes act for oblate grains as the magnetization of the grain stays constant in amplitude, but changes its direction within the grain body. An additional relaxation process was introduced in \cite{LazarianHoang:2019} for irregular wobbling grains and it involves the change of the magnetization amplitude.  A study in \cite{LazarianHoang:2019} has shown that the for a limited range of scales this process is dominant. However, as this does not radically change our estimates, for the same of simplicity, we consider only the internal relaxation of oblate grains. 

The internal relaxation significantly changes the processes of alignment and the spin-up. For instance, it was shown in \cite{HoangLazarian:2009b} that the alignment of grains with only low-J attractor points happens with the long grain axes parallel to either ${\bf k}$ or ${\bf B}$ depending whether $\Omega_k$ or $\Omega_B$ are larger. As it was shown in \cite{Lazarian:2020} the alignment of carbonaceous grains can also happen with respect to ${\bf E}$ if $\Omega_E>\Omega_B$, and $\Omega_E>\Omega_k$. Then the alignment will happen still with respect to magnetic field, but the alignment will be perpendicular to that expected for the classical alignment for which $\Omega_E$ is disregarded.

\section{E: "Wrong" alignment}
\label{app:E}

The RAT alignment is significantly affected by the internal relaxation. This was revealed in \cite{HoangLazarian:2009a} where the alignment on the time scales less than the internal relaxation was considered. The condition considered there was $\tau_{\rm int}>\tau_{\rm damp}$ and therefore the internal alignment was dynamically irrelevant. For this setting, it was shown that the alignment with the low-$J$ attractors happens in a "wrong way", i.e. with the long grain axes being aligned parallel to the alignment axes, which for $\Omega_B> \Omega_k$ corresponds to the magnetic field direction. 

The alignment with low-$J$, as we discussed earlier, is unstable and we expect significant fluctuations of the grain axes to take place. This can make the "wrong alignment" difficult to observe via polarization measurements. We, however, want to point out that in the presence of additional Purcell's torques (see Section \ref{sec:pinwheel}) the rotation with about the axis of minimal inertia can be stabilized.

In addition, even $\tau_{\rm int}<\tau_{\rm damp}$ the transient fast RAT alignment can take place on the time $\sim  \Omega_k^{-1}$ as described in LH07. Such an alignment will happen with grain low-$J$ attractor points corresponding to long grain axes along the magnetic field. Similar to the earlier case, Purcell's torques can stabilize the direction of the "wrong alignment". On the time scales larger than $\tau_{\rm int}$ the alignment gets into the conventional regime of the "right alignment", i.e. in the alignment with long grain axes perpendicular to the alignment axis.

The suppression of Purcell's torques is important both from the point of view of their joint action with RATs in terms of grain alignment as well as the grain disruption (see Section \ref{sec:pinwheel}).

\section{F: Grain rotation induced by isotropic and anisotropic torques}
\label{app:F}

Grain can be spun up both by isotropic and anisotropic radiation fluxes. The numerical calculations show that the torques induced by isotropic fluxes are reduced by a factor of $\sim 100$ compared to the torques
arising from unidirectional flow. Therefore, unless the radiation field is highly isotropic, the most important for grain dynamics are the torques arising from anisotropic radiation flux. 

Originally it was considered that RATs are similar to the pinwheel torques that we discuss in the next section (see Section \ref{sec:pinwheel}), i.e. that RATs produce grain rotation that can allow paramagnetic torques to align grains with magnetic field (\citealt{DraineWeingartner:1996}). In fact, RATs can be subdivided into anisotropic and isotropic components. These components arise from the grain interacting, respectively, with anisotropic and isotropic radiation fields. The isotropic RATs are, indeed, similar to the pinwheel torques, while the anisotropic RATs are different.

Later research has revealed a complex dynamics of grains subjected to anisotropic RATs. In particular, it was found that the rate of grain rotation and the degree of grain alignment by anisotropic RATs are interrelated (LH07). On the contrary, the amplitude of isotropic torques does not depend on grain alignment, but the rates of rotation induced by these torques are significantly smaller than those that can be achieved from the action of anisotropic torques.

METs introduced in analogy with RATs in are anisotropic in its nature. The role analogous to the isotropic mechanical torques are played by pinwheel torques introduced by \cite{Purcell:1979} (see Section \ref{sec:pinwheel}). 

For most settings the effect
of isotropic torques RATs is negligible compared to their anisotropic counterparts. Therefore, for the rest of the paper we do not consider torques arising from the isotropic illumination of an irregular grain or due to the mechanical bombardment by atoms of a grain in rest with respect to the gas. The isotropic METs, i.e. pinwheel torques are discussed in the next section.

\section{G: Properties of Pinwheel Torques and resurfacing}
\label{sec:prop_purcell}
The properties of Purcell's torques depend on the properties of grain surface. This surface is subject is resurfacing which can change the amplitude and the direction of the torques. For instance, the catalytic sites of H$_2$ formation can be "poisoned" while new sites can be created. The corresponding {\it resurfacing} process was considered by \cite{SpitzerMcGlynn:1979} who showed that this process can decrease the grain rotational rate if the time scale of resurfacing $t_{\rm resurf}$ is less than the damping time $\tau_{\rm damp}$.\footnote{They also showed that the paramagnetic alignment of grains subject to Purcell's torques is significantly reduced. The latter conclusion was modified later in \cite{LazarianDraine:1997}.}

The time-scale for grain resurfacing $t_{\rm resurf}$ is not well defined with designated laboratory experiments required. In \cite{SpitzerMcGlynn:1979} it was assumed that that $t_{\rm resurf}$ is of the order of $\tau_{\rm damp}$. This, however, is not a necessary requirement and grain shape, as we can see on the example of METs can play also an important role for determining the life-time of pinwheel torques. Indeed, a grain with the uniform surface accommodation coefficient but with a cavity on its surface will interact with the bombarding gaseous molecules similar to the grain with the variation of the accommodation coefficient. However the $t_{\rm resurf}$ in this case would require the change of the grain shape, which can take significantly longer than the $\tau_{\rm damp}$. In other words, the pinwheel torques can be long-lived. The argument against this can be that the the long-lived Purcell's torques should induce the efficient alignment of grains of different sizes, including small grains. This, however, is not what is observations tell us (see \citealt{Anderssonetal:2015}). Therefore, in what follows, we will assume that the $t_{\rm resurf}$ is limited by $\tau_{\rm damp}$. In reality, there can be other factors that prevent small grains from alignment, one of them is fast flipping of sufficiently small grains that we discuss further.


\end{document}